\documentclass[final,1p,11pt]{elsarticle}

\usepackage{epsfig}
\usepackage{array,tabularx,epsfig,mathrsfs,graphicx,rotating}
\usepackage{ifthen}
\usepackage{amsfonts}
\usepackage{ragged2e}
\PassOptionsToPackage{hyphens}{url}
\usepackage[hyphens]{url}
\usepackage{hyperref}
\usepackage{listings}
\usepackage{lineno}
\usepackage{subfigure}
\usepackage{epstopdf}
\usepackage{color}
\usepackage{float}
\usepackage{verbatim}
\usepackage{color,soul}

\usepackage[normalem]{ulem} 
\usepackage{soul} 
\usepackage{amsmath,amssymb}

\let\originallesssim\lesssim
\let\originalgtrsim\gtrsim

\DeclareRobustCommand{\lesssim}{%
  \mathrel{\mathpalette\lowersim\originallesssim}%
}
\DeclareRobustCommand{\gtrsim}{%
  \mathrel{\mathpalette\lowersim\originalgtrsim}%
}

\makeatletter
\newcommand{\lowersim}[2]{%
  \sbox\z@{$#1<$}%
  \raisebox{-\dimexpr\height-\ht\z@}{$\m@th#1#2$}%
}
\makeatother

\hypersetup{
  colorlinks=true,
  linkcolor=blue,
  citecolor=blue,
  urlcolor=blue
}

\graphicspath{{figs/}}

\newcommand{\beq}{\begin{equation}}
\newcommand{\eeq}{\end{equation}}

\chardef\til=126
\newcommand{\GEANTfour} {\textsc{geant4}}
\newcommand{\pythia} {\textsc{Pythia8~}}
\newcommand{\pt}{\ensuremath{p_{\mathrm{T}}}}

\journal{ANL-HEP-149528}

\begin{document}
\hfill ANL-HEP-149528
\definecolor{mygreen}{rgb}{0,0.6,0} \definecolor{mygray}{rgb}{0.5,0.5,0.5} \definecolor{mymauve}{rgb}{0.58,0,0.82}

\lstset{ %
 backgroundcolor=\color{white},   
 basicstyle=\footnotesize,        
 breakatwhitespace=false,         
 breaklines=true,                 
 captionpos=b,                    
 commentstyle=\color{mygreen},    
 deletekeywords={...},            
 escapeinside={\%*}{*)},          
 extendedchars=true,              
 keepspaces=true,                 
 frame=tb,
 keywordstyle=\color{blue},       
 language=Python,                 
 otherkeywords={*,...},            
 rulecolor=\color{black},         
 showspaces=false,                
 showstringspaces=false,          
 showtabs=false,                  
 stepnumber=2,                    
 stringstyle=\color{mymauve},     
 tabsize=2,                        
 title=\lstname,                   
 numberstyle=\footnotesize,
 basicstyle=\small,
 basewidth={0.5em,0.5em}
}

\begin{frontmatter}

\title{
Studies of granularity of a hadronic calorimeter for tens-of-TeV jets  at a 100~TeV $pp$ collider 
}

\author[add3]{C.-H. Yeh}
\ead{a9510130375@gmail.com}

\author[add1]{S.V.~Chekanov}
\ead{chekanov@anl.gov}

\author[addDuke]{A.V.~Kotwal}
\ead{ashutosh.kotwal@duke.edu}

\author[add1]{J.~Proudfoot}
\ead{proudfoot@anl.gov}

\author[addDuke]{S.~Sen}
\ead{sourav.sen@duke.edu}

\author[add2]{N.V.~Tran}
\ead{ntran@fnal.gov}

\author[add3]{S.-S.~Yu}
\ead{syu@cern.ch}

\address[add3]{
Department of Physics and Center for High Energy and High Field Physics, 
National Central University, Chung-Li, Taoyuan City 32001, Taiwan
}

\address[add1]{
HEP Division, Argonne National Laboratory,
9700 S.~Cass Avenue,
Argonne, IL 60439, USA. 
}

\address[addDuke]{
Department of Physics, Duke University, USA
}

\address[add2]{
Fermi National Accelerator Laboratory
}

\begin{abstract}
Jet substructure variables for hadronic jets with transverse momenta in the range from 2.5~TeV to 20~TeV
were studied using several designs for the spatial size of calorimeter cells. The studies  used 
the full Geant4 simulation 
of calorimeter response combined with realistic reconstruction of calorimeter clusters.
In most cases, the results indicate that the performance of jet-substructure 
reconstruction improves with reducing cell size of a hadronic calorimeter 
from $\Delta \eta \times \Delta \phi = 0.087\times0.087$,
which are similar to the cell sizes of the calorimeters of LHC experiments, by a factor of four, to  $0.022\times0.022$.

\end{abstract}


\end{frontmatter}

\section{Introduction}
Particle collisions at energies  beyond those attained at the LHC will lead to many challenges for detector technologies.
Future circular $pp$ colliders~\cite{Mangano:2018mur} such as the European initiatives, FCC-hh~\cite{Benedikt:2018csr}, high-energy LHC (HE-LHC)~\cite{Zimmermann:2018wdi}, and the Chinese initiative, SppC~\cite{Tang:2015qga} will measure high-momentum bosons ($W$, $Z$, $H$) and top quarks with highly-collimated decay products that form jets. 
Jet substructure techniques are used
to identify such boosted particles, and thus can maximize the physics potential of the future colliders.

The reconstruction of jet substructure  variables for collimated jets with transverse momenta above 10 TeV 
requires an appropriate detector design. The most important detector systems for reconstruction of such jets are tracking and calorimetry.
Recently, a number of studies \cite{Calkins:2013ega,Chekanov:2015ihl,Coleman:2017fiq} 
have been discussed using various fast simulation tools, such as 
Delphes  \cite{deFavereau:2013fsa}, in which momenta of particles
are smeared to mimic detector response. 

A major step towards the usage of full Geant4 simulation to verify the granularity requirements 
for calorimeters was made in Ref.~\cite{Chekanov:2016ppq}.
These studies have illustrated a significant impact 
of granularity of electromagnetic (ECAL) and hadronic (HCAL) calorimeters on the
cluster separation between two particles. It was concluded that high granularity is essential 
in resolving two close-by particles for energies above 100 GeV. 

This paper takes the next step in understanding this problem in terms of high-level quantities typically used in physics analyses. Similar to the studies presented in Ref.~\cite{Chekanov:2016ppq}, this paper is based on a full
Geant4 simulation with realistic jet reconstruction.

\section{Simulation of detector response}
\label{sec:sim}

The description of the detector and software used for this study is discussed in Ref.~\cite{Chekanov:2016ppq}.
We use the SiFCC detector geometry with a software package that provides a versatile environment for simulations
of detector performance, testing new technology options, and event reconstruction techniques for future
100~TeV colliders.

The baseline detector discussed in Ref.~\cite{Chekanov:2016ppq}
includes a silicon-tungsten electromagnetic calorimeter with a transverse cell size of 2~$\times$~2~cm$^2$, a steel-scintillator hadronic calorimeter with a transverse cell size of 5~$\times$~5~cm$^2$, and a solenoid outside the ECAL and HCAL that provides a 5 T magnetic field. The studies presented in this paper focus on the performance of  the 
baseline HCAL with the cell size of 5~$\times$~5~cm$^2$, which corresponds to $\Delta \eta \times \Delta \phi = 0.022\times0.022$, where $\eta$ is the pseudorapidity, $\eta \equiv -\ln\tan(\theta/2)$, and $\phi$ is the azimuthal angle.
The depth of the HCAL in the barrel region is 11.25 interaction lengths ($\lambda_I$).
The HCAL has 64 longitudinal layers in the barrel and the endcap regions.

In addition to the baseline HCAL geometry,
two geometry variations were considered without changing other settings. 
We used the HCAL with transverse cell size of
20~$\times$~20~cm$^2$ and  1~$\times$~1~cm$^2$.
In the terms of $\Delta \eta \times \Delta \phi$,  such cell sizes correspond to
$0.087\times0.087$ and  $0.0043\times0.0043$, respectively.

The \GEANTfour\ (version 10.3)~\cite{Allison2016186} simulation of calorimeter response was followed by the full reconstruction of calorimeter clusters formed by the Pandora algorithm~\cite{Charles:2009ta,Marshall:2013bda}. The criteria for clustering in the calorimeter were discussed in Ref.~\cite{THOMSON200925}. We use the same criteria as those in the SiD detector design~\cite{Behnke:2013lya}, which were optimized for a high-granularity HCAL with a cell size of 1~$\times$~1~cm$^2$.
Calorimeter clusters were built from calorimeter hits in the  ECAL and HCAL after applying the corresponding sampling fractions. No other corrections are applied.
Hadronic jets were 
reconstructed with the {\sc FastJet} package~\cite{fastjet} using the anti-$k_T$ algorithm \cite{Cacciari:2008gp}
with  a distance parameter of 0.5. 

In the following discussion, we use the simulations of a heavy $Z'$ boson, 
a hypothetical gauge boson  that arises from extensions of the electroweak symmetry of the Standard Model.
The $Z'$ bosons were simulated with the masses $M=5$, $10$, $20$ and $40$~TeV. The lowest value 
represents a typical mass that is within the reach of the LHC experiments. The  resonance mass of $40$~TeV 
represents the physics reach for a  100~TeV collider. The $Z'$ bosons are forced to decay to two light-flavor quarks ($q\bar{q}$)~\cite{Langacker:2008yv}, $W^+W^-$~\cite{Leike:1998wr} or $t\bar{t}$~\cite{Rosner:1996eb} final states, where the 
$W$ bosons and $t$ quarks decay hadronically. In these scenarios, two highly-boosted
jets are produced,  which are typically back-to-back in the laboratory frame.
The typical transverse momenta of the jets are $\simeq M/2$.
The main difference between the considered decay modes lies in the different jet substructures. In the case of the $q\bar{q}$ decays,
jets do not have any internal structure. In the case of the $W^+W^-$ final state, each jet has two subjets  because of the decay $W\rightarrow q\bar{q}$. In the case of hadronic top decays, jets have three subjets due
to the decay $t \rightarrow  W^+\>b \rightarrow q\bar{q} b$. We use the $Z' \to q\bar{q} \to$~jets process to model the background from QCD jets with approximately the same energy as 
 the $W$ bosons and top quarks.   
The signal events were generated using the \pythia generator~\cite{Sjostrand:2006za} with the default settings,
ignoring interference with SM processes.
The event samples used in this paper are  available from the
HepSim database~\cite{Chekanov:2014fga}.

\section{Studies of jet properties}
\label{sec:jets}

We consider several variables that characterize jet substructure using different calorimeter granularities. The question we want to answer is, how closely the reconstructed
jet substructure variables reflect the input ``truth'' values  that are reconstructed using particles directly from the \pythia generator.

In this study we use the jet effective radius and jet splitting scales as benchmark variables
to study jet substructure properties with the signal process $Z'\rightarrow WW$ only. 
The effective radius is the average of the energy-weighted radial distance $\delta R_i$ in $\eta-\phi$ space of jet constituents.
It is defined as $(1/E) \sum_i e_i \delta R_i$, where $E$ is the energy of the jet and $e_i$ is the energy of a calorimeter 
constituent cluster $i$ at the distance $\delta R_i$ from the jet center. The sum runs over all constituents of the jet. 
This variable has been studied for multi-TeV jets in Ref.~\cite{Auerbach:2014xua}.
A jet $k_T$ splitting scale~\cite{Butterworth:2002tt} is defined as a distance measure
used to form jets by the $k_T$ recombination
algorithm~\cite{Catani1993187,Ellis:1993tq}.
This variable has been studied by ATLAS~\cite{ATLAS:2012am}, and more recently in the context of 100 TeV physics~\cite{Auerbach:2014xua}.
The splitting scale is defined as $\sqrt{d_{12}}=\min(p_T^1,p_T^2) \times \delta R_{12}$~\cite{ATLAS:2012am} 
 at the final stage of the $k_T$ clustering, where two subjets are merged into the final jet.

\begin{figure}
\begin{center}
   \subfigure[$M(Z')=5$~TeV] {
   \includegraphics[width=0.46\textwidth]{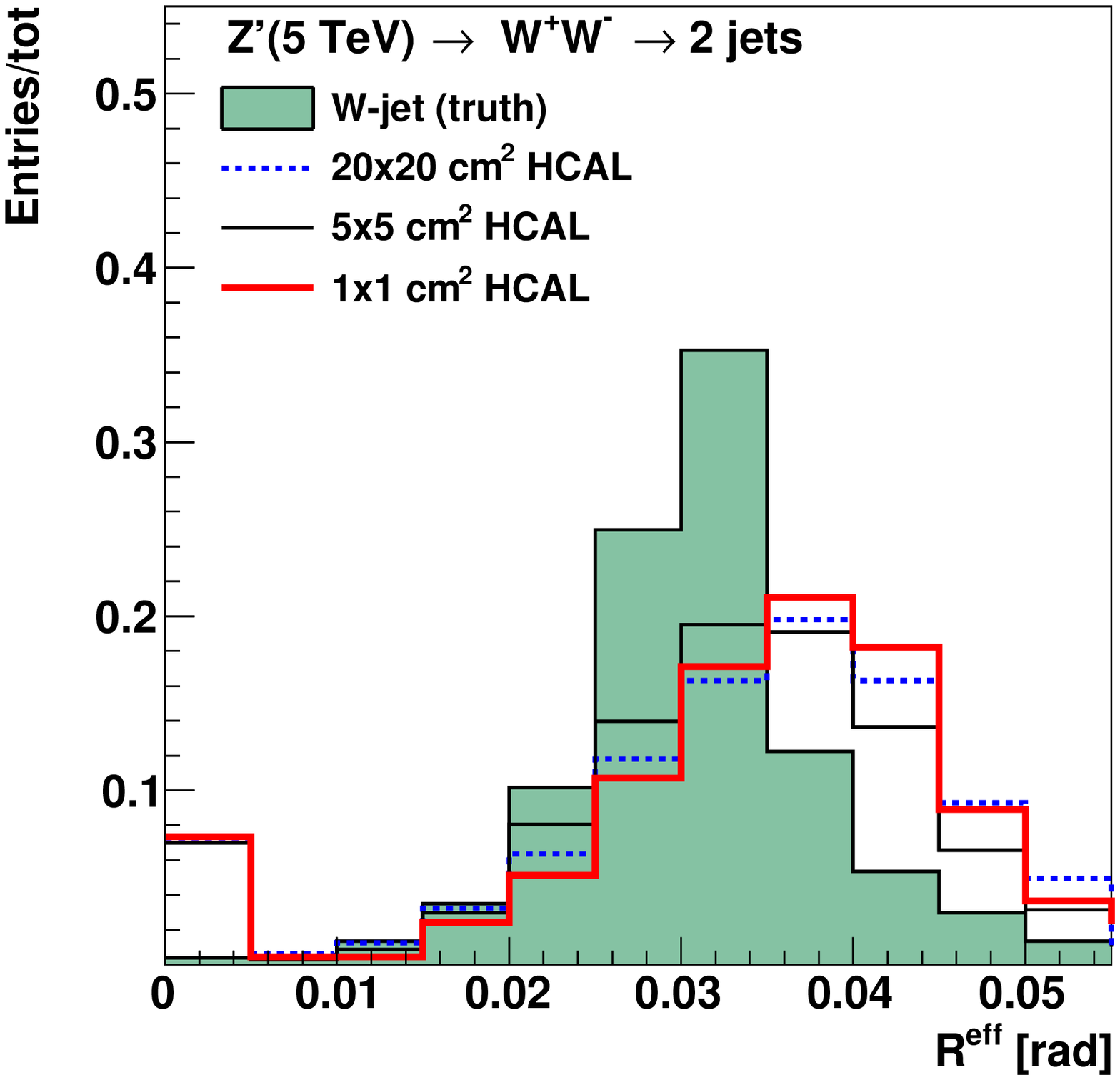}\hfill
   }
   \subfigure[$M(Z')=10$~TeV] {
   \includegraphics[width=0.46\textwidth]{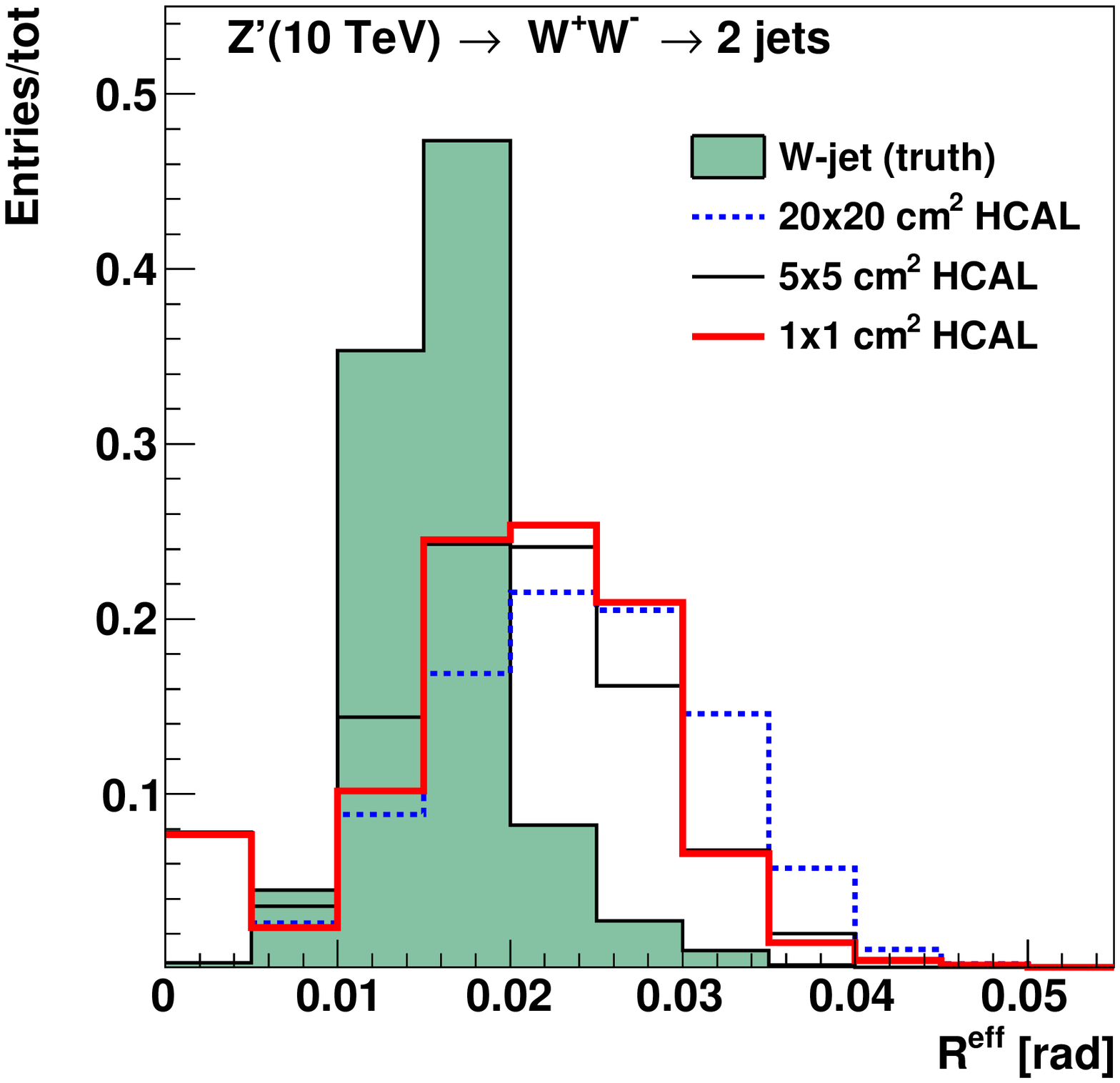}
   }
   \subfigure[$M(Z')=20$~TeV] {
   \includegraphics[width=0.46\textwidth]{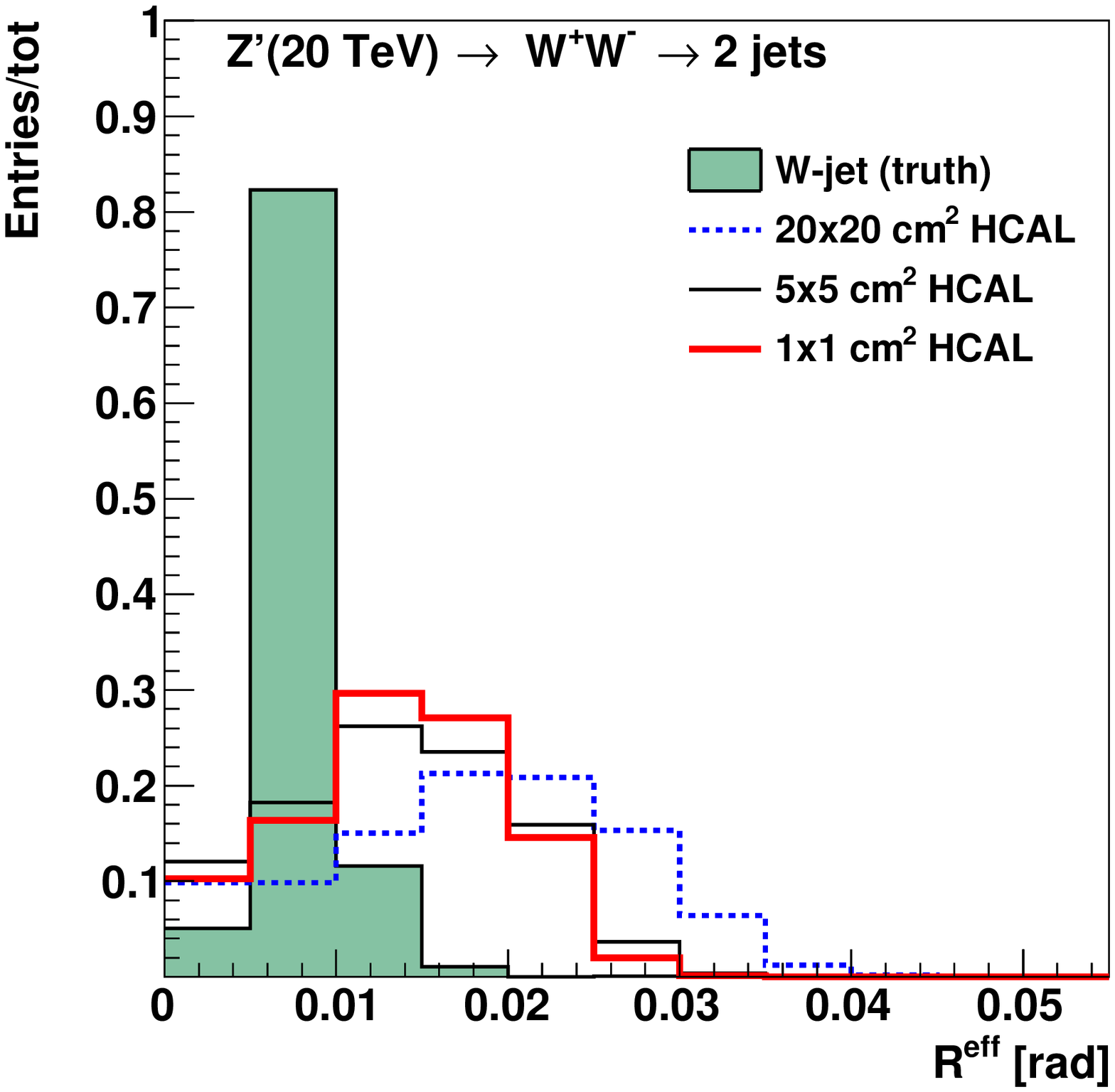}
   }
   \subfigure[$M(Z')=40$~TeV] {
   \includegraphics[width=0.46\textwidth]{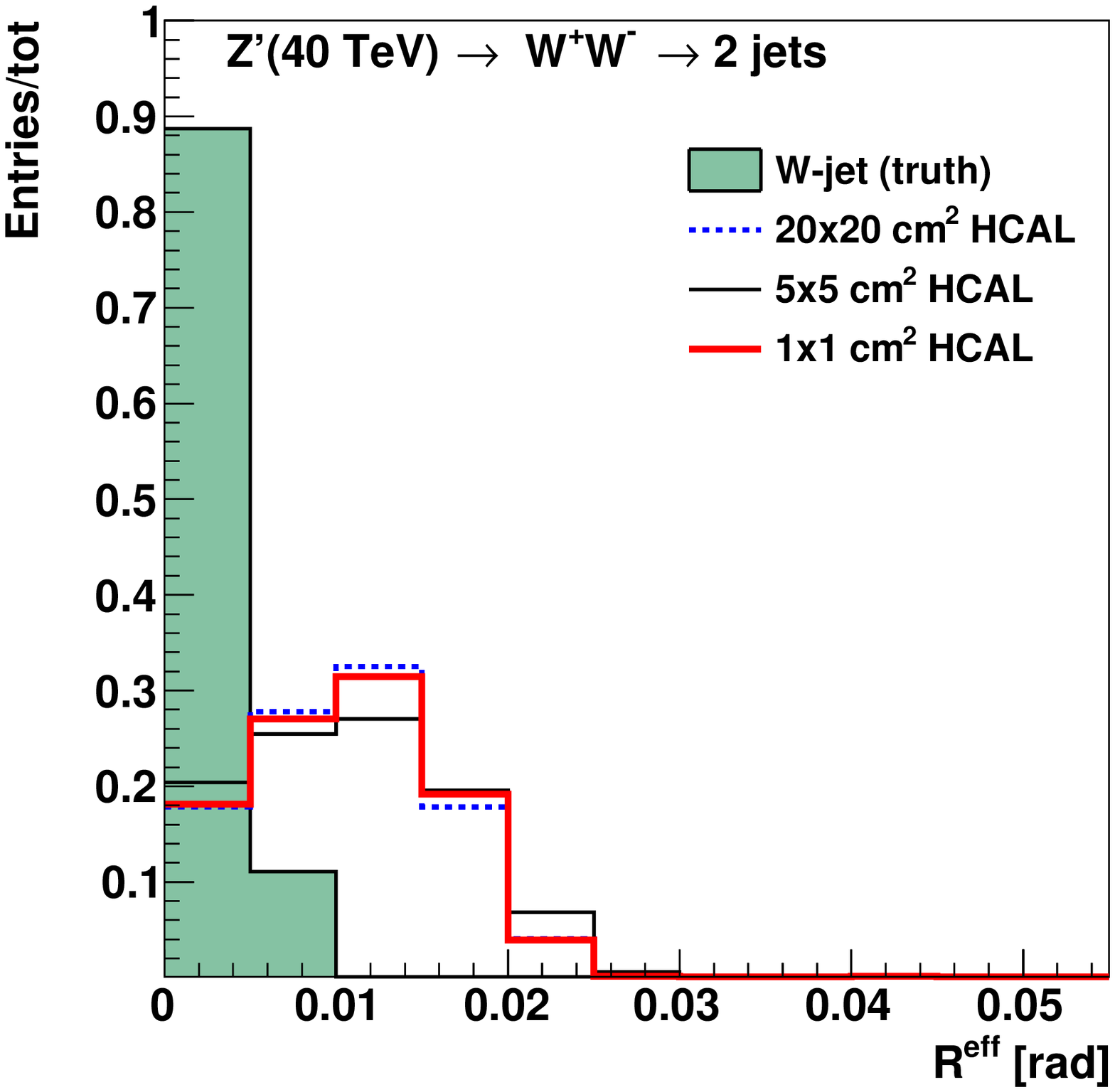}
   }
\end{center}
\caption{Jet effective radius for different jet transverse momenta and HCAL granularities.}
\label{fig:eff_rad}
\end{figure}

\begin{figure}
\begin{center}
   \subfigure[$M(Z')=5$~TeV] {
   \includegraphics[width=0.46\textwidth]{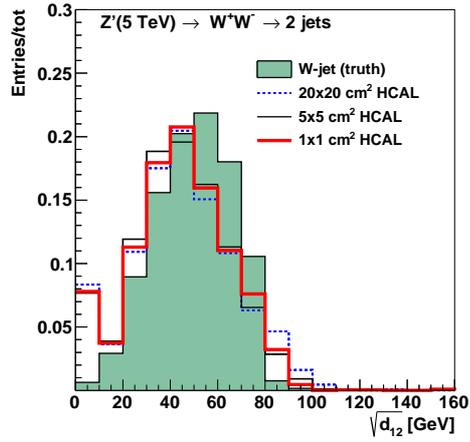}\hfill
   }
   \subfigure[$M(Z')=10$~TeV] {
   \includegraphics[width=0.46\textwidth]{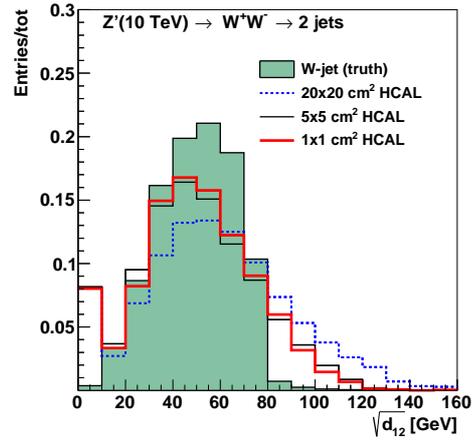}
   }
   \subfigure[$M(Z')=20$~TeV] {
   \includegraphics[width=0.46\textwidth]{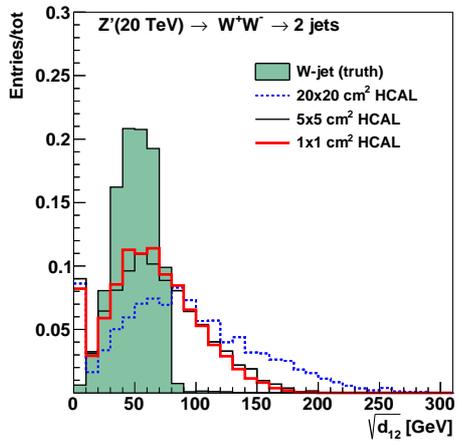}
   }
   \subfigure[$M(Z')=40$~TeV] {
   \includegraphics[width=0.46\textwidth]{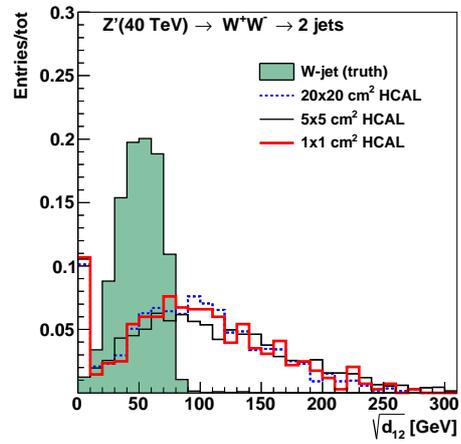}
   }
\end{center}
\caption{Jet splitting scale for different jet transverse momenta and HCAL granularities.}
\label{fig:d12}
\end{figure}

Figures~\ref{fig:eff_rad} and~\ref{fig:d12} show the distributions of 
the jet effective radius and jet splitting scale for  different jet transverse momenta and HCAL granularities. 
The reconstructed-level distributions  disagree significantly with the distributions  
reconstructed using truth-level particles. The distributions reconstructed with 1~$\times$~1~cm$^2$ or 5~$\times$~5~cm$^2$ cells 
 are generally closer to the truth-level variables, than the distributions 
reconstructed using 20~$\times$20~cm$^2$ cells, particularly for resonance masses in the 10-20 TeV range. In these cases, there is not much  difference between the 
 5~$\times$~5~cm$^2$ and  1~$\times$~1~cm$^2$ cell sizes. The extreme case with $M(Z')=40$ TeV corresponds to very boosted jets with 
 $p_T \simeq 20$~TeV.  This case does not show 
differences between the different HCAL configurations.

This study confirms the baseline SiFCC detector geometry~\cite{Chekanov:2016ppq} that uses $5 \times 5$~cm$^2$ HCAL cells,
corresponding to $\Delta \eta \times \Delta \phi = 0.022\times0.022$.
Similar HCAL cell sizes,  $0.025\times0.025$,  were recently adopted for the baseline FCC-hh detector~\cite{Benedikt:2018csr,fcc1,fcc2} planned at CERN.
Before the publication~\cite{Chekanov:2016ppq},   such a choice for the HCAL cells   
was motivated by the studies of jet substructure  using a fast detector simulation of boosted jets.
In addition to the improvements in physics performance, the smaller HCAL cells 
reduce the required dynamic range for 
signal reconstruction~\cite{Chekanov:2015ihl}, and thus can simplify the calorimeter readout.

It should be noted that the ATLAS and CMS detectors at the LHC use HCAL cell sizes in the barrel region which are close to 
$\Delta \eta \times \Delta \phi = 0.087\times 0.087$. These experiments focus on jet substructure variables for jets
with $p_T \lesssim 4$~TeV. Our studies indicate that the future experiments, which will
measure jets with significantly greater transverse momenta, require an HCAL with higher granularity in order to achieve optimal performance for jet substructure variables. In the following sections we consider several other physics-motivated variables that can shed light on the performance of the HCAL for tens-of-TeV jets.

\section{Detector performance with soft drop mass}

In this section, we use the jet mass computed with a specific algorithm, soft 
drop declustering, to study the performance with various detector 
cell sizes and resonance masses. 

\subsection{The technique of soft drop declustering}
The soft drop declustering~\cite{Larkoski:2014wba} is a grooming method 
that removes soft wide-angle radiation from a jet. The constituents of a jet 
$j_0$ are first reclustered using the Cambridge-Aachen
 (C/A) algorithm~\cite{Dokshitzer:1997in,Wobisch:1998wt}. Then, the jet $j_0$ 
is broken into two subjets $j_1$ and $j_2$ by undoing the last stage of C/A 
clustering.
If the subjets pass the following soft drop condition, jet $j_0$ is the final 
soft-drop jet. Otherwise, the algorithm redefines $j_0$ to be the subjet with 
larger $p_T$ (among $j_1$ and $j_2$) and iterates the procedure. The condition is, 
\begin{equation} \label{eq:soft-drop}
\frac{\mathrm{min}(p_{T1},p_{T2})}{p_{T1}+p_{T2}}>z_\mathrm{cut}(\frac{\Delta R_{12}}{R_{0}})^{\beta},
\end{equation}
where $p_{T1}$ and $p_{T2}$ are the transverse momenta of the two subjets, 
$z_\mathrm{cut}$ is soft drop threshold, 
$\Delta R_{12}$ is the distance between the two subjets in the 
rapidity-azimuthal plane ($y$-$\phi$), $R_0$ is the characteristic radius 
of the original jet, and $\beta$ is the angular exponent.

In our study, we compare the HCAL performance for  the soft drop mass with 
$\beta=0$  and $\beta=2$. For $\beta=0$~\cite{CMS:2017wyc,Tripathee:2017ybi}, the soft drop condition 
depends only on the $z_\mathrm{cut}$ and is angle-independent. At the parton level, this condition is infrared safe. For $\beta=2$~\cite{Aaboud:2017qwh}, the condition depends on both 
the angular distance between the two subjets and $z_\mathrm{cut}$, making the 
algorithm become both infrared and collinear safe at the parton level. Upon calorimeter clustering, the two $\beta$ values give  different sensitivities to large-angle radiation.

\subsection{Analysis method \label{sec:massana}}
We employ the following method to quantify the detector performance and 
determine the cell size that gives the best separation between  
signal and background. For each configuration of detector and resonance mass, 
we draw the receiver operating characteristic (ROC) curves in which the $x$-axis
 is the signal efficiency ($\epsilon_\mathrm{sig}$) and $y$-axis is the inverse 
of the background efficiency ($1/\epsilon_\mathrm{bkg}$). 
In order to scan the efficiencies of soft drop mass cuts, we vary the mass 
window as follows. We center the initial window on the median of the signal histogram, and increase its width symmetrically left and right in bins of 5~GeV. 
If one side of the mass window reaches the boundary 
of the mass histogram, we increase the width on the other side. For each mass window, the corresponding efficiencies 
$\epsilon_\mathrm{sig}$ and $\epsilon_\mathrm{bkg}$ give a point on 
the ROC curve.

\subsection{Results and conclusion}\label{Rebin_section}

\begin{figure}
\begin{center}
   \subfigure[20$\times$20 cm$^2$] {
   \includegraphics[ width=0.3\textwidth]{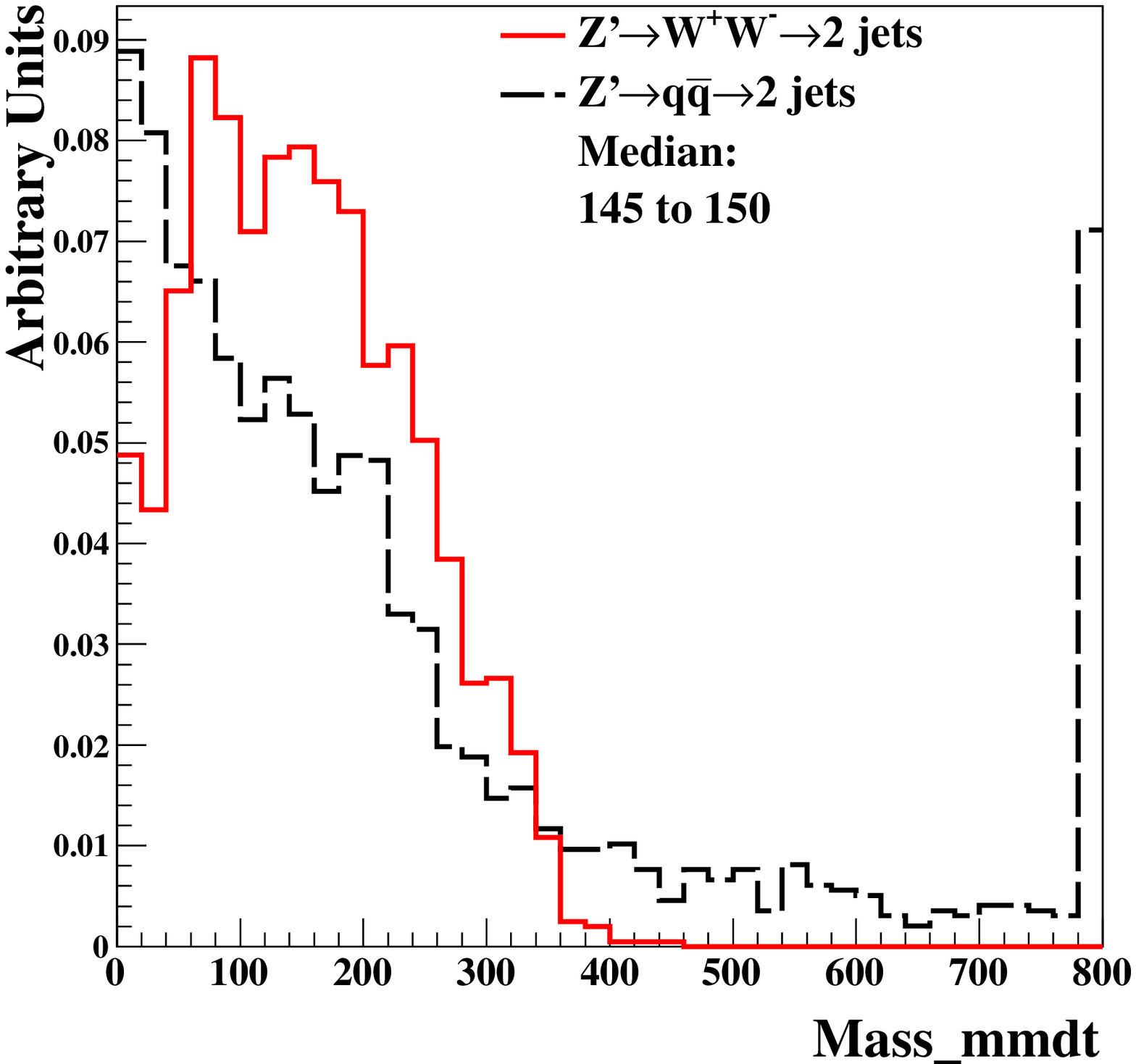}
   }
      \subfigure[5$\times$5 cm$^2$] {
   \includegraphics[ width=0.3\textwidth]{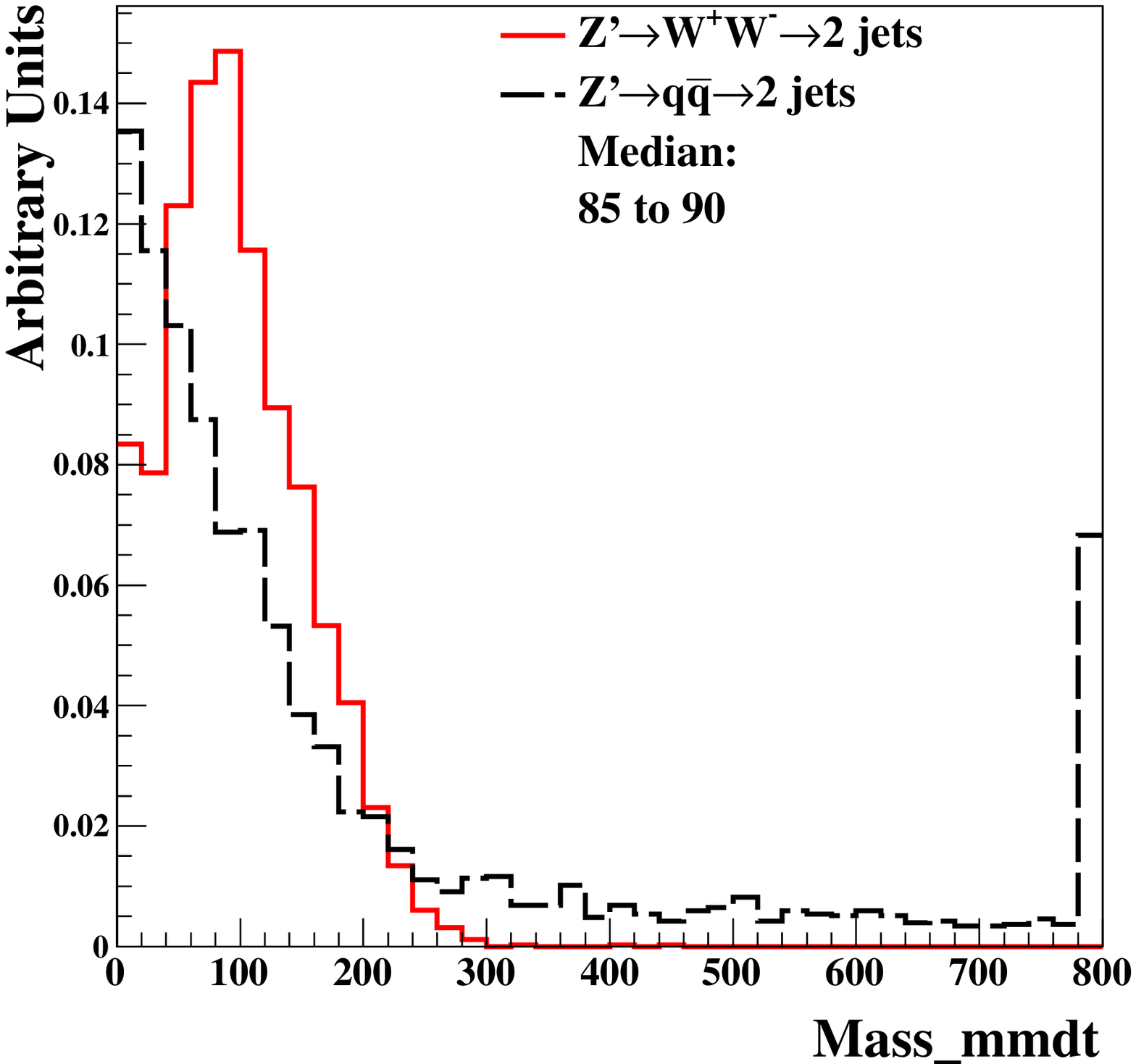}\hfill
   }
   \subfigure[1$\times$1 cm$^2$] {
   \includegraphics[ width=0.3\textwidth]{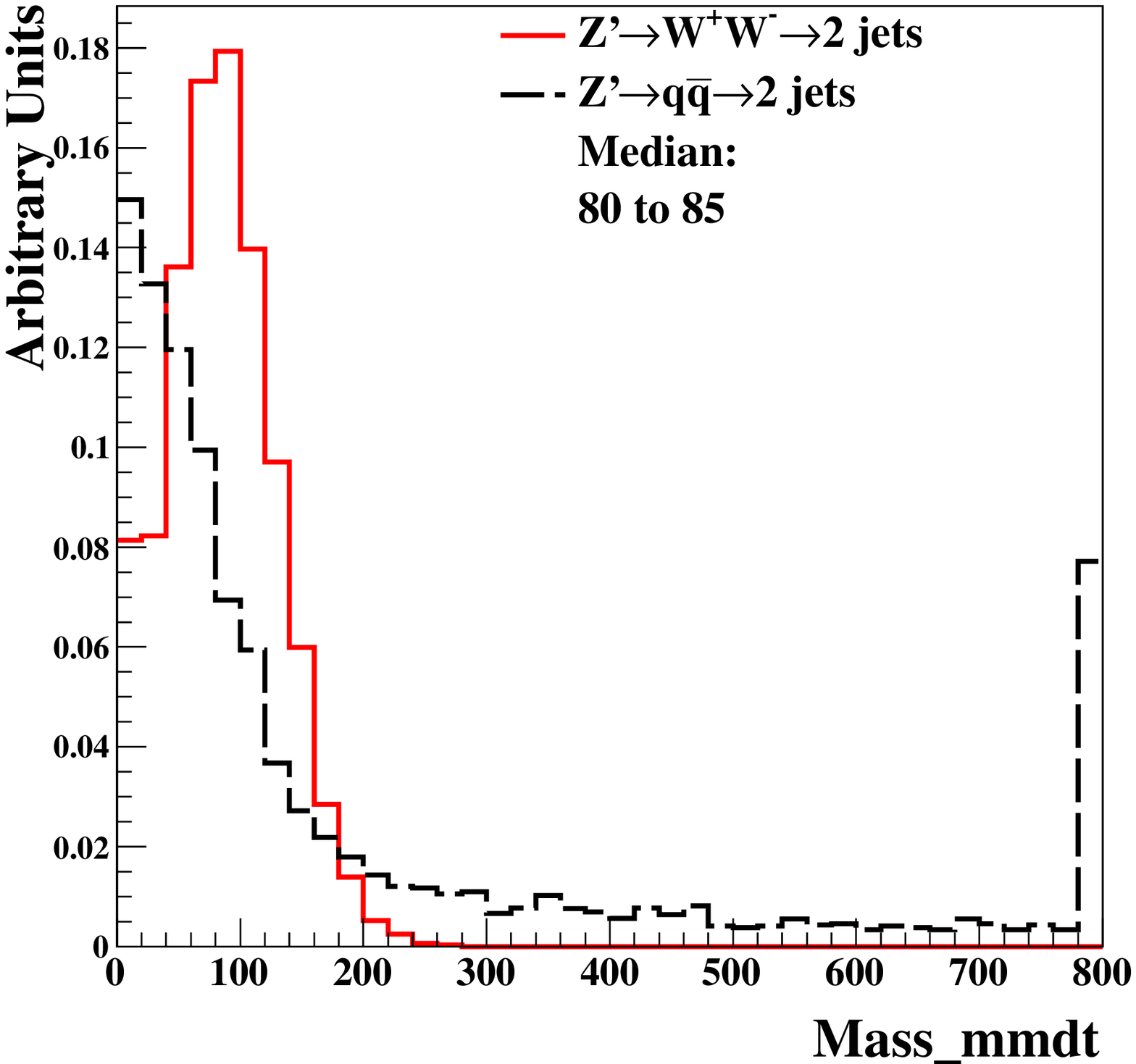}\hfill
   }
\end{center}
\caption{Distributions of soft drop mass for $\beta$=0, with $M(Z') = 20$~TeV and three different detector cell sizes: 20$\times$20, 
5$\times$5 and 1$\times$1 cm$^2$. The signal (background) process is 
$Z' \rightarrow WW$ ($Z'\rightarrow q\bar{q}$).
\label{fig:cluster_mass_mmdt_ww}}
\end{figure}

\begin{figure}
\begin{center}
  \subfigure[$M(Z')=5$~TeV] {
  \includegraphics[  width=0.45\textwidth]{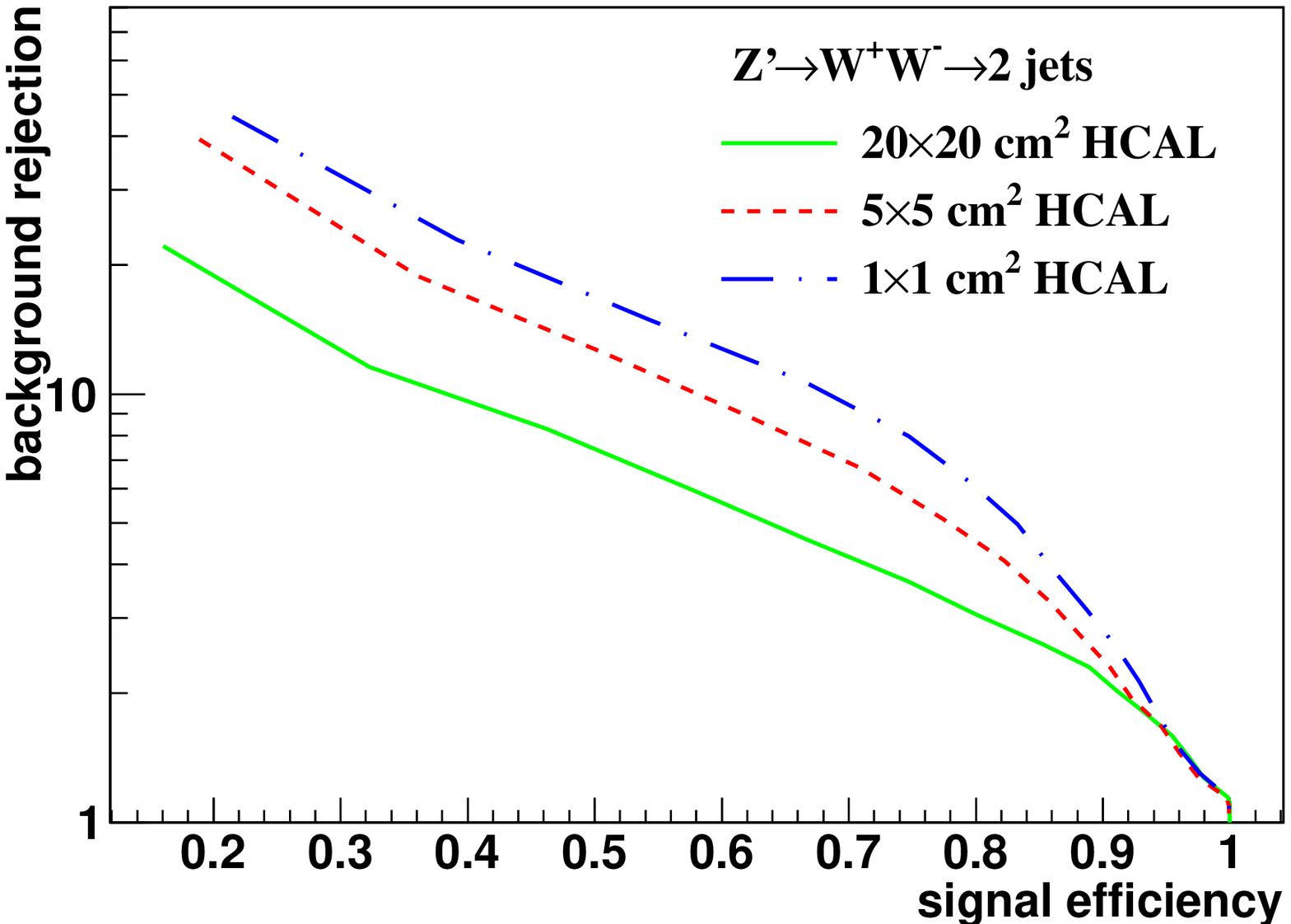}
  }
  \subfigure[$M(Z')=10$~TeV] {
  \includegraphics[  width=0.45\textwidth]{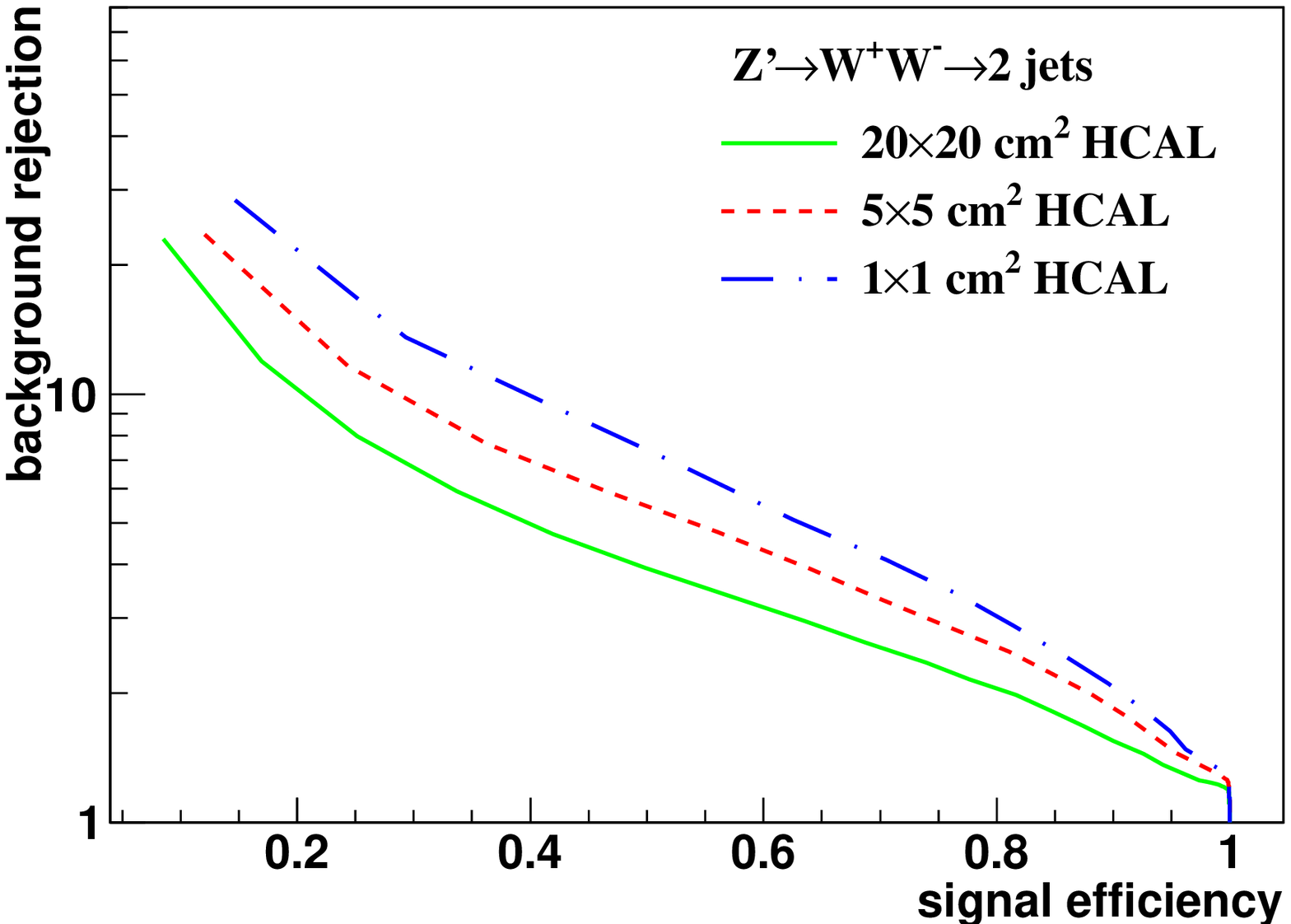}
  }
 \subfigure[$M(Z')=20$~TeV] {
 \includegraphics[  width=0.45\textwidth]{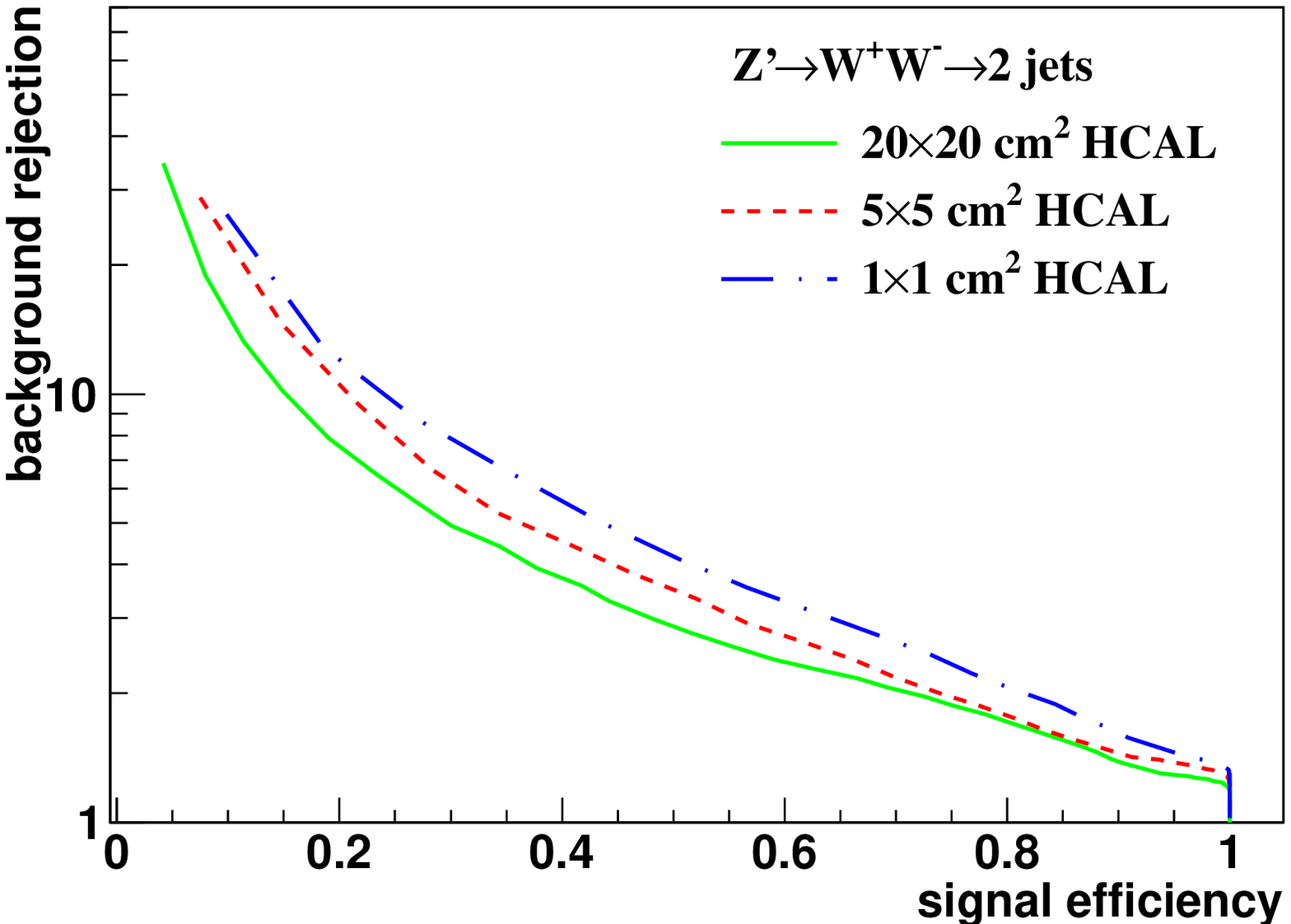}
 }
 \subfigure[$M(Z')=40$~TeV] {
 \includegraphics[  width=0.45\textwidth]{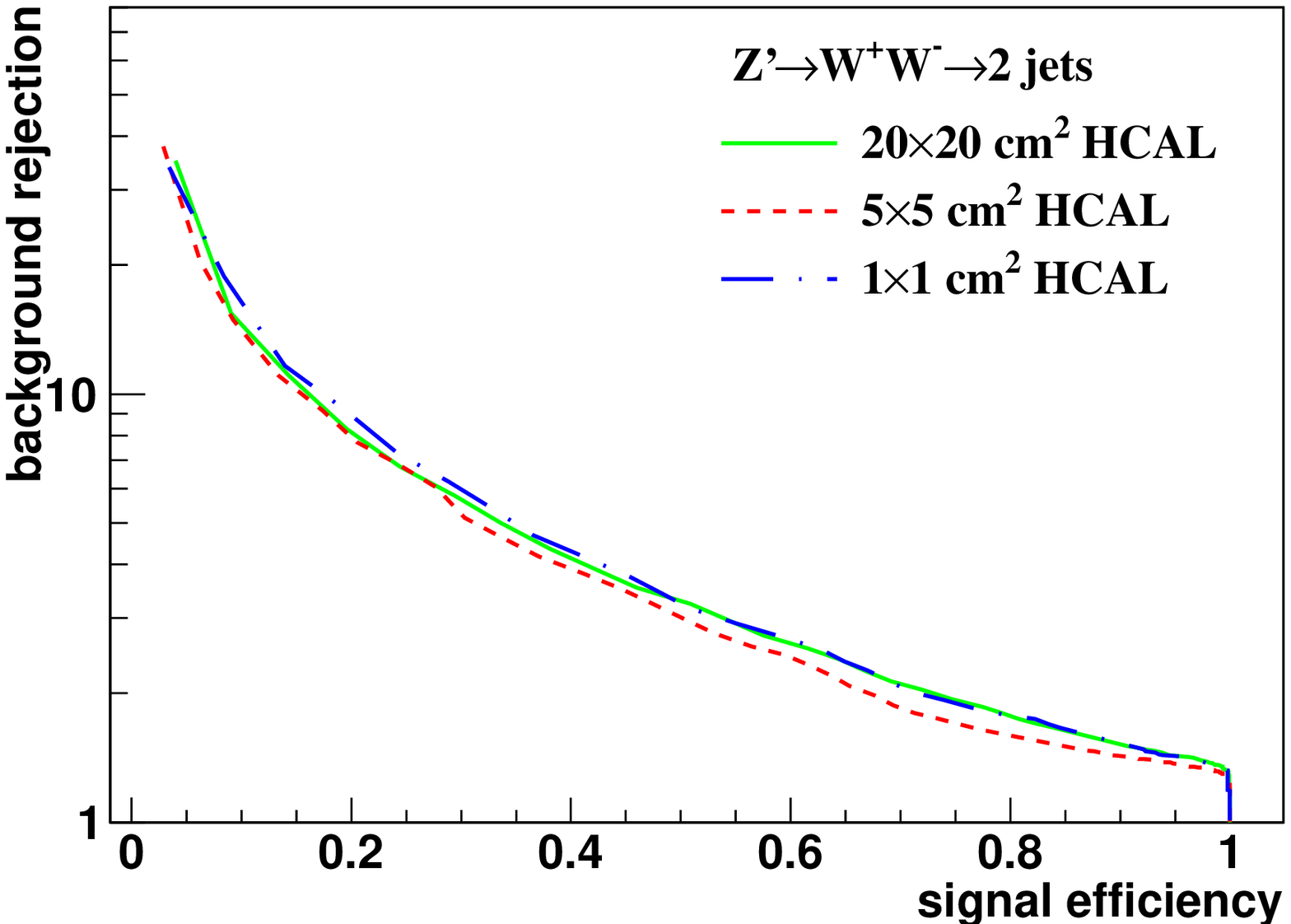}
 }
\end{center}
\caption{The ROC curves of soft drop mass selection for $\beta$=0 
with resonance masses of 5, 10, 20 and 40 TeV. 
Three different detector cell sizes are compared: 20~$ \times $~20, 
5~$ \times $~5, and 1~$ \times $~1 cm$^2$. 
The signal (background) process is $Z'\rightarrow WW$ 
($Z' \rightarrow q\bar{q}$).}
\label{fig:cluster_mass_mmdt_ww_ROC}
\end{figure}

\begin{figure}
\begin{center}
   \subfigure[5~$\times$~5 cm$^2$] {
   \includegraphics[   width=0.3\textwidth]{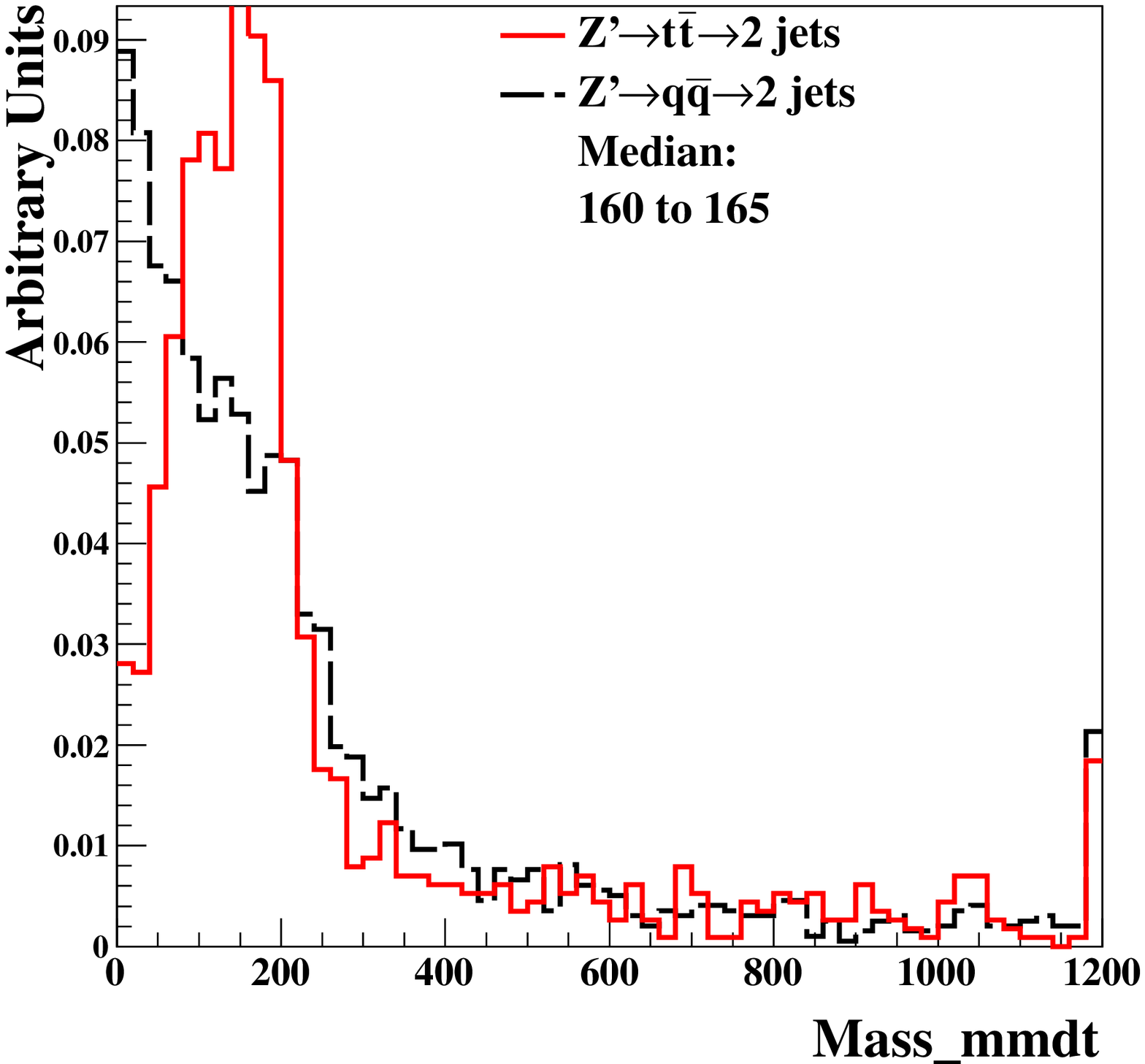}
   }
   \subfigure[20~$\times$~20 cm$^2$] {
   \includegraphics[   width=0.3\textwidth]{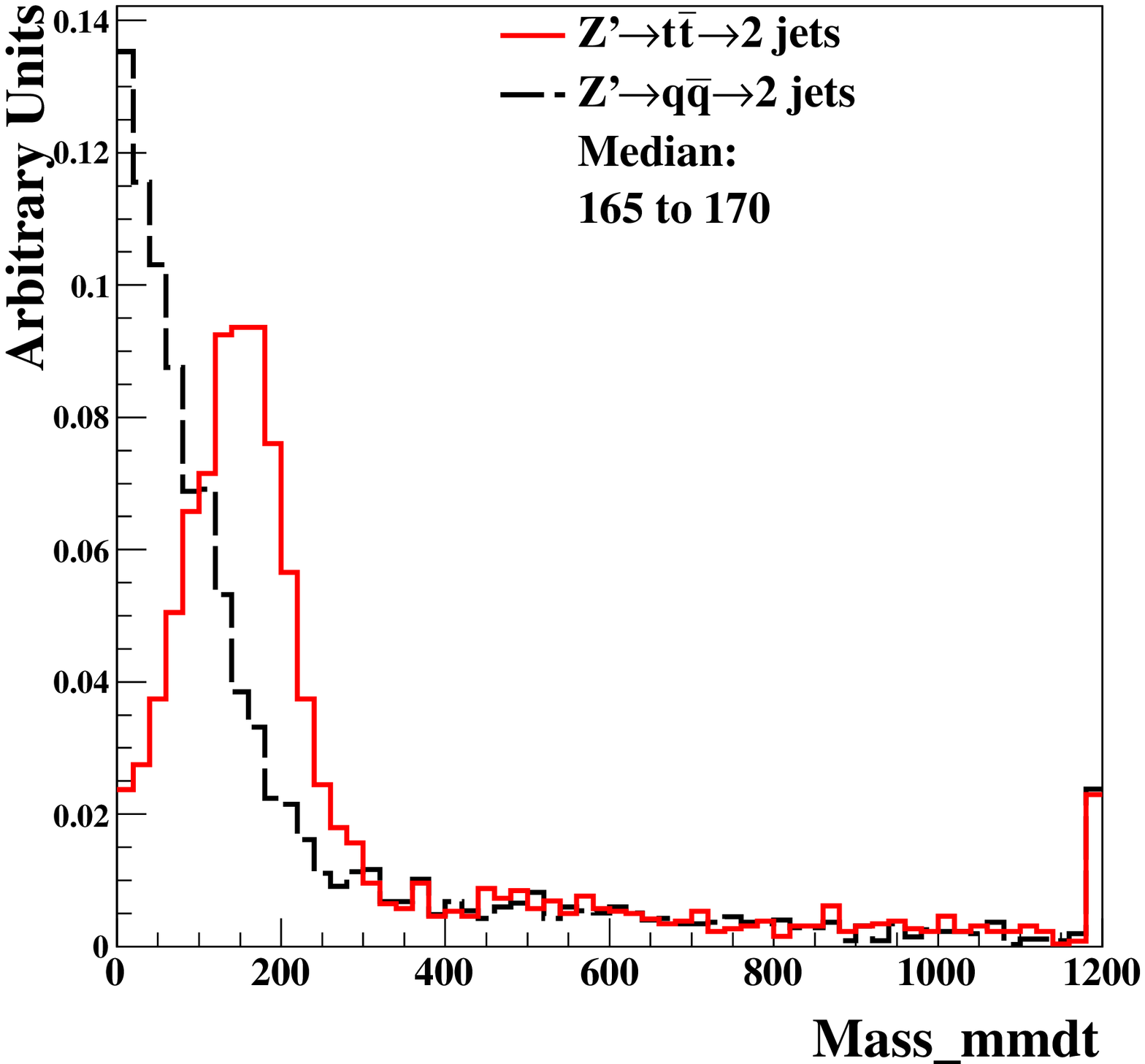}\hfill
   }
   \subfigure[1~$\times$~1 cm$^2$] {
   \includegraphics[   width=0.3\textwidth]{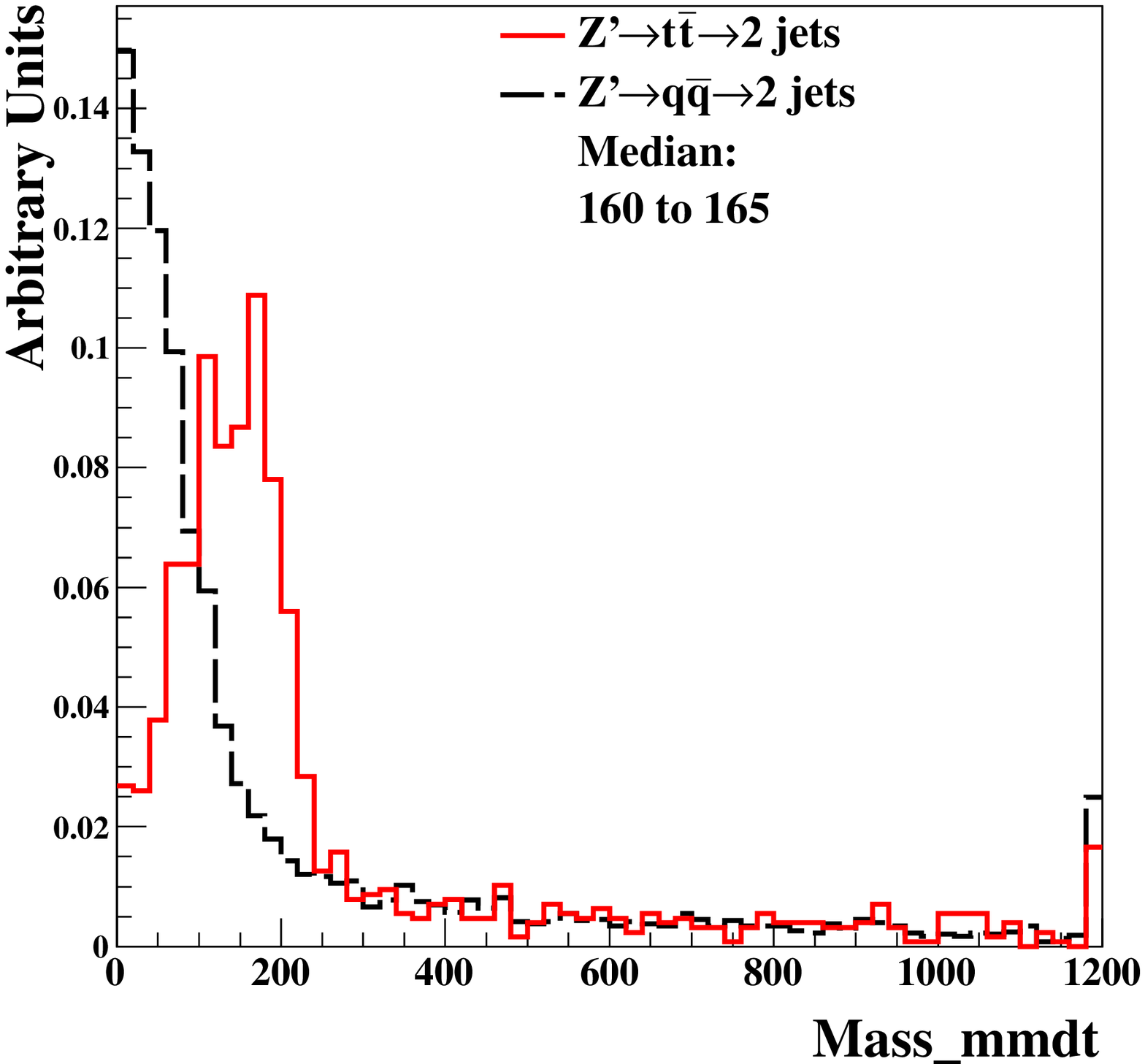}\hfill
   }
\end{center}
\caption{
Distributions of soft drop mass for $\beta$=0, with $M(Z') = 20$~TeV  and three different detector cell sizes: 20~$\times$~20, 
5~$\times$~5, and 1~$\times$~1 cm$^2$. The signal (background) process is 
$Z' \rightarrow t\bar{t}$ ($Z'\rightarrow q\bar{q}$).
}
\label{fig:cluster_mass_mmdt_tt}
\end{figure}

\begin{figure}
\begin{center}
  \subfigure[$M(Z')=5$~TeV] {
  \includegraphics[  width=0.45\textwidth]{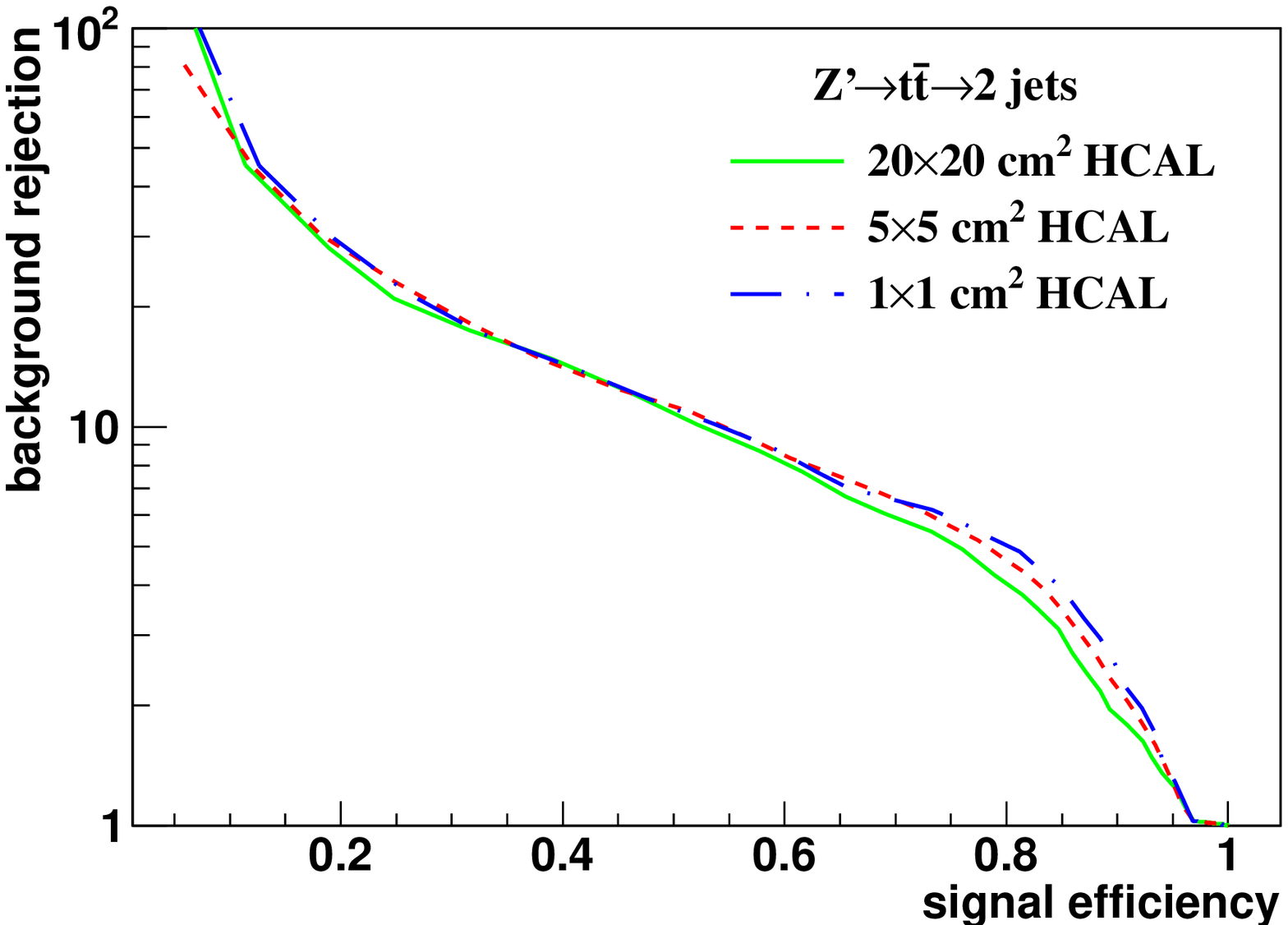}
  }
  \subfigure[$M(Z')=10$~TeV] {
  \includegraphics[  width=0.45\textwidth]{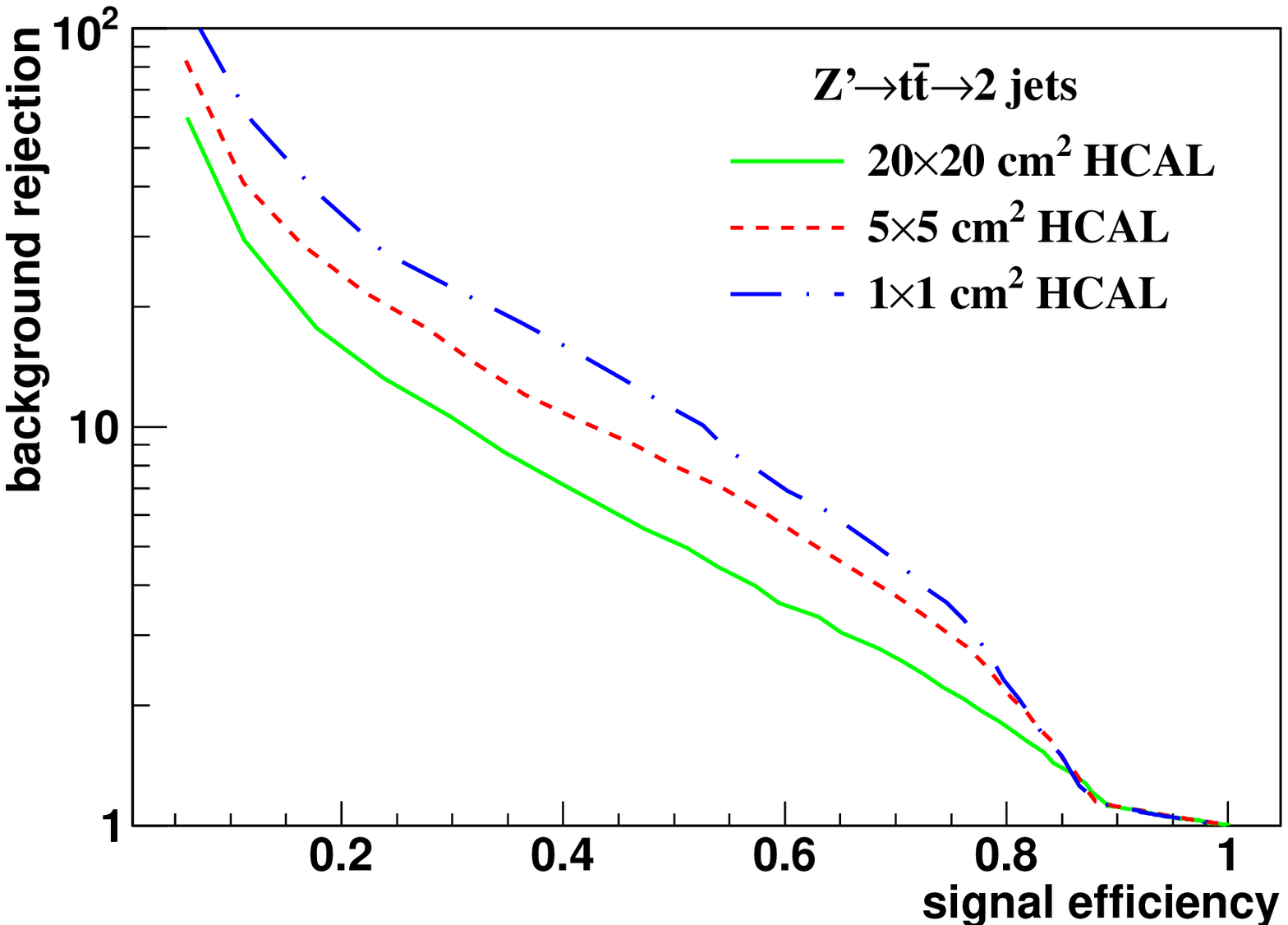}
  }
 \subfigure[$M(Z')=20$~TeV] {
 \includegraphics[  width=0.45\textwidth]{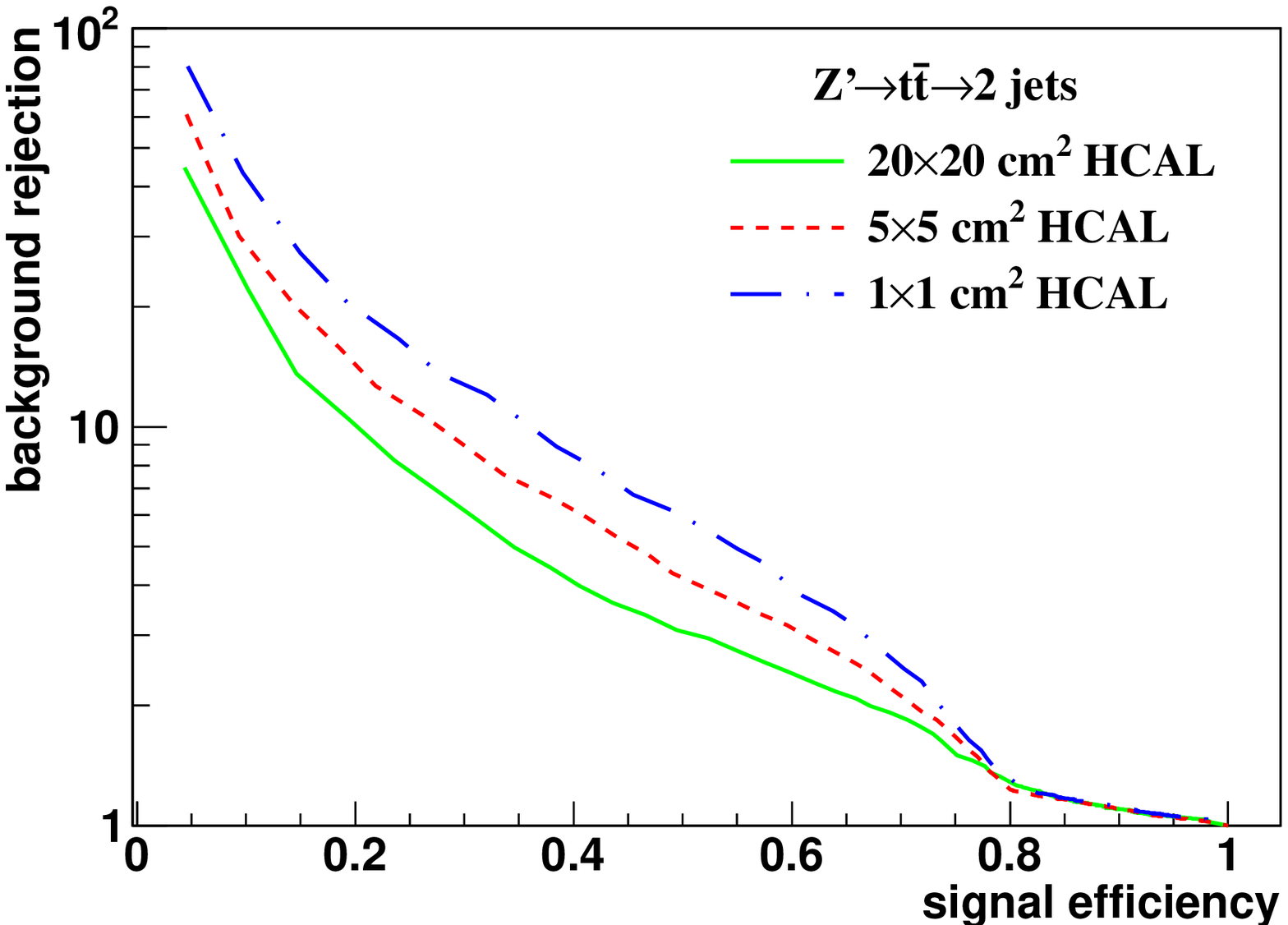}
 }
 \subfigure[$M(Z')=40$~TeV] {
 \includegraphics[  width=0.45\textwidth]{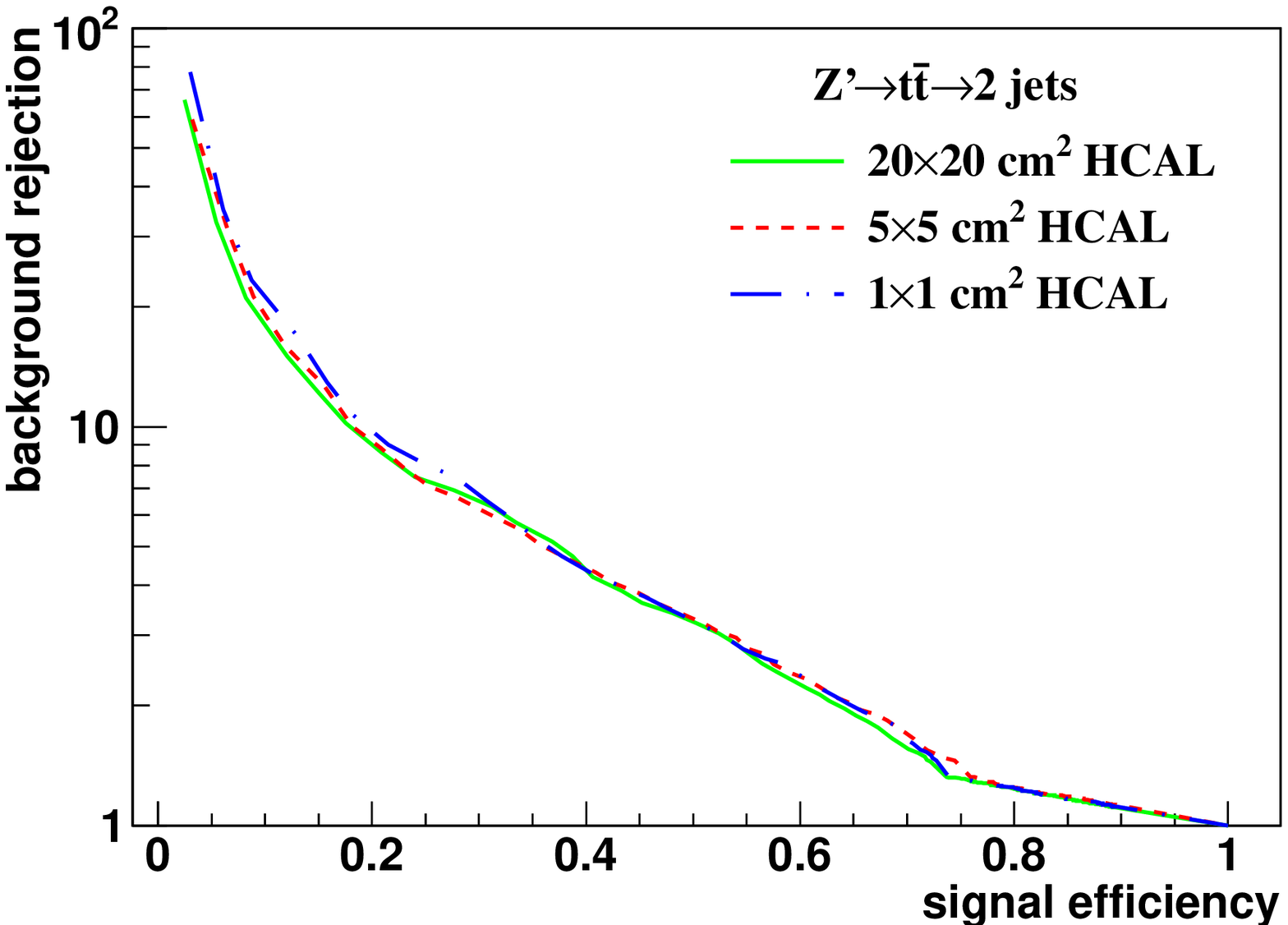}
 }
\end{center}
\caption{
The ROC curves of soft drop mass selection for $\beta$=0 
with resonance masses of 5, 10, 20 and 40 TeV. 
Three different detector cell sizes are compared: 20~$\times$~20, 
5~$\times$~5, and 1~$\times$~1 cm$^2$. 
The signal (background) process is $Z'\rightarrow t\bar{t}$
($Z' \rightarrow q\bar{q}$).
}
\label{fig:cluster_mass_mmdt_tt_ROC}
\end{figure}

\begin{figure}
\begin{center}
   \subfigure[20~$\times$~20 cm$^2$] {
   \includegraphics[   width=0.3\textwidth]{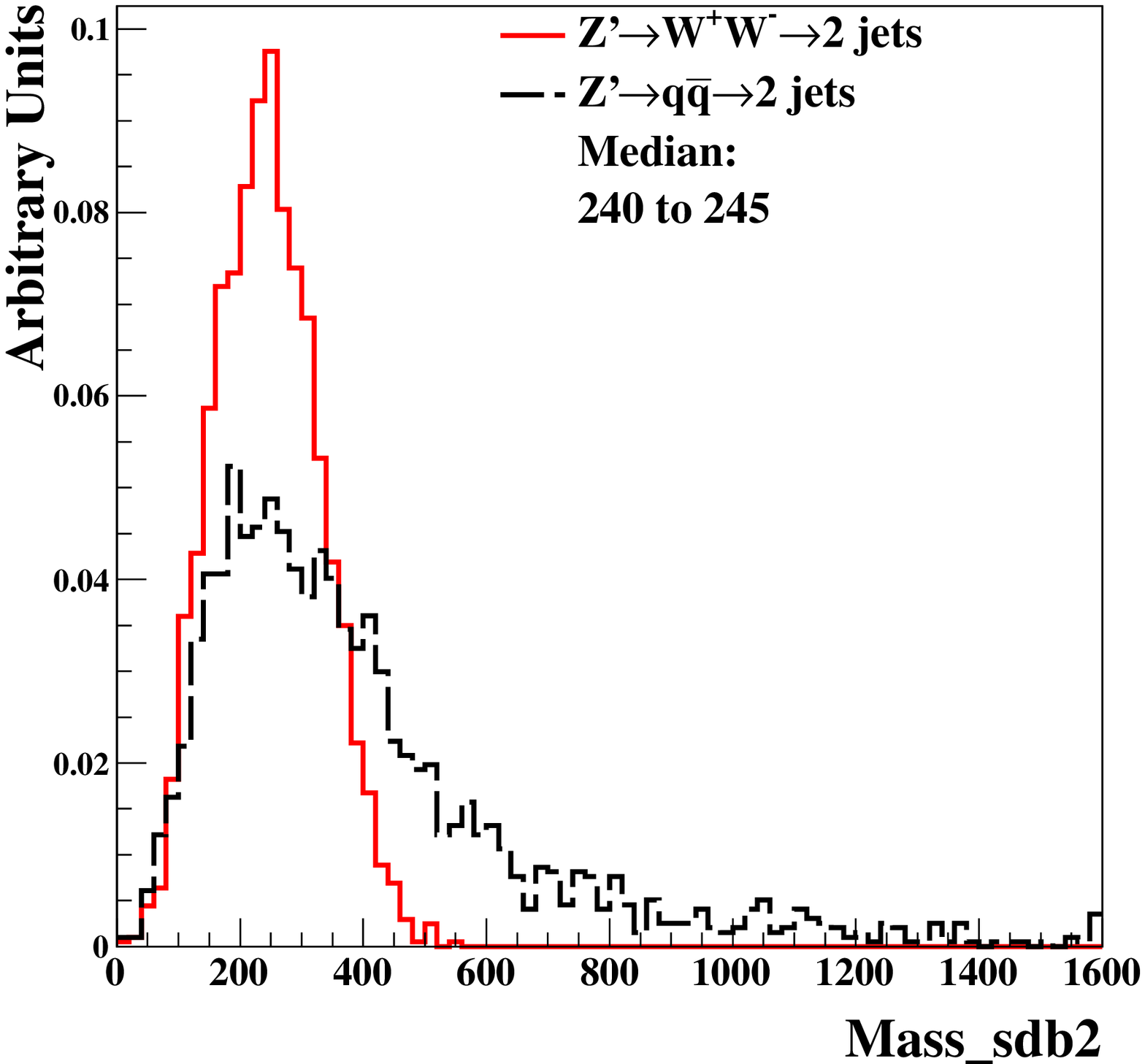}
   }
    \subfigure[5~$\times$~5 cm$^2$] {
   \includegraphics[   width=0.3\textwidth]{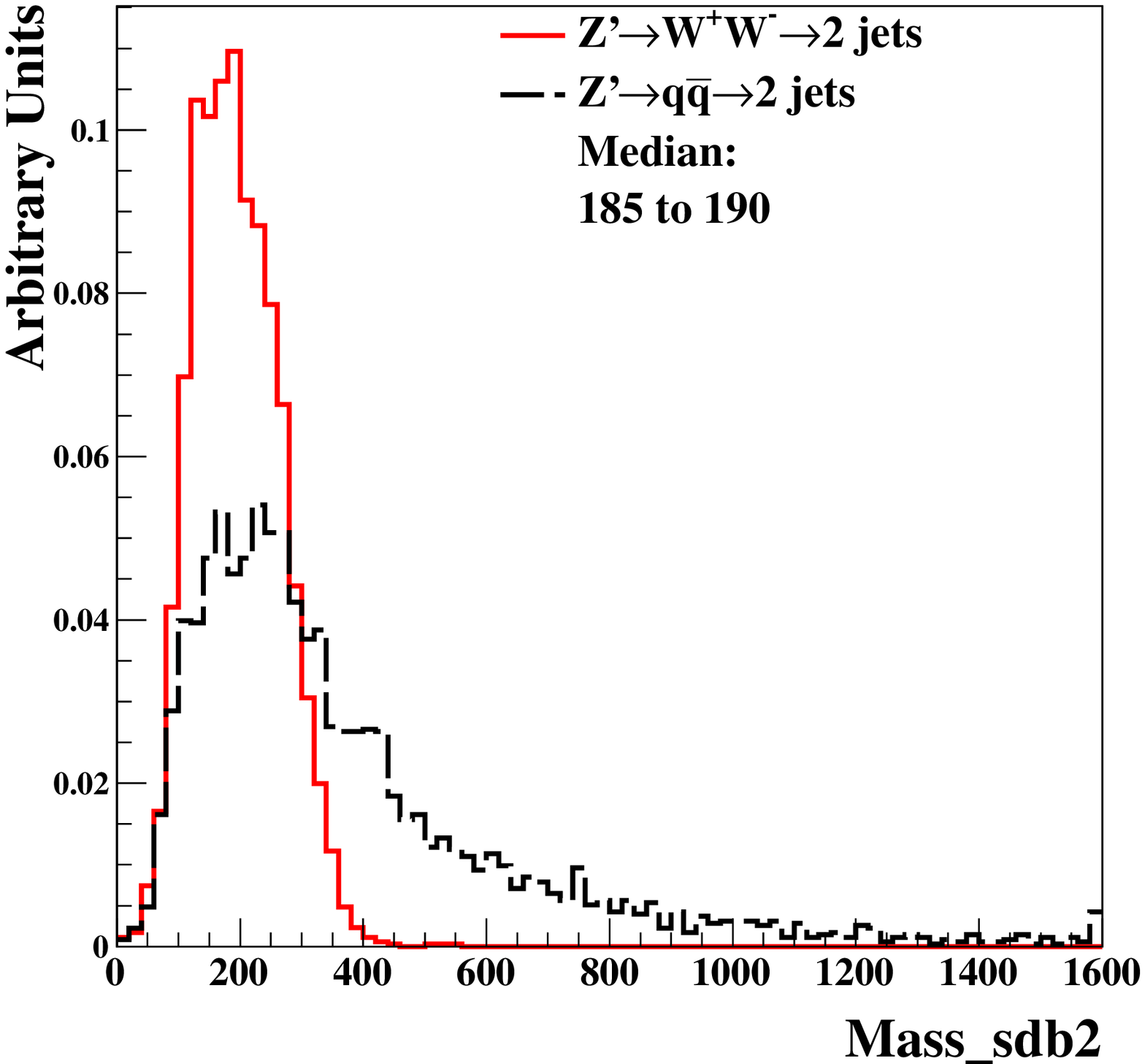}\hfill
   }
   \subfigure[1~$\times$~1 cm$^2$] {
   \includegraphics[   width=0.3\textwidth]{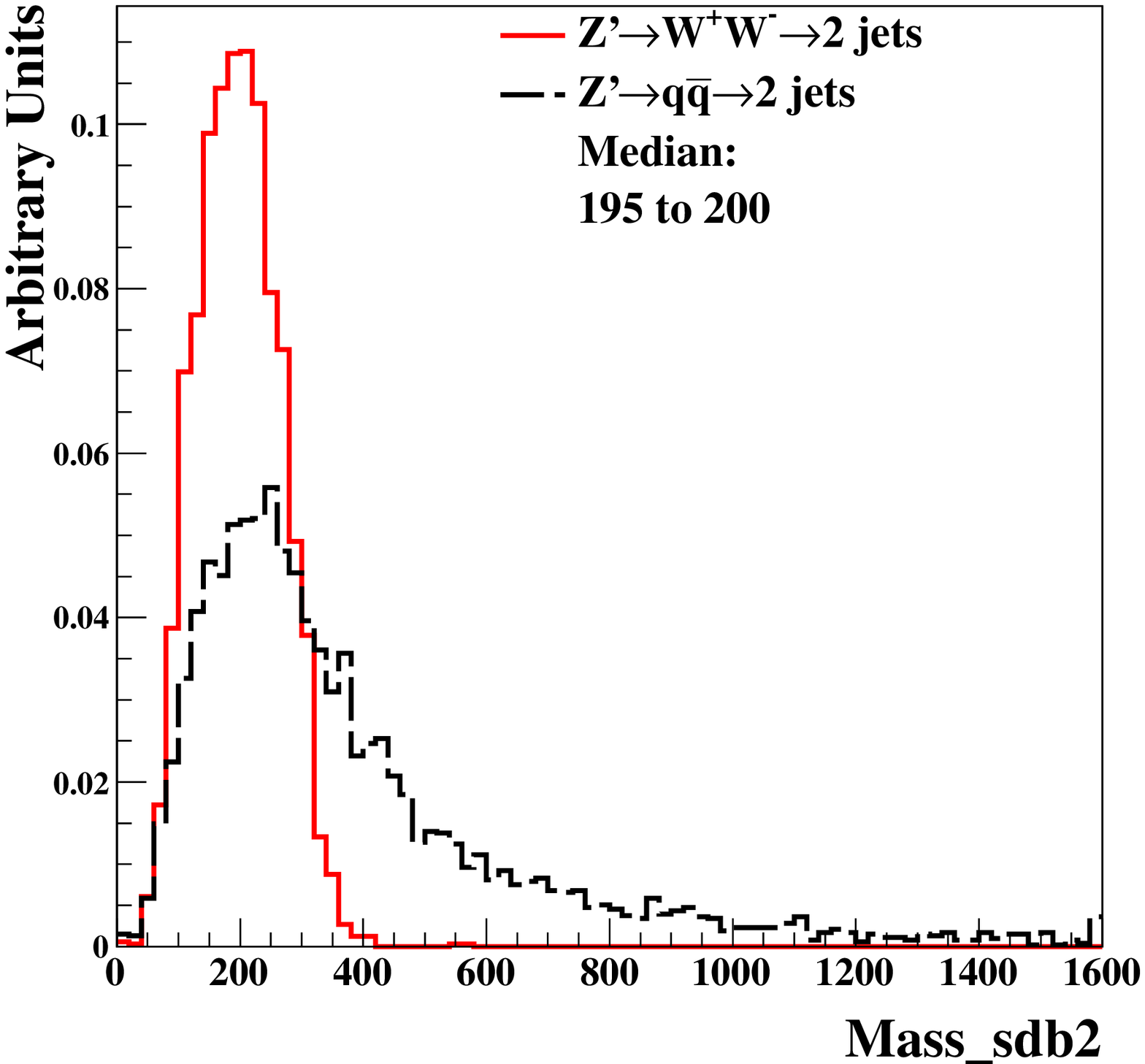}\hfill
   }
\end{center}
\caption{
Distributions of soft drop mass for $\beta=2$, with $M(Z') = 20 $~TeV  and three different detector cell sizes: 20~$\times$~20, 
5~$\times$~5 and 1~$\times$~1 cm$^2$. The signal (background) process is 
$Z'\rightarrow WW$ ($Z'\rightarrow q\bar{q}$).
}
\label{fig:cluster_mass_sdb2_ww}
\end{figure}

\begin{figure}
\begin{center}
  \subfigure[$M(Z')=5$~TeV] {
  \includegraphics[  width=0.45\textwidth]{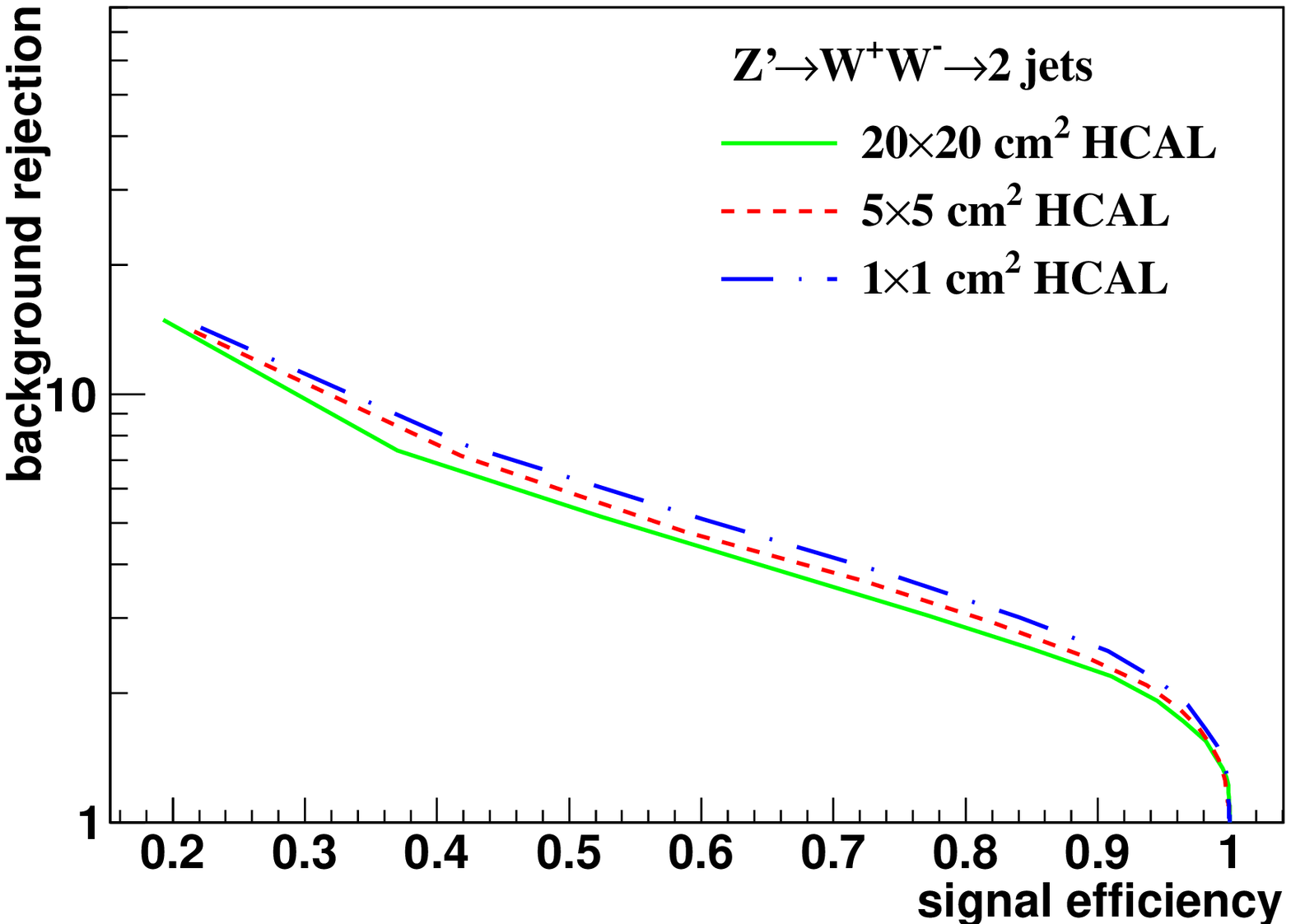}
  }
  \subfigure[$M(Z')=10$~TeV] {
  \includegraphics[  width=0.45\textwidth]{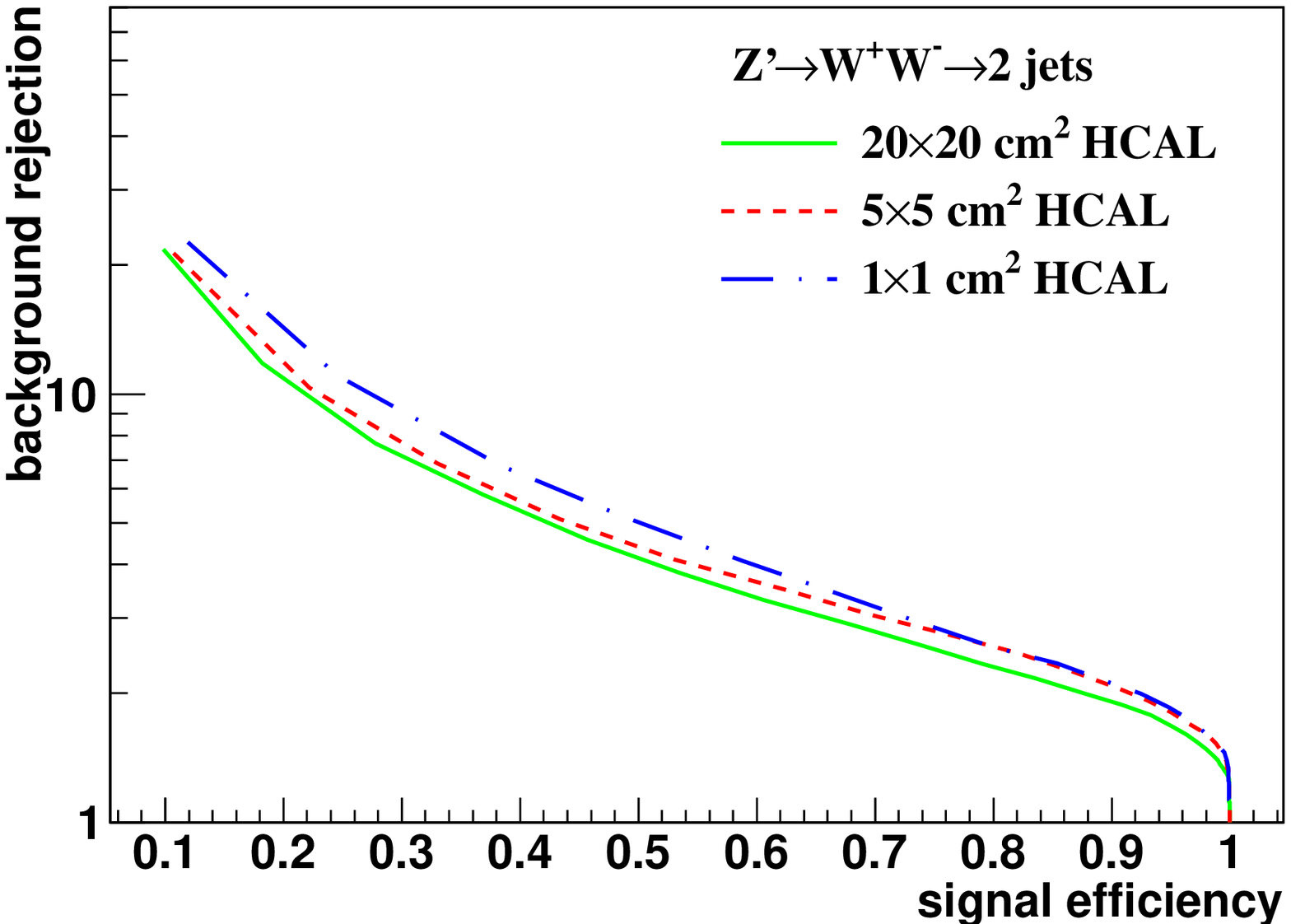}
  }
 \subfigure[$M(Z')=20$~TeV] {
 \includegraphics[  width=0.45\textwidth]{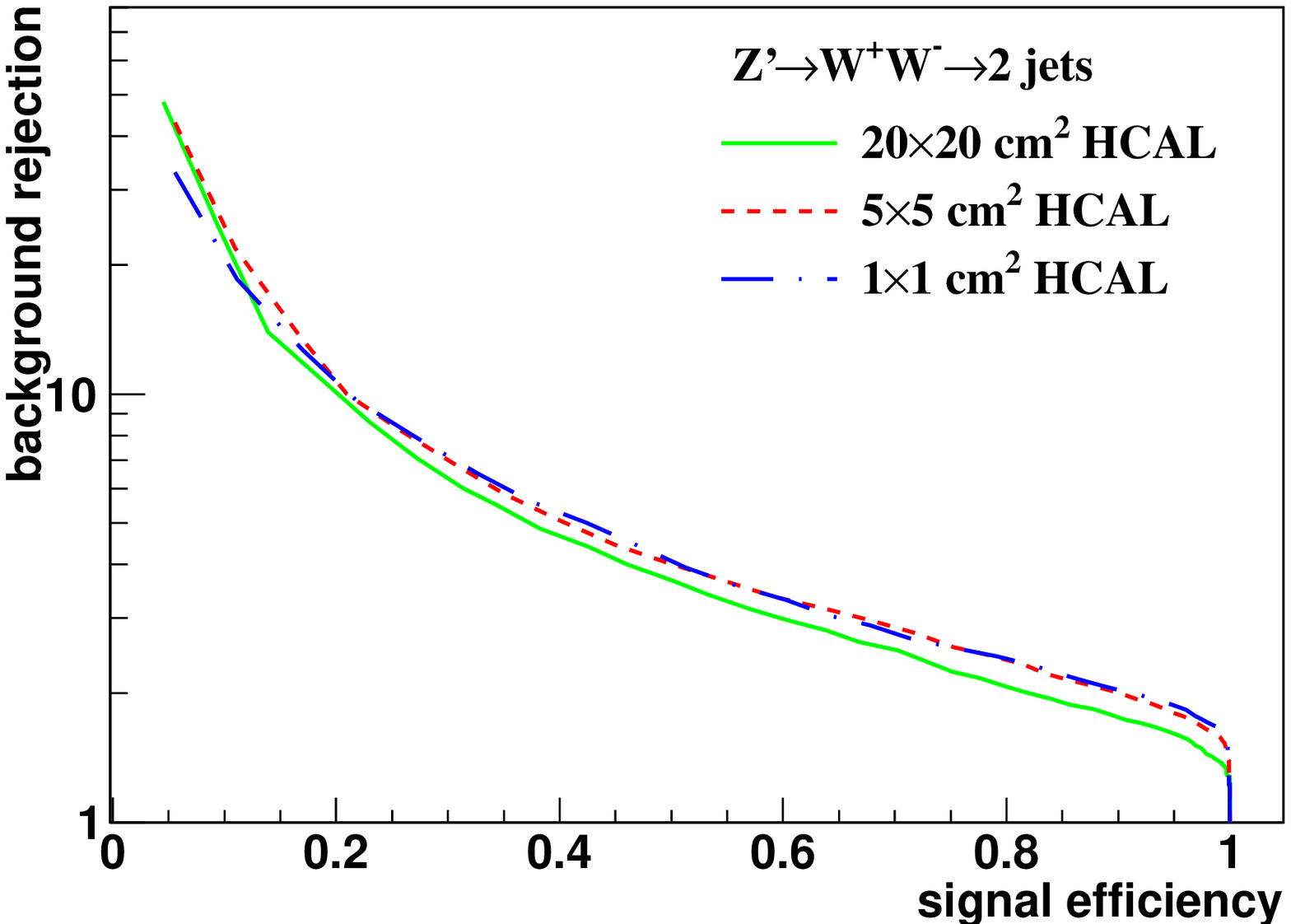}
 }
 \subfigure[$M(Z')=40$~TeV] {
 \includegraphics[  width=0.45\textwidth]{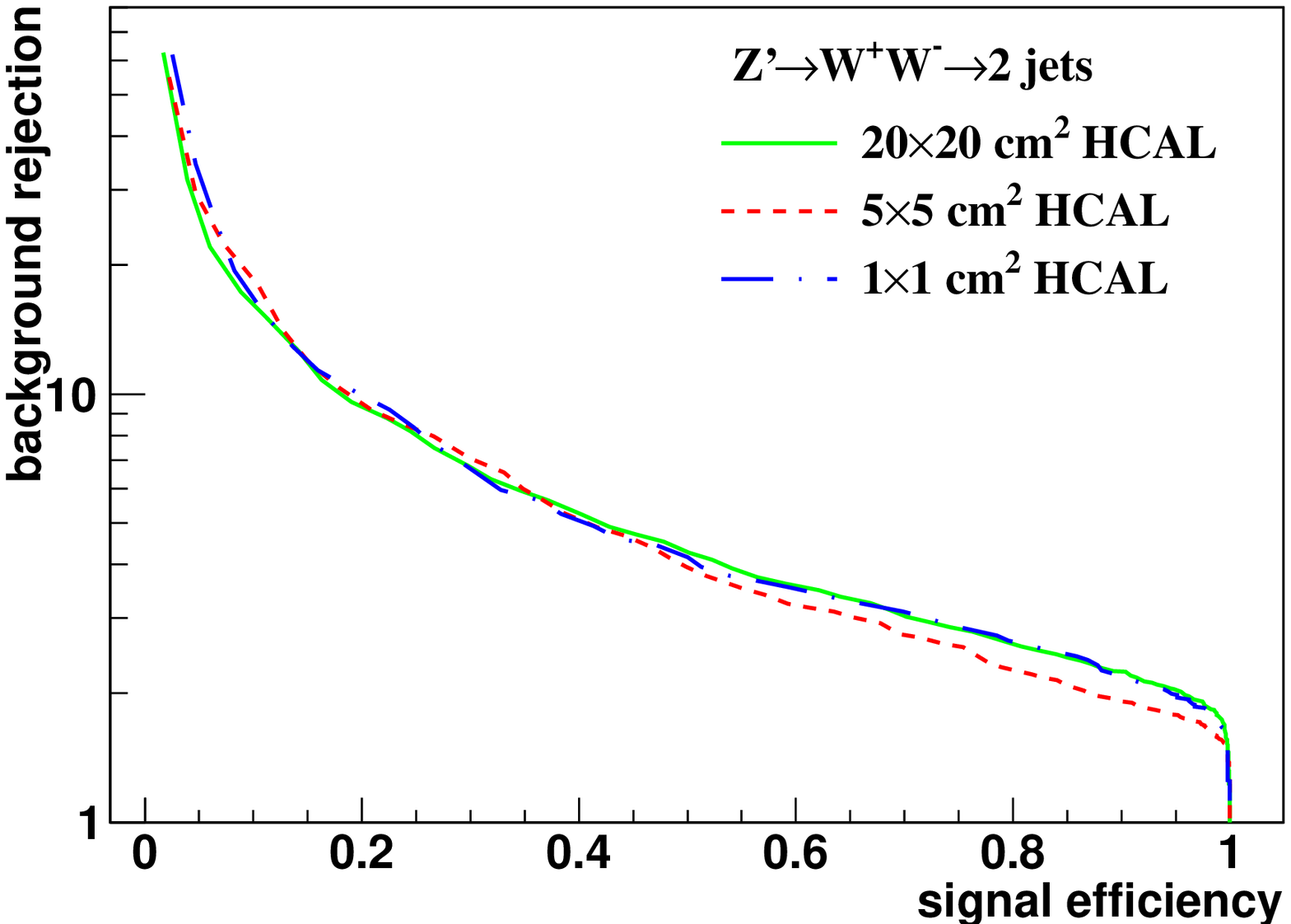}
 }
\end{center}
\caption{
The ROC curves of soft drop mass selection for $\beta=2$
with resonance masses of 5, 10, 20 and 40 TeV. 
Three different detector cell sizes are compared: 20~$\times$~20, 
5~$\times$~5, and 1~$\times$~1 cm$^2$. 
The signal (background) process is $Z'\rightarrow WW$ 
($Z'\rightarrow q\bar{q}$).}
\label{fig:cluster_mass_sdb2_ww_ROC}
\end{figure}

\begin{figure}
\begin{center}
   \subfigure[20~$\times$~20 cm$^2$] {
   \includegraphics[   width=0.3\textwidth]{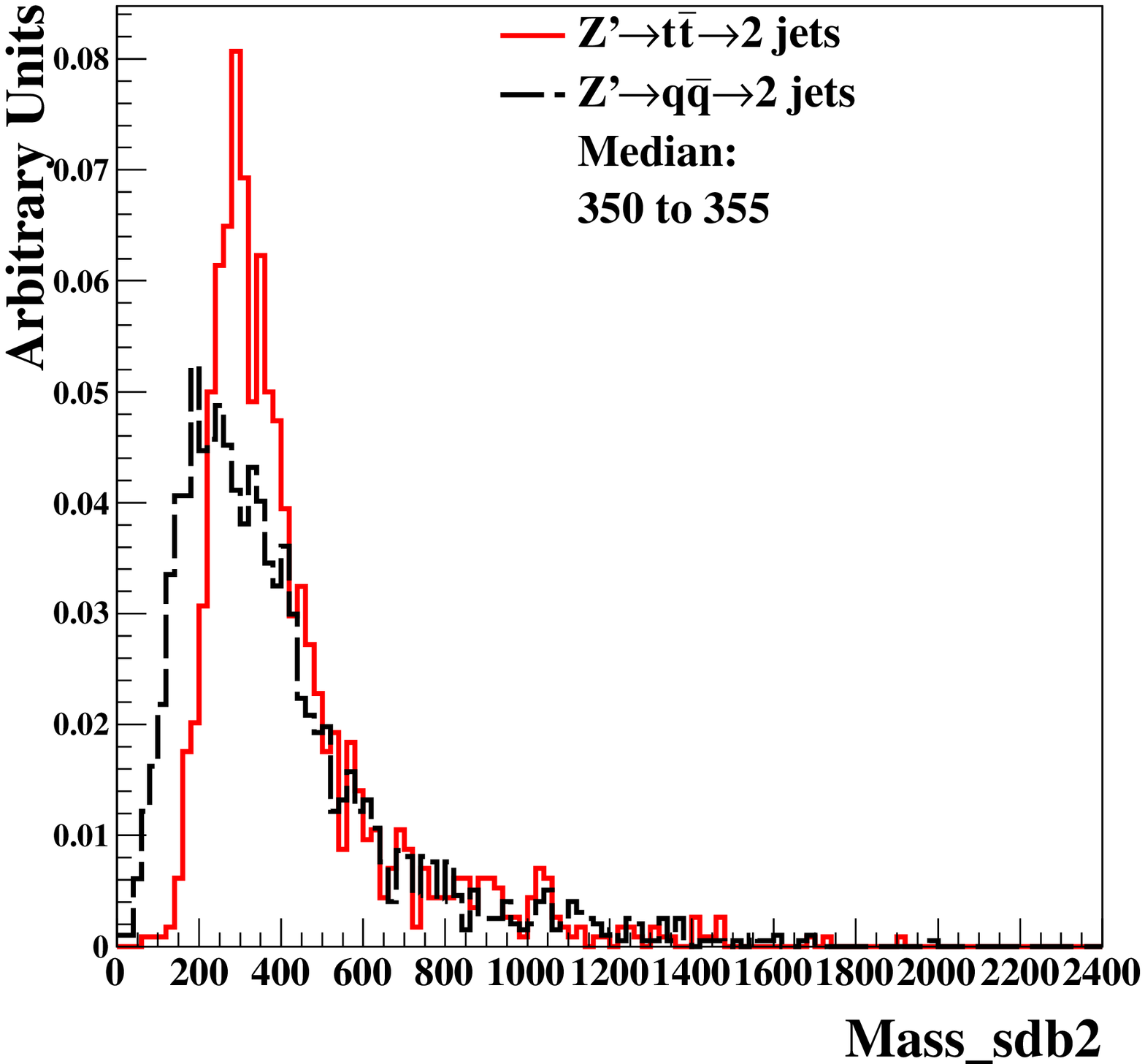}
   }
   \subfigure[5~$\times$~5 cm$^2$] {
   \includegraphics[   width=0.3\textwidth]{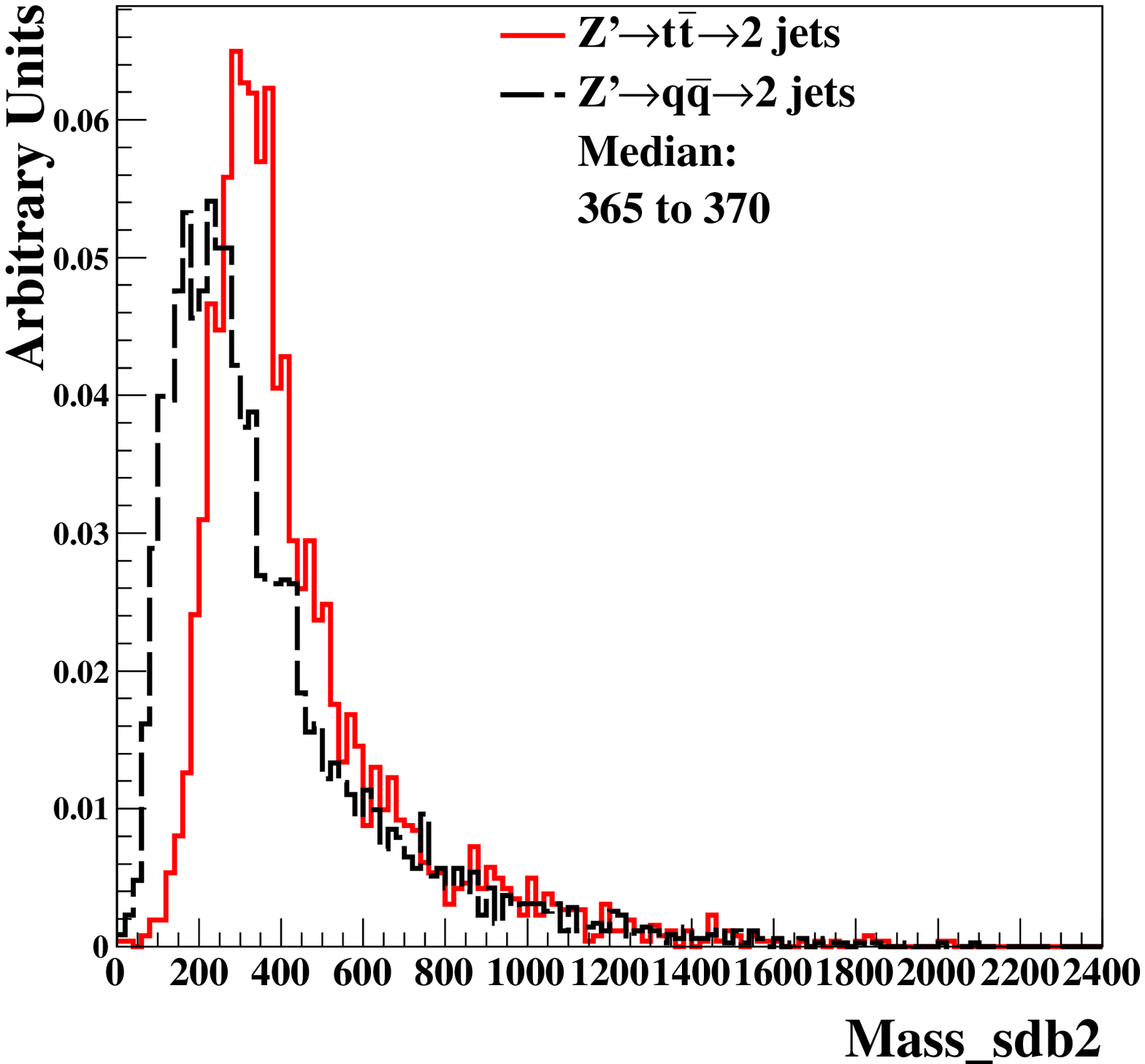}\hfill
   }
   \subfigure[1~$\times$~1 cm$^2$] {
   \includegraphics[   width=0.3\textwidth]{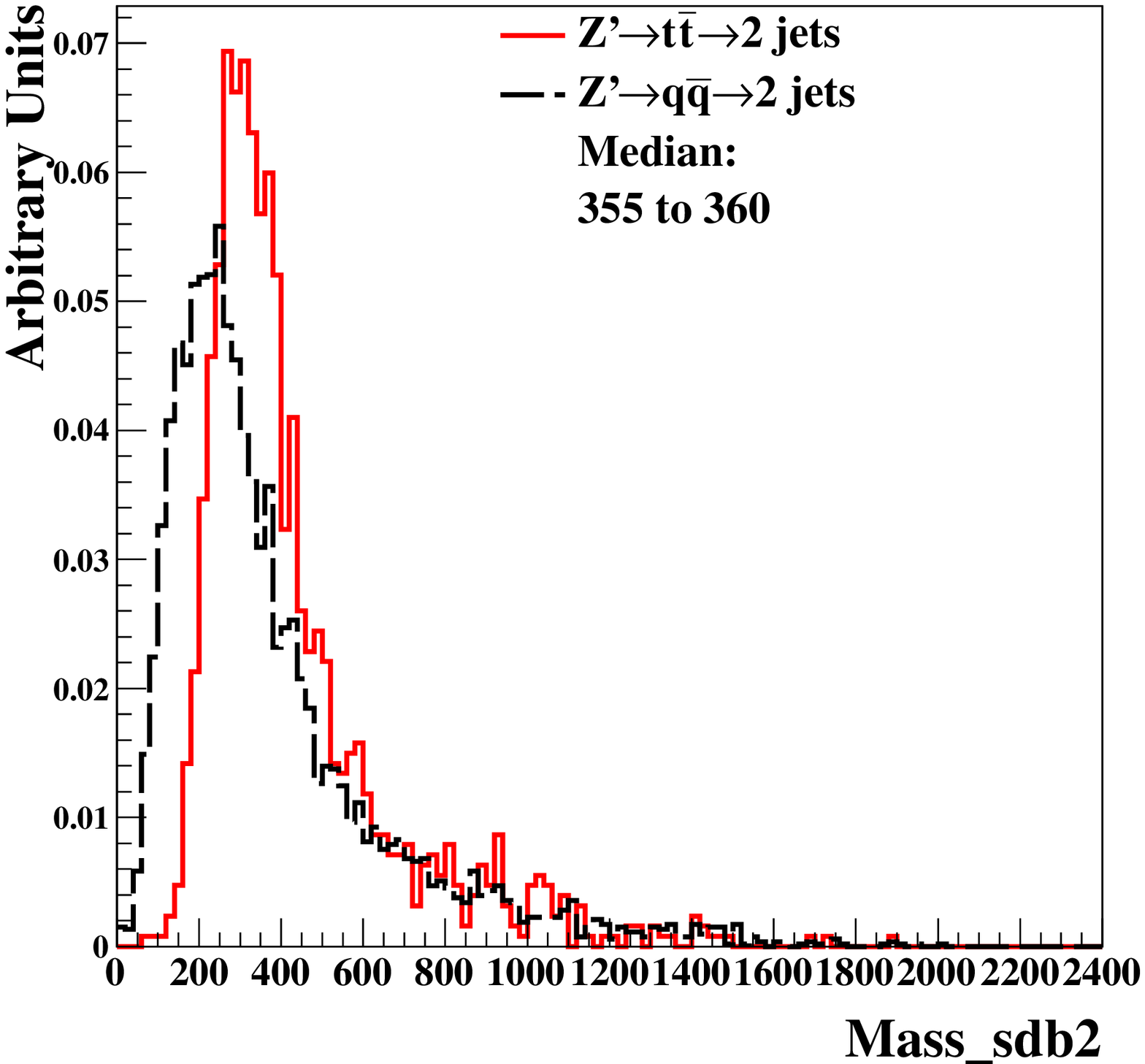}\hfill
   }
\end{center}
\caption{
Distributions of soft drop mass for $\beta =2$, with $M(Z') =20$~TeV  and three different detector cell sizes: 20~$\times$~20, 
5~$\times$~5, and 1~$\times$~1 cm$^2$. The signal (background) process is  
$Z'\rightarrow t\bar{t}$ ($Z'\rightarrow q\bar{q}$).
}
\label{fig:cluster_mass_sdb2_tt}
\end{figure}

\begin{figure}
\begin{center}
  \subfigure[$M(Z')=5$~TeV] {
  \includegraphics[  width=0.45\textwidth]{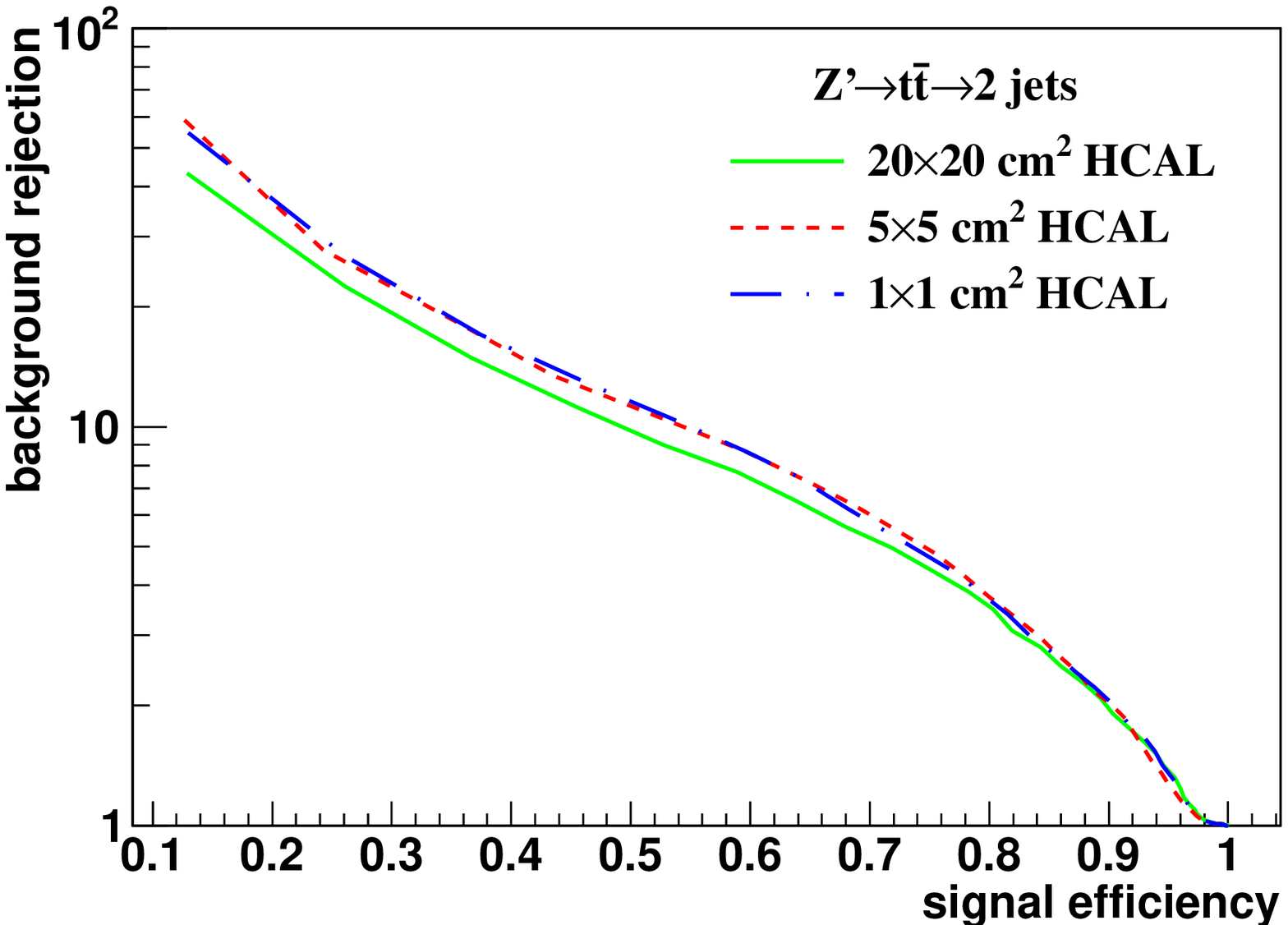}
  }
  \subfigure[$M(Z')=10$~TeV] {
  \includegraphics[  width=0.45\textwidth]{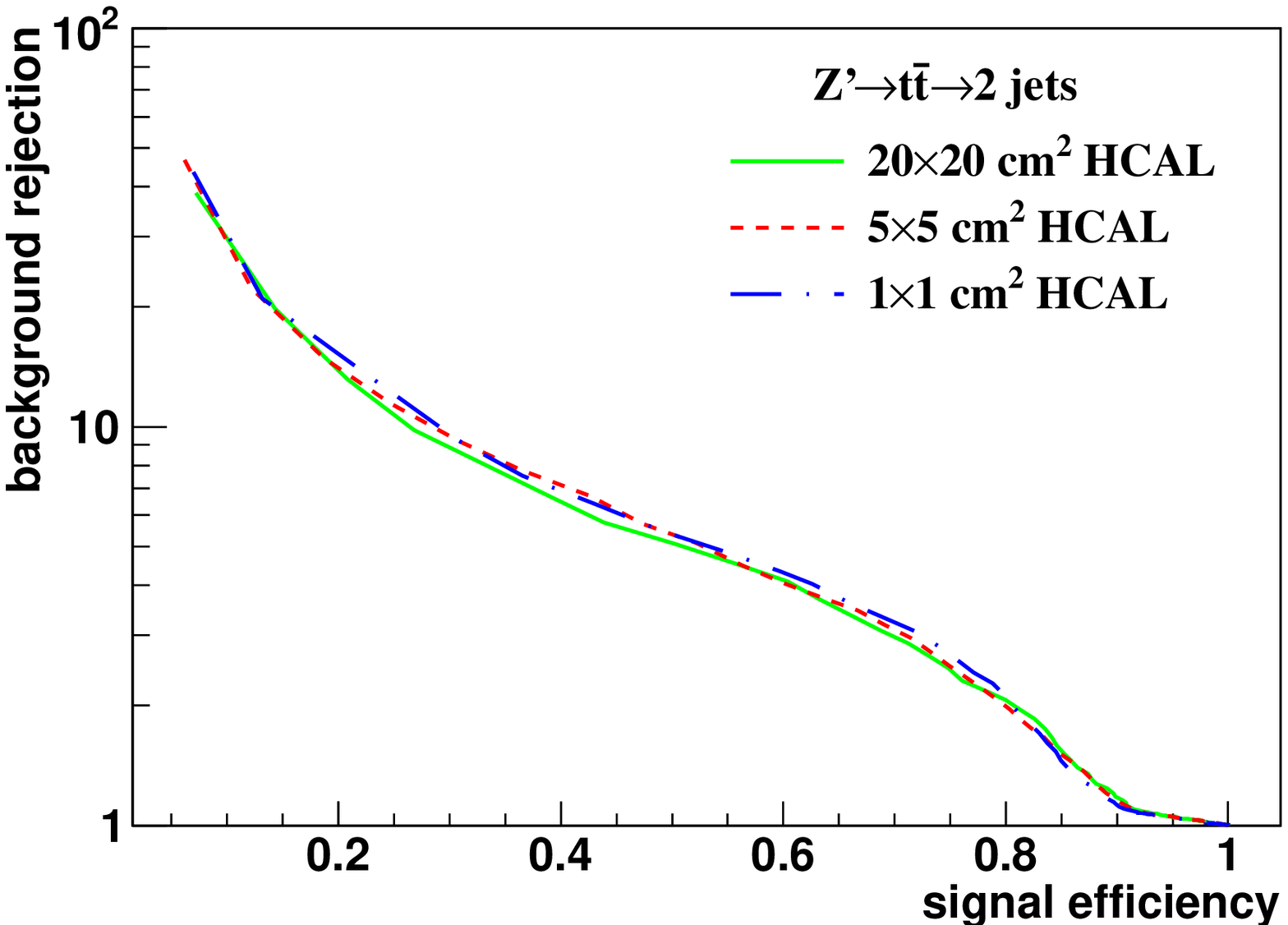}
  }
 \subfigure[$M(Z')=20$~TeV] {
 \includegraphics[  width=0.45\textwidth]{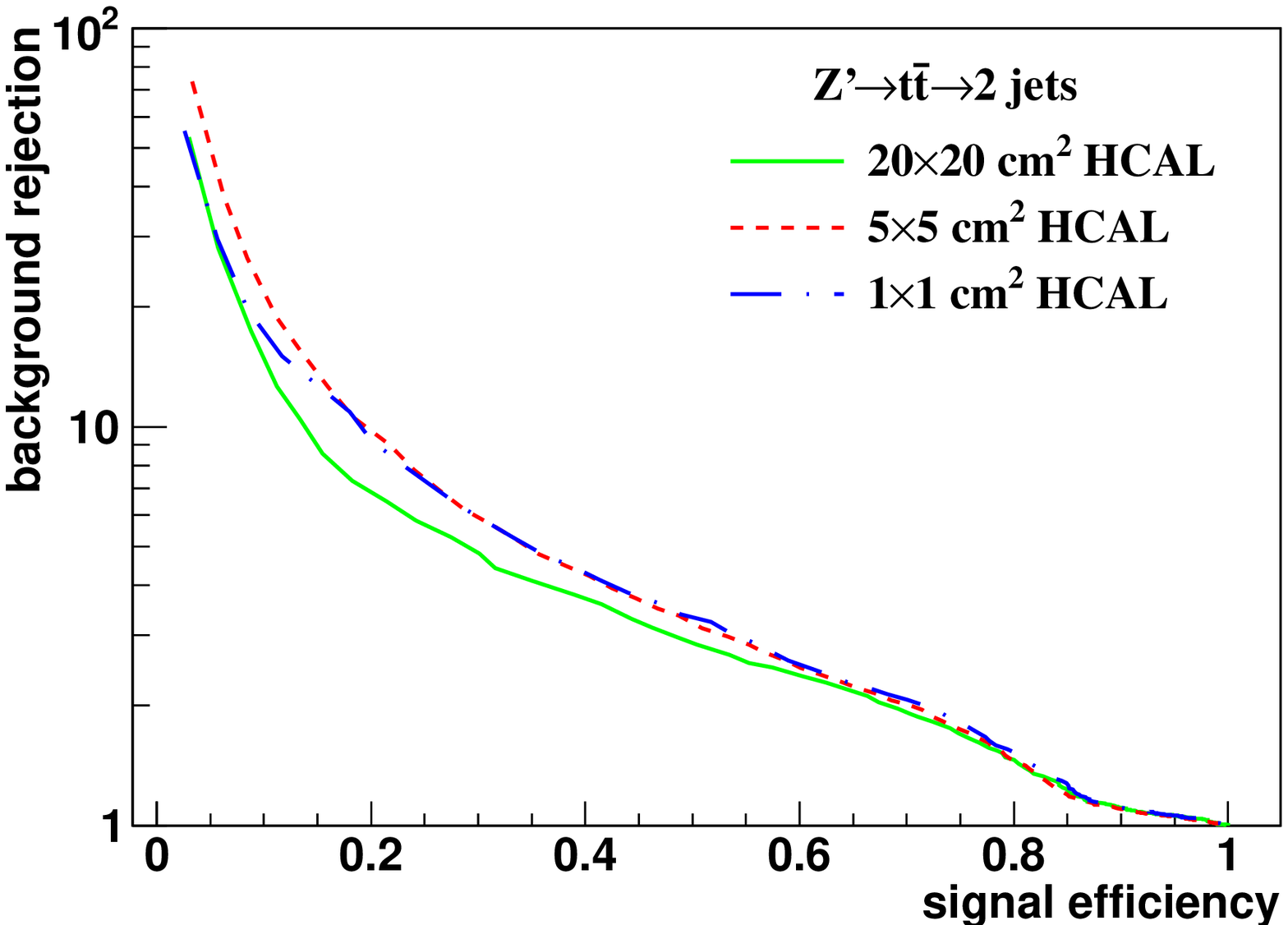}
 }
 \subfigure[$M(Z')=40$~TeV] {
 \includegraphics[  width=0.45\textwidth]{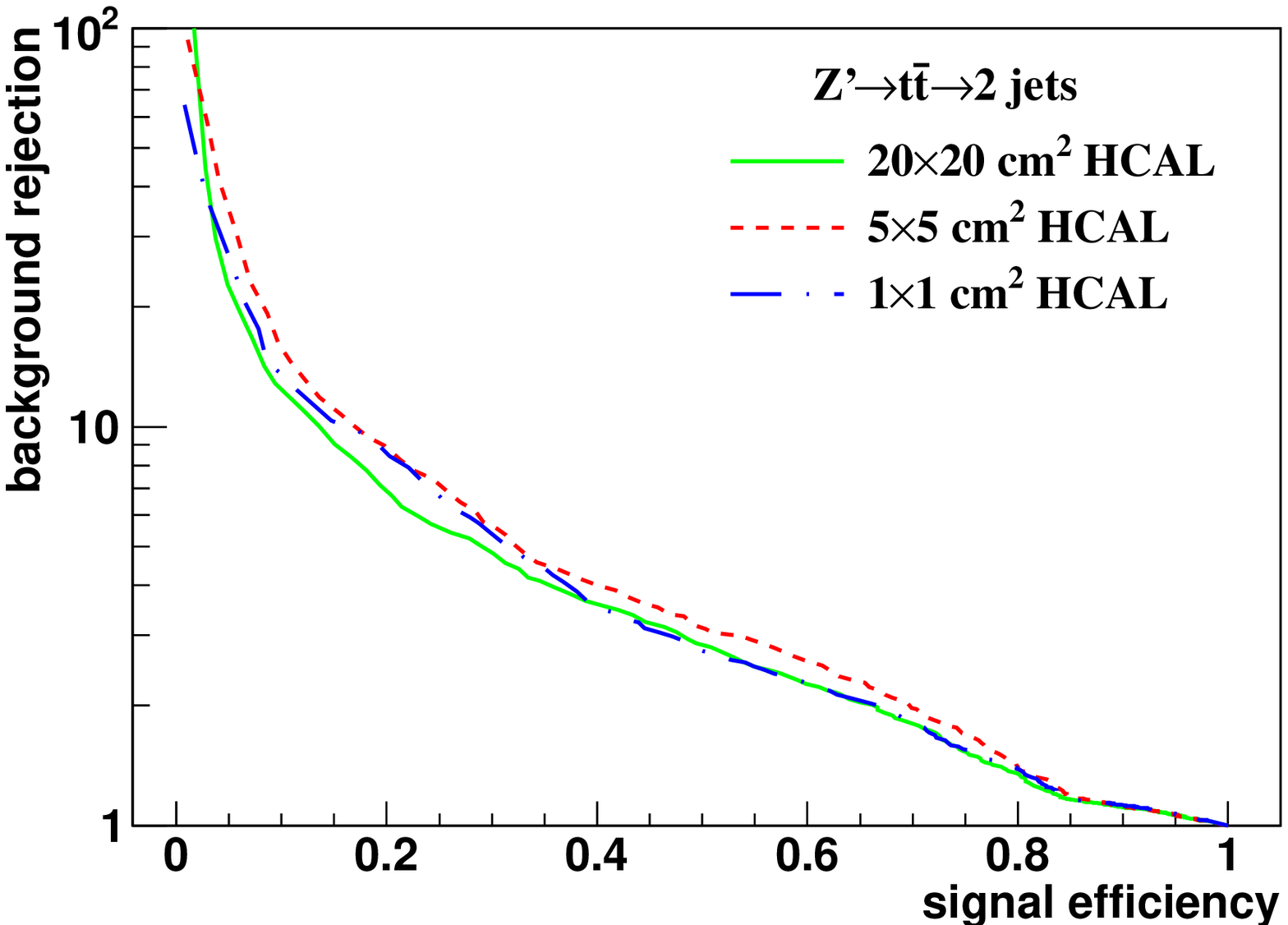}
 }
\end{center}
\caption{
The ROC curves of soft drop mass selection for $\beta=2$
with resonance masses of 5, 10, 20 and 40 TeV. 
Three different detector cell sizes are compared: 20~$\times$~20, 
5~$\times$~5 and 1~$\times$~1 cm$^2$. 
The signal (background) process is $Z'\rightarrow t\bar{t}$
($Z' \rightarrow q\bar{q}$).
}
\label{fig:cluster_mass_sdb2_tt_ROC}
\end{figure}

Figures~\ref{fig:cluster_mass_mmdt_ww},~\ref{fig:cluster_mass_mmdt_tt},~\ref{fig:cluster_mass_sdb2_ww} and~\ref{fig:cluster_mass_sdb2_tt} 
show the distributions for the soft drop mass for $\beta=0$ and $\beta=2$ with  
different resonance masses and detector cell sizes; the signals considered are 
the $Z'\rightarrow WW$ and $Z'\rightarrow t\bar{t}$ processes. 
Figures~\ref{fig:cluster_mass_mmdt_ww_ROC},~\ref{fig:cluster_mass_mmdt_tt_ROC},~\ref{fig:cluster_mass_sdb2_ww_ROC} and~\ref{fig:cluster_mass_sdb2_tt_ROC} 
show the corresponding ROC curves for different detector cell sizes and resonance masses.
The ROC curves are computed with finely-binned histograms; the latter are rebinned coarsely for display purpose only.

These studies show that the reconstruction of soft drop mass improves with decreasing  HCAL cell sizes.
Figures~\ref{fig:cluster_mass_mmdt_ww_ROC} and 
\ref{fig:cluster_mass_mmdt_tt_ROC} show that for $\beta=0$ the 
smallest detector cell size, 
$1\times1$~cm$^2$, has the best separation power at 
resonance masses of 5, 10, and 20~TeV when the signal is the $Z' \rightarrow WW$ process,  and 
at  resonance masses of 10 and 20 TeV when the signal is the $Z' \rightarrow t\bar{t}$ process.
However, for $\beta=2$, Figs.~\ref{fig:cluster_mass_sdb2_ww_ROC} and \ref{fig:cluster_mass_sdb2_tt_ROC} 
show that the smallest detector cell size 
does not have improvements in the separation power when compared with 
larger cell sizes. In fact, the performance for the three cell sizes is  
similar. 

Note that the separation between ROC curves depends 
on the physics variable and on the boost of the top quarks or the $W$ bosons. For example, 
the similarity between the ROC curves shown in Fig.~\ref{fig:cluster_mass_mmdt_tt_ROC}(a)
is due to the insufficient boost of the top quarks, 
 where even the largest cell size provides adequate discrimination from unstructured jets.
On the other hand, Fig.~\ref{fig:cluster_mass_mmdt_tt_ROC}(d) does not show a difference 
between the ROC curves because the boost is too high,  where even the smallest  cell size is not small enough, or the lateral spreading of the particle showers 
 prevents discrimination from unstructured jets. For both $Z' \to WW$ and $Z' \to t \bar{t}$ processes at $M(Z’) = 40$~TeV, the typical opening angle between the daughter jets 
 is 17 mrad or less; the smallest cell size we consider (1~$\times$~1~cm$^2$ or $\Delta \eta \times \Delta \phi = 0.0043 \times 0.0043$) 
 is not able to distinguish the substructure at this angular scale.  

We also find that the  soft drop mass with $\beta=0$ has better 
performance for distinguishing signal from background than with  $\beta=2$. Therefore, we will 
apply requirements on the soft drop mass with $\beta=0$  when studying the other jet substructure 
variables.

\section{Detector performance with jet substructure variables}\label{sec:Jsubvar}
In this section, we use several jet substructure variables to study the performance with various detector cell sizes and resonance masses.

\subsection{$N$-subjettiness \label{sec:nsub}}
The variable $N$-subjettiness~\cite{Thaler:2010tr}, denoted by $\tau_N$, is designed to 
``count'' the number of subjet(s) in a large radius jet in order to separate 
signal jets from decays of heavy bosons and background jets from QCD processes. 
 $\tau_N$ is the $\pt$-weighted angular distance between each jet 
constituent and the closest subjet axis: 
\begin{equation}\label{eq:Nsub_1}
\tau_{N}=\frac{1}{d_{0}}\sum_{k}p_{T,k} \; \mathrm{min}\{\Delta R_{1,k},\Delta R_{2,k},.....\Delta R_{N,k}\},
\end{equation}
with a normalization factor $d_0$: \[d_{0}=\sum_{k}p_{T,k} R_{0}.\] 
The $k$ index runs over all constituent particles in a given large radius jet, 
$p_{T,k}$ is the transverse momentum of each individual constituent, 
$\Delta R_{j,k}=\sqrt{(\Delta y)^{2}+(\Delta \phi)^{2}}$ is the distance 
between the constituent $k$ and the candidate subjet axis $j$ in the 
$y-\phi$ plane. $R_{0}$ is the characteristic jet radius used in 
the anti-$k_t$ jet algorithm. 

This analysis uses the jet reconstruction described in Sect.~\ref{sec:sim}. 
The subjet axes are obtained by running the 
exclusive $k_{t}$ algorithm~\cite{Catani:246812} and reversing the last $N$ clustering steps. 
Namely, when $\tau_N$ is computed, the $k_{t}$ algorithm is forced to return 
exactly $N$ jets. If a large radius jet has $N$ subjet(s), its $\tau_{N}$ is 
smaller than $\tau_{N-1}$. Therefore, in our analysis, 
the ratios $\tau_{21} \equiv \tau_{2}/\tau_{1}$ and $\tau_{32} \equiv \tau_{3}/\tau_{2}$ 
are used to distinguish the one-prong background jets and 
the two-prong jets from $W$ boson decays or the three-prong jets from top quark decays. 

Following the suggestion of Ref.~\cite{Dreyer:2018tjj}, the requirement on the 
soft drop mass with $\beta=0$ is applied before the study of $N$-subjettiness. 
For each detector configuration and resonance mass, the soft drop mass prerequisite window  
is determined as follows. The window is initialized by the median bin of the soft drop 
mass histogram from simulated signal events. Comparing the adjacent bins, the bin with the larger number of events is included to extend the mass window iteratively. The procedure is 
repeated until the prerequisite mass window cut reaches a signal  efficiency of 75\%.

\begin{figure}
\begin{center}
   \subfigure[20~$\times$~20 cm$^2$] {
   \includegraphics[width=0.3\textwidth]{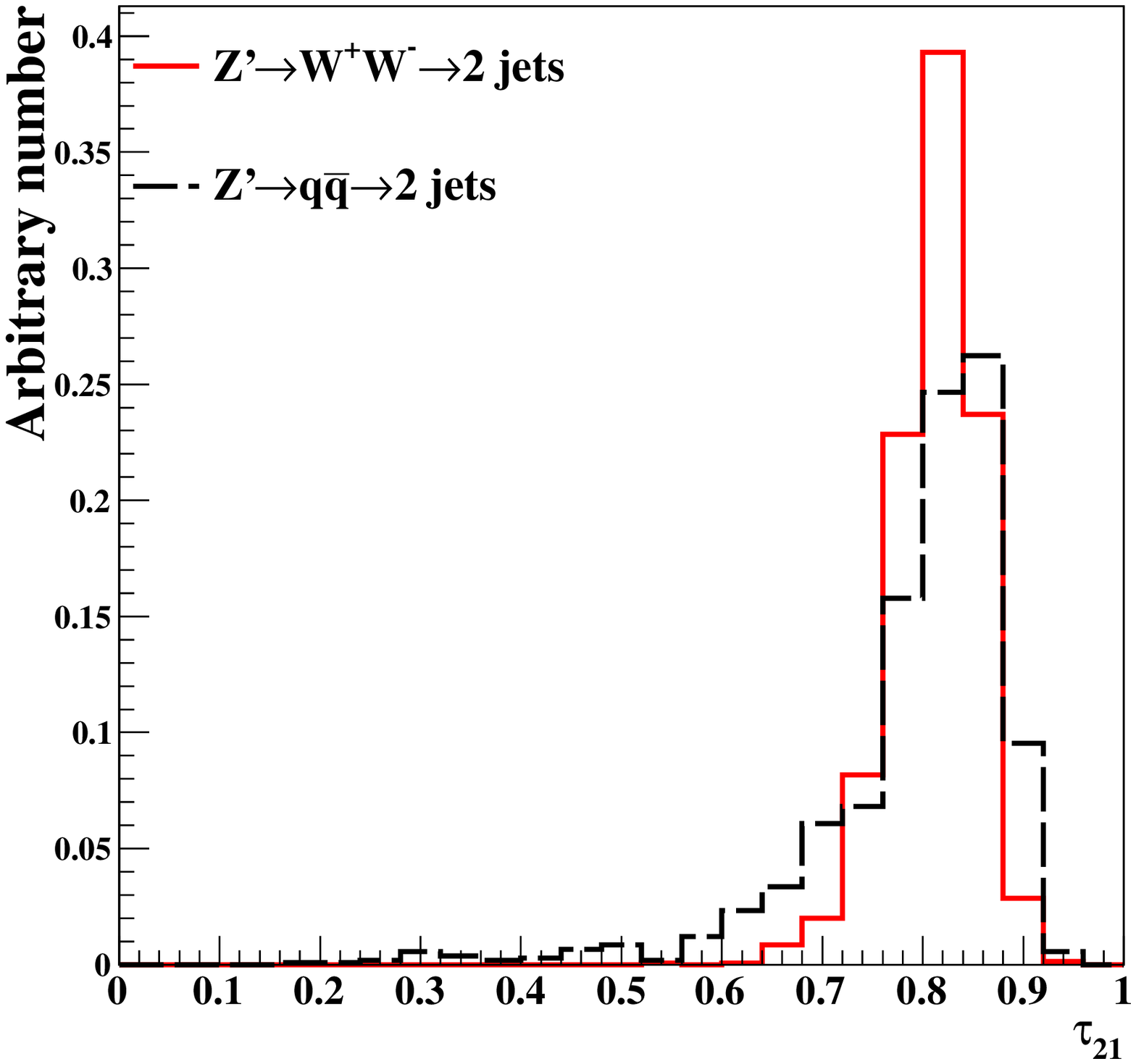}
   }
   \subfigure[5~$\times$~5 cm$^2$] {
   \includegraphics[width=0.3\textwidth]{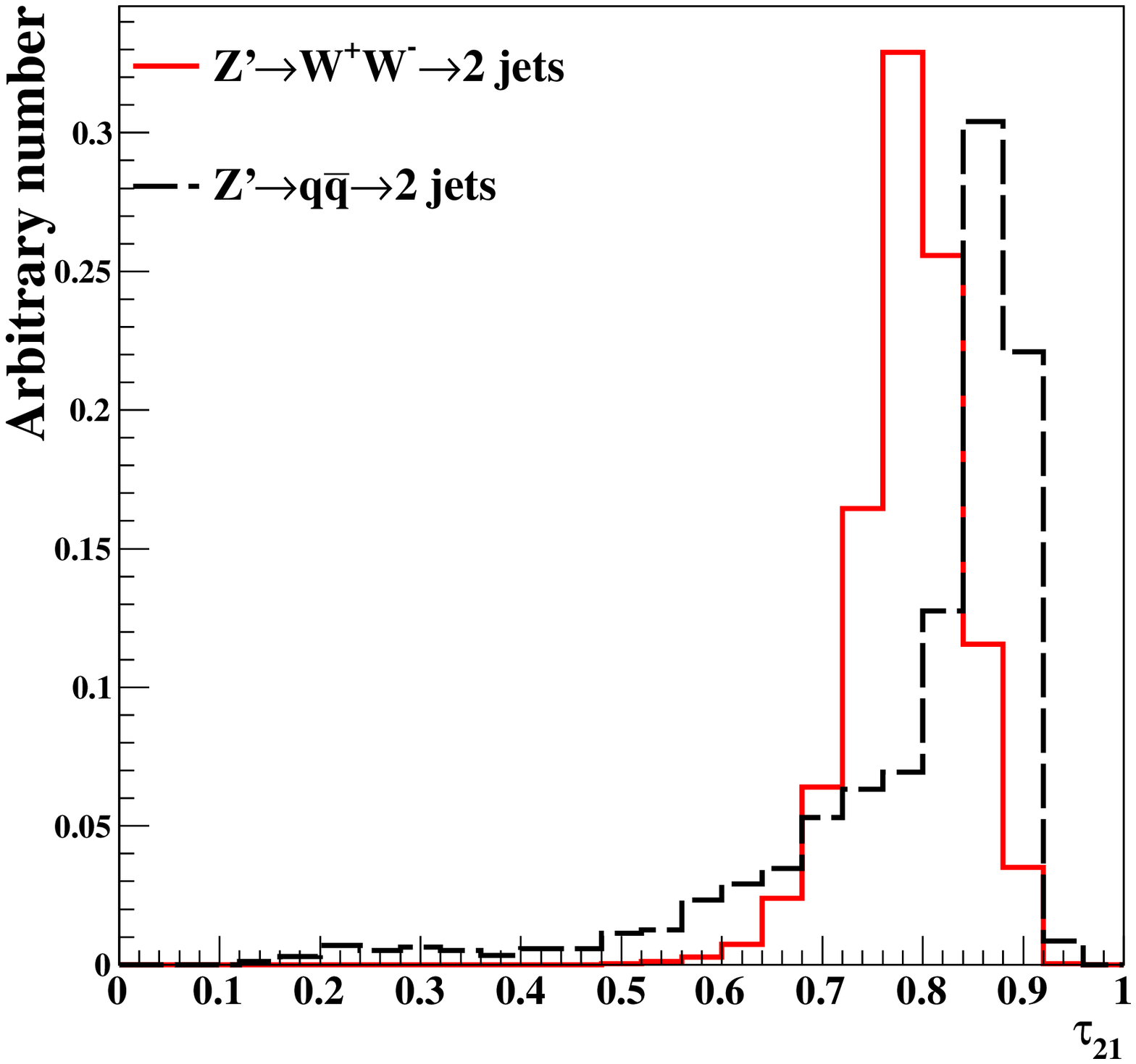}
   }
   \subfigure[1~$\times$~1 cm$^2$] {
   \includegraphics[width=0.3\textwidth]{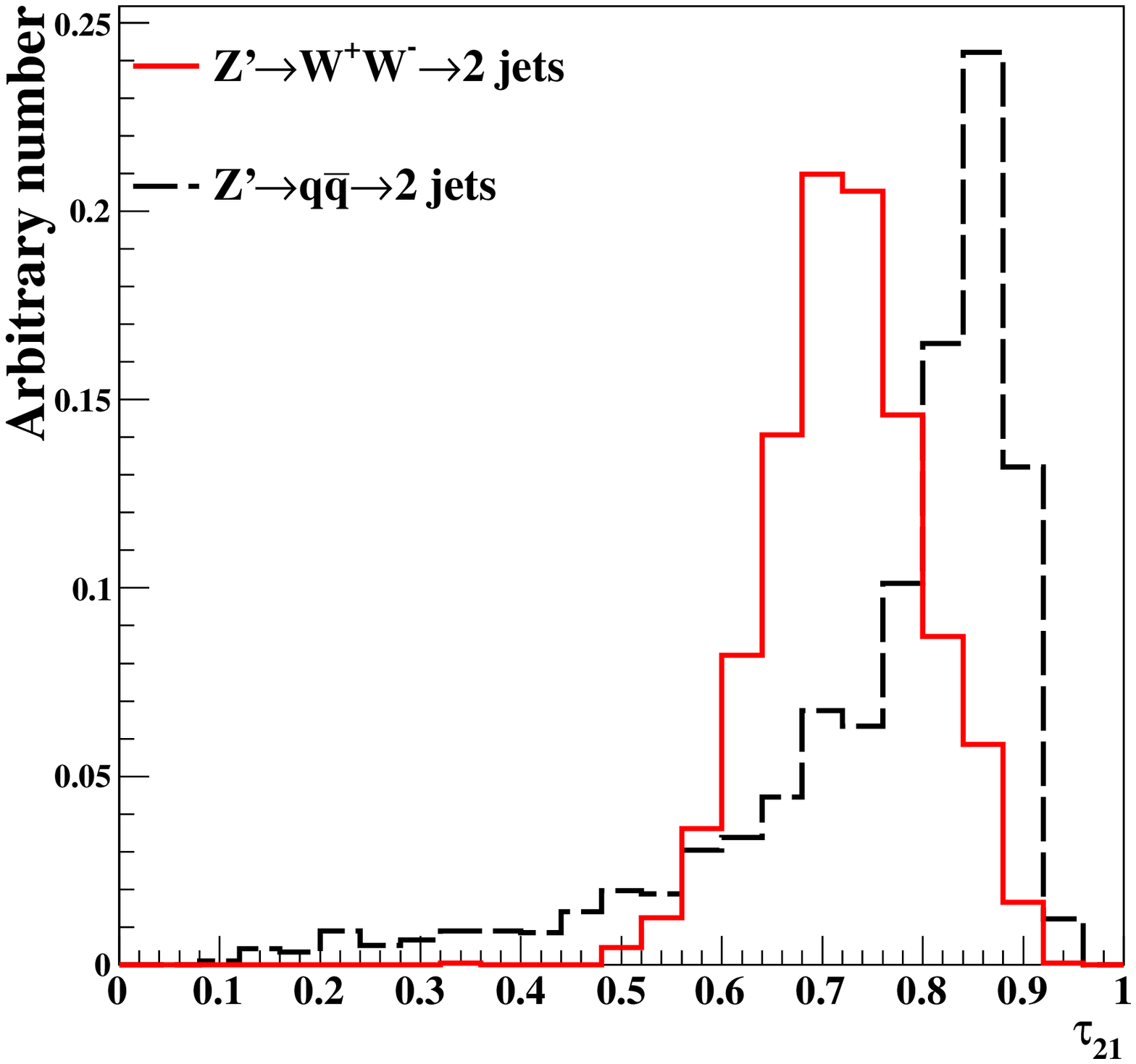}
   }
\end{center}
\caption{Distributions of $\tau_{21}$ for $M(Z') = 20$~TeV for different 
detector granularities. Cell sizes of 20~$\times$~20, 5~$\times$~5, and 1~$\times$~1~cm$^2$ 
are shown here. \label{fig:Rawhit_05GeV_tau21_Dis}}
\end{figure}

\begin{figure}
\begin{center}
   \subfigure[$M(Z')=5$~TeV] {
   \includegraphics[width=0.43\textwidth]{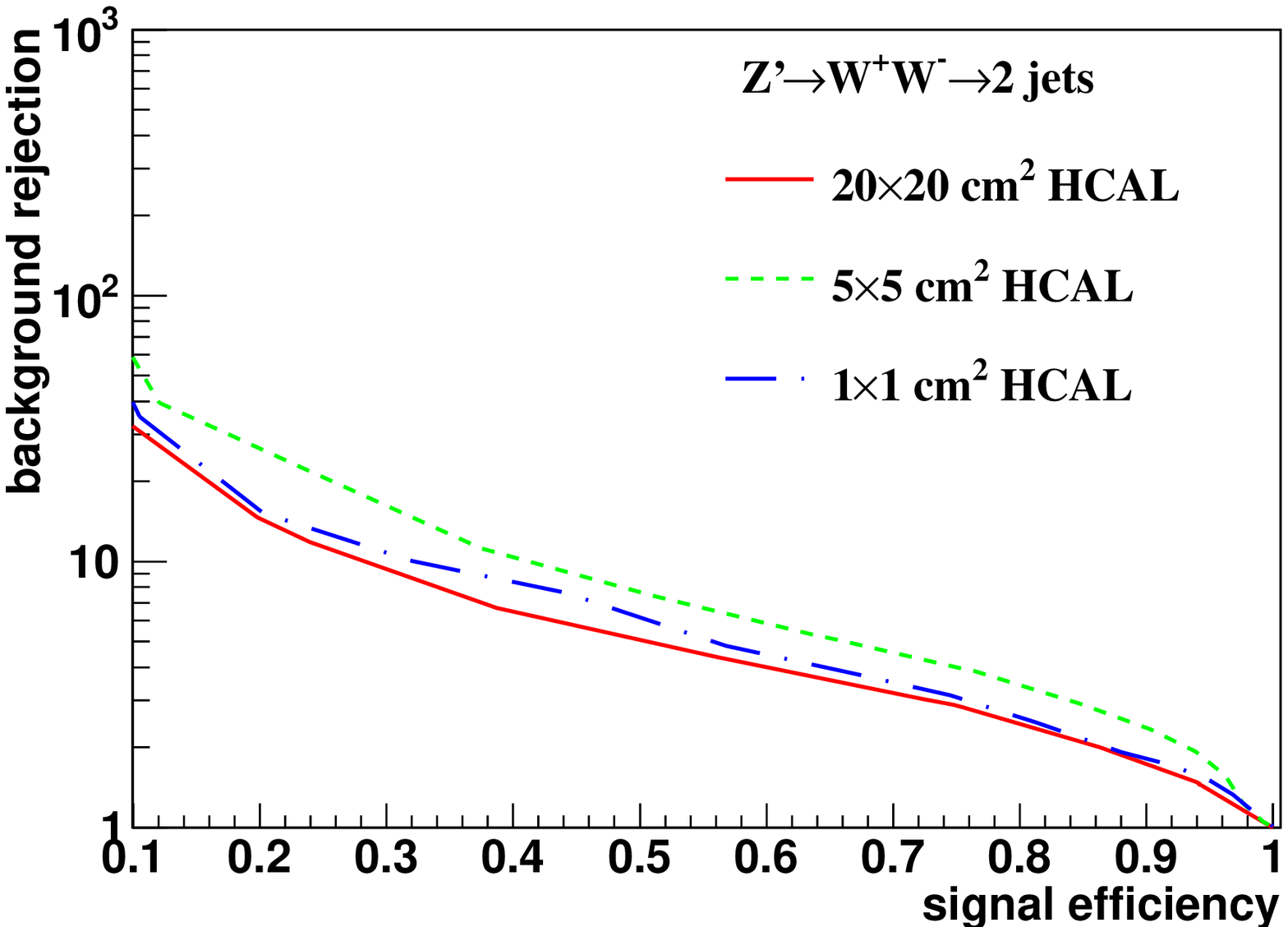}
   }
   \subfigure[$M(Z')=10$~TeV] {
   \includegraphics[width=0.43\textwidth]{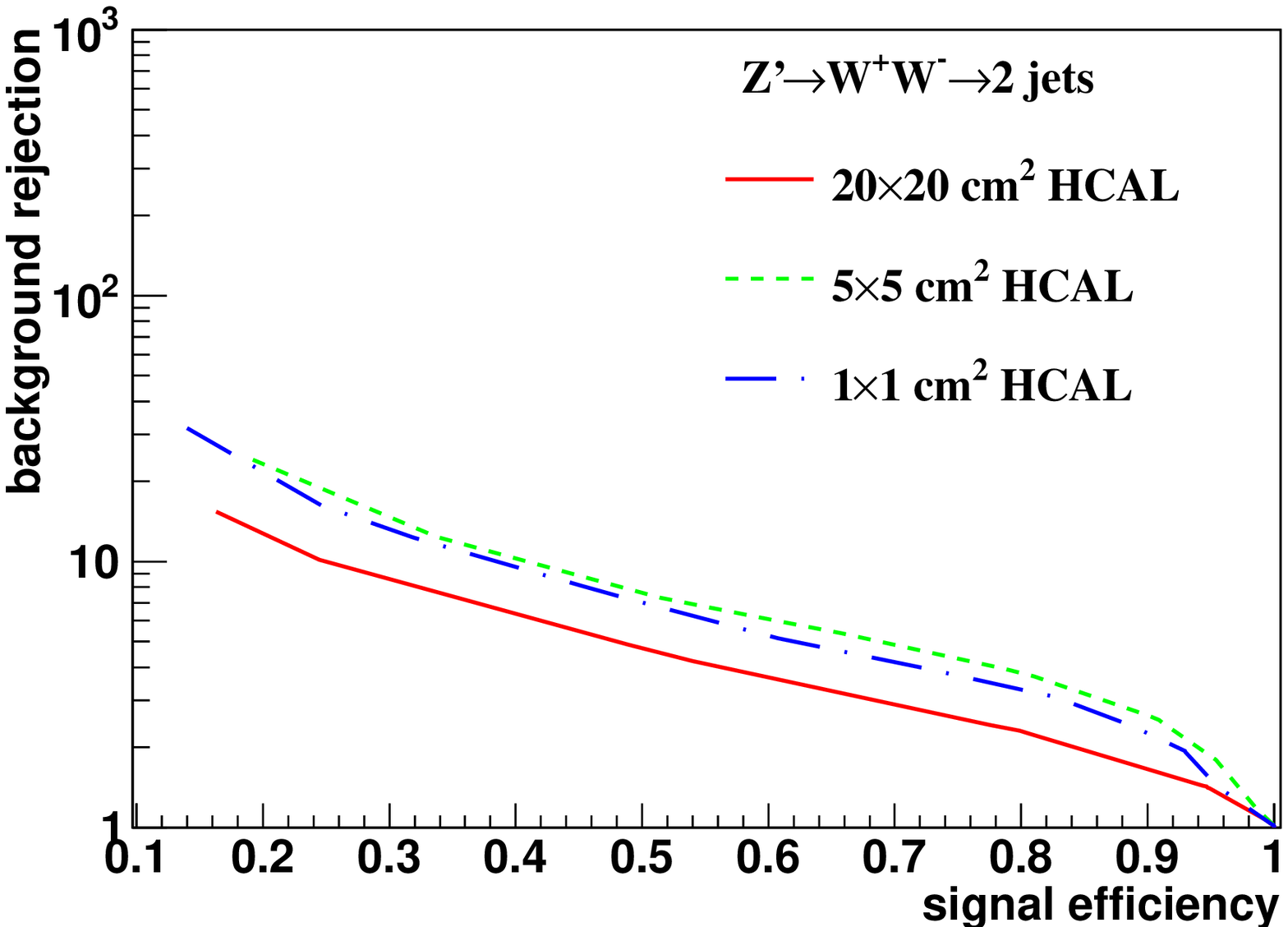}
   }
   \subfigure[$M(Z')=20$~TeV] {
   \includegraphics[width=0.43\textwidth]{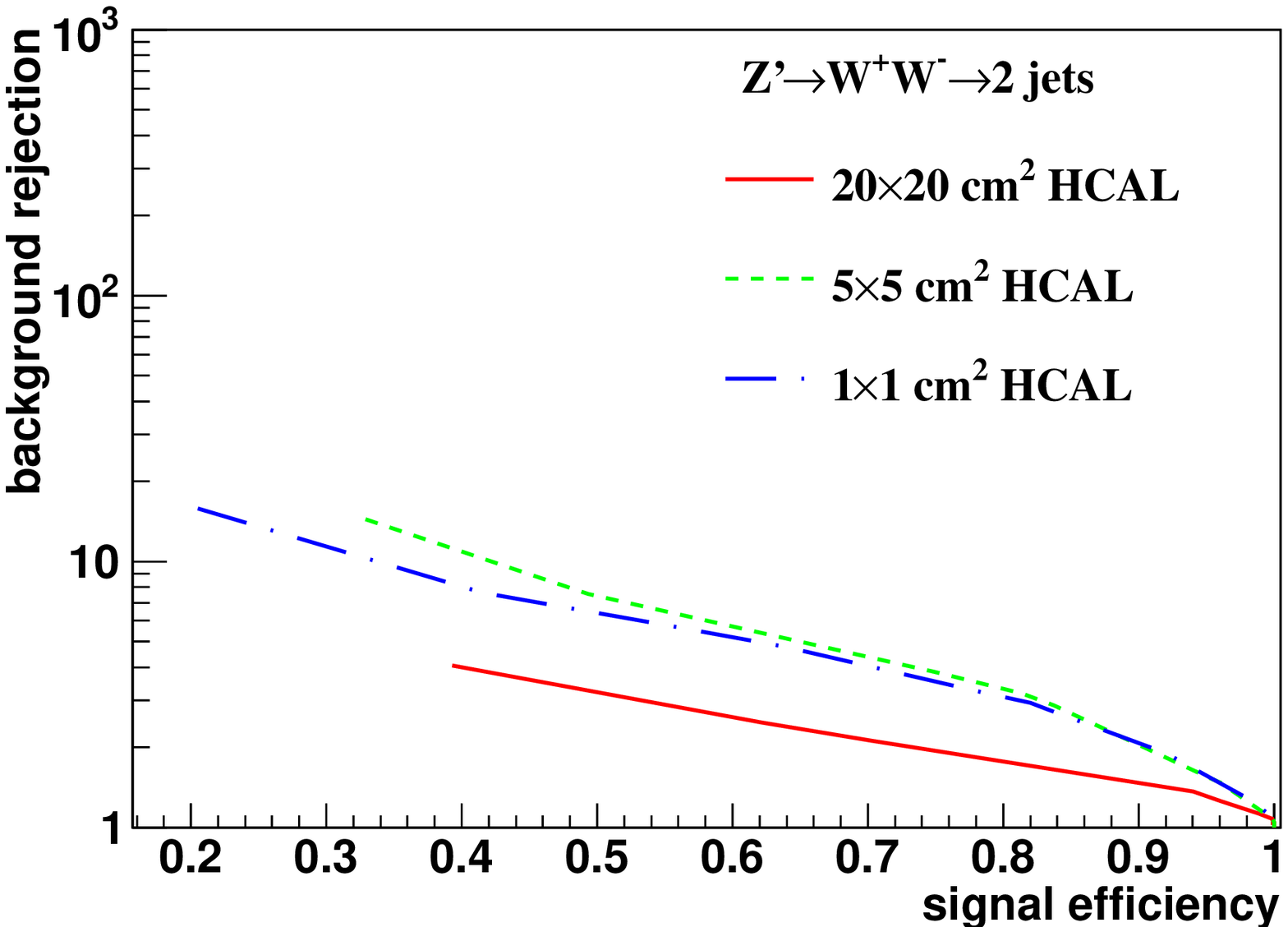}
   }
   \subfigure[$M(Z')=40$~TeV] {
   \includegraphics[width=0.43\textwidth]{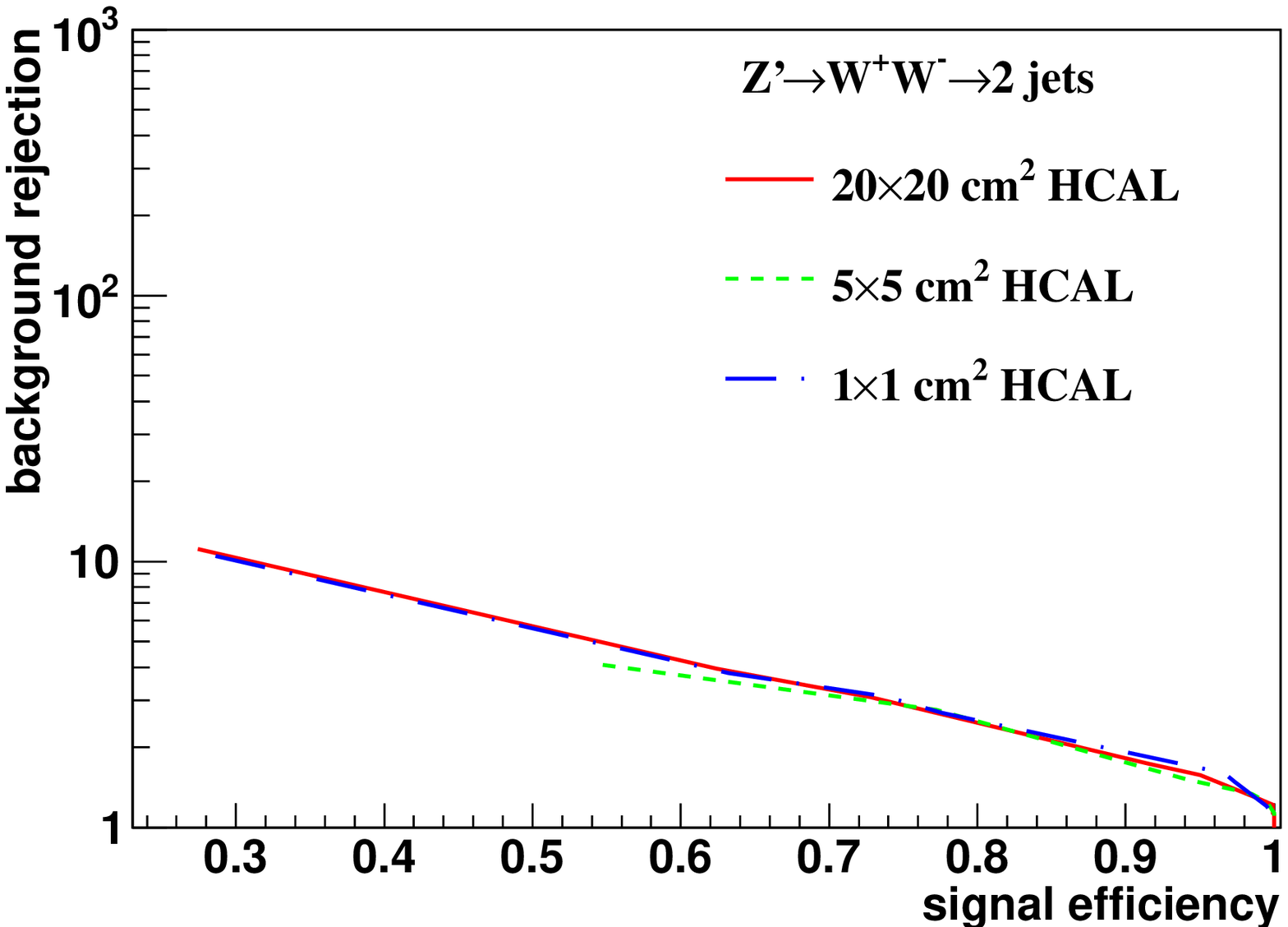}
   }
\end{center}
\caption{Signal efficiency versus background rejection rate using $\tau_{21}$.
Resonance masses of (a) 5~TeV, (b) 10~TeV, (c) 20~TeV and (d) 40~TeV are shown 
here. 
In each figure, the three ROC curves correspond to different cell sizes.
\label{fig:Rawhit_05GeV_tau21_ROC}
}
\end{figure}

With this {\it a-priori} mass window pre-selection, the signal and background efficiencies of 
various $\tau_{21}$ and $\tau_{32}$ window cuts are scanned. 
Since some of the background distributions have long tails and leak into the 
signal-dominated region, we use the following method based on the 
Neyman-Pearson lemma to determine the $\tau$ windows. 
First, we take the ratio of the signal to background $\tau_{21}$ (or $\tau_{32}$) 
histograms. The window is initialized by the bin with the maximum signal to 
background ratio (S/N).  
Comparing the adjacent bins,  the bin with the larger S/N is included  to extend the $\tau_{21}$ (or $\tau_{32}$) 
selection window iteratively.  Every window has its corresponding $\epsilon_\mathrm{sig}$ and 
1/$\epsilon_\mathrm{bkg}$ and an ROC curve is mapped out. 

\begin{figure}
\begin{center}
   \subfigure[20$\times$20 cm$^2$] {
   \includegraphics[width=0.3\textwidth]{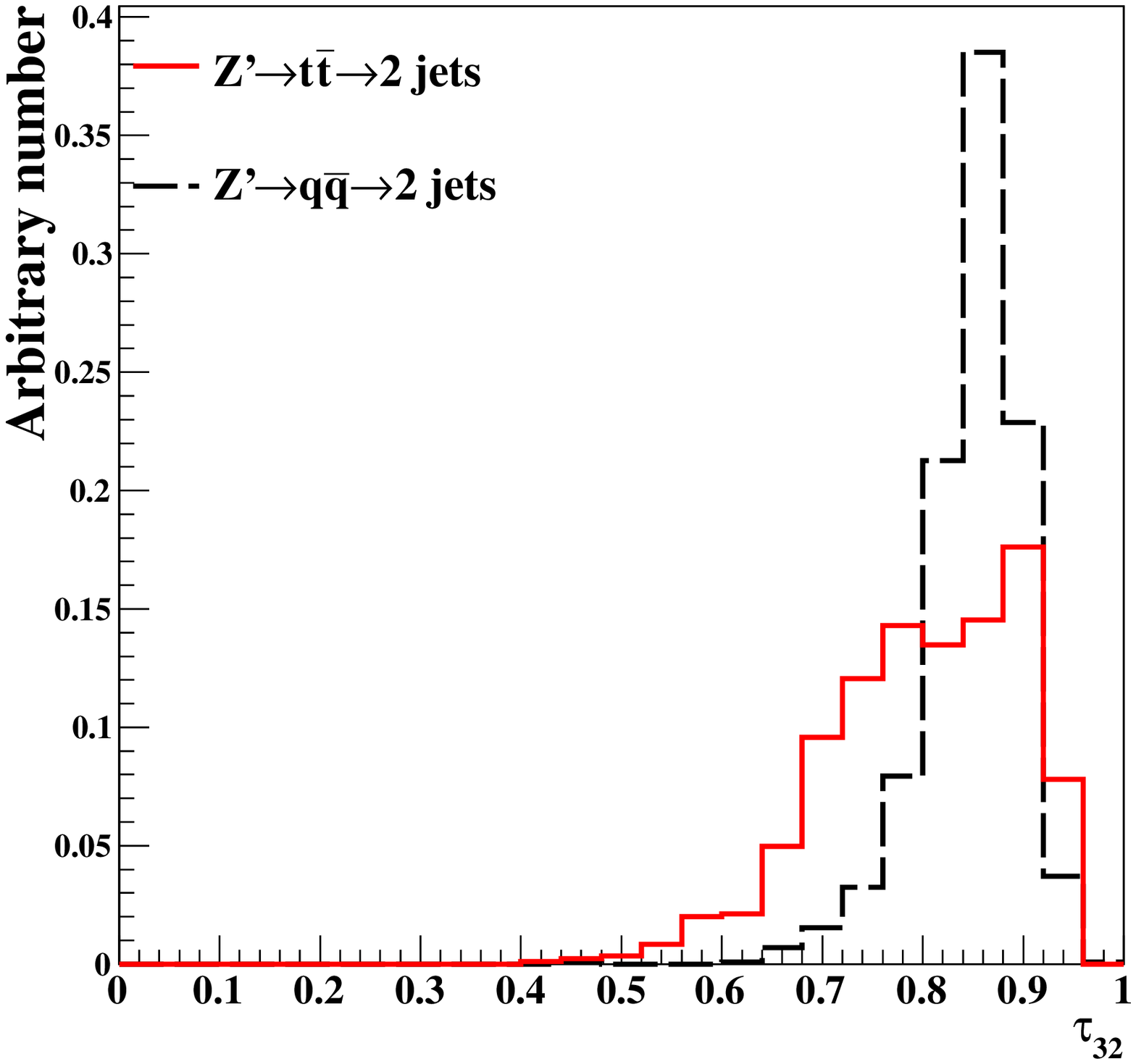}
   }
   \subfigure[5$\times$5 cm$^2$] {
   \includegraphics[width=0.3\textwidth]{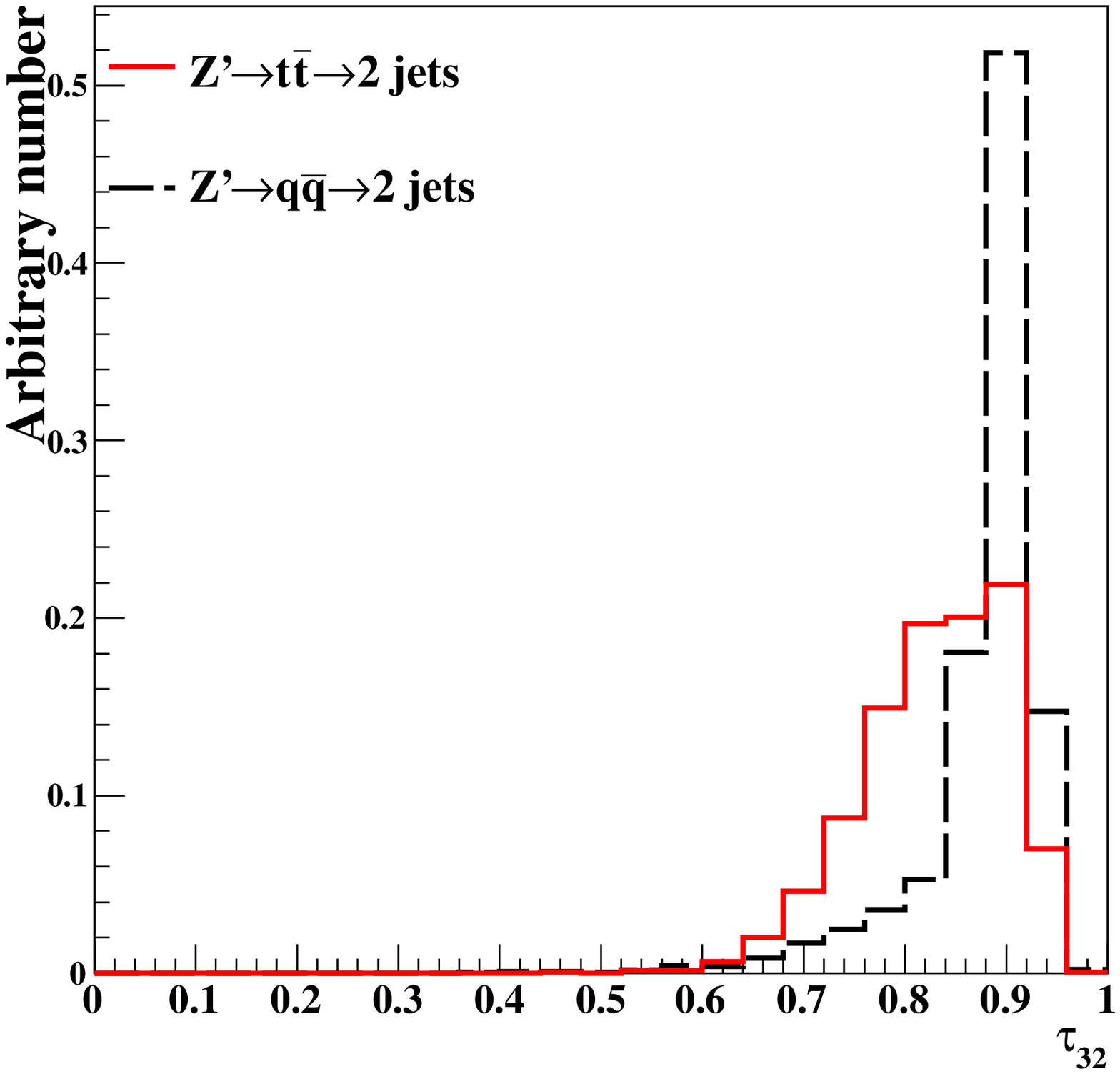}
   }
   \subfigure[1$\times$1 cm$^2$] {
   \includegraphics[width=0.3\textwidth]{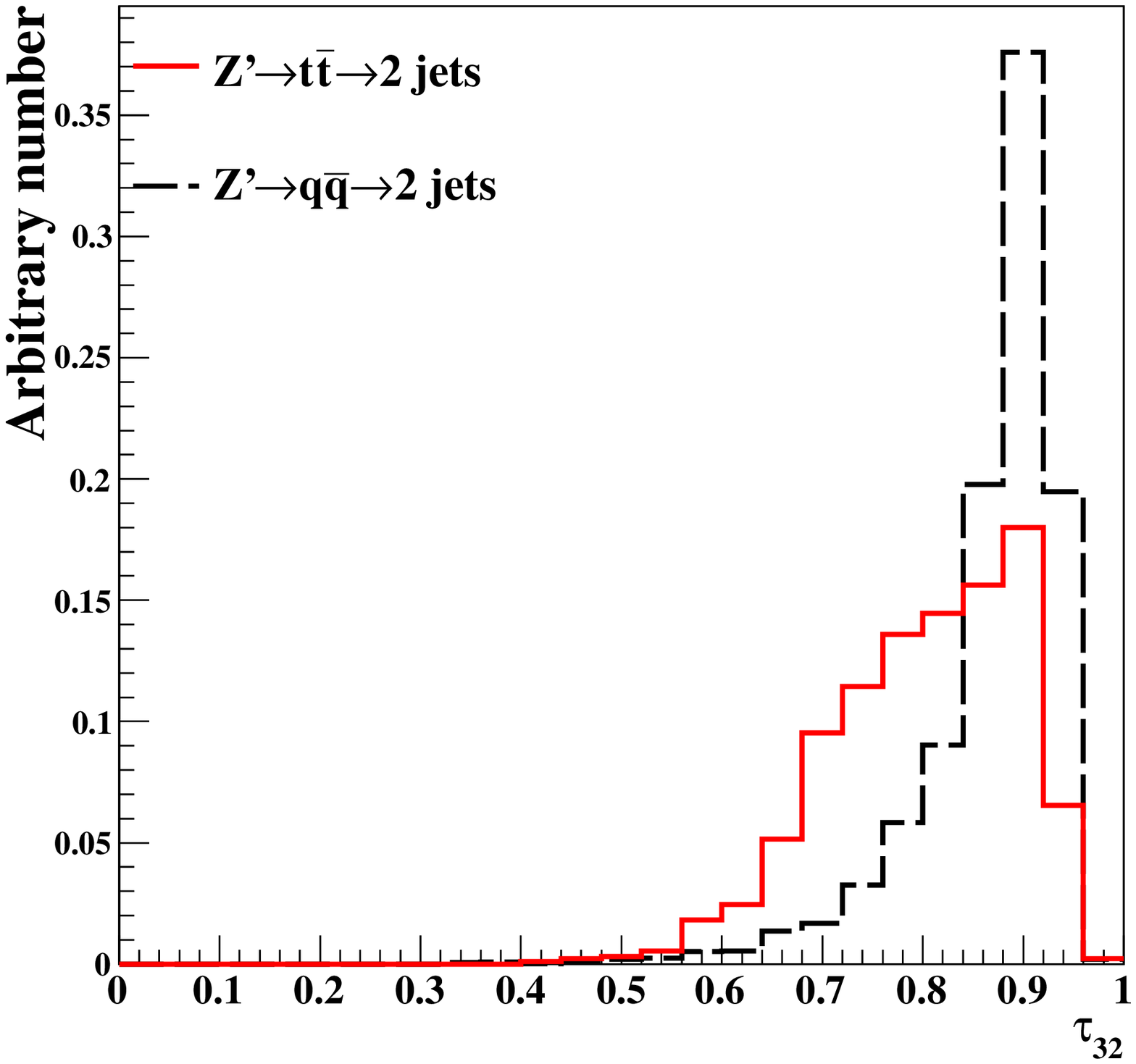}
   }
\end{center}
\caption{Distributions of $\tau_{32}$ for $M (Z') =20$~TeV for different 
detector granularities. Cell sizes of 20~$\times$~20, 5~$\times$~5, and 1~$\times$~1~cm$^2$ 
are shown here.
\label{fig:Rawhit_05GeV_tau32_Dis}
}
\end{figure}
\begin{figure}
\begin{center}
   \subfigure[$M(Z') = 5$~TeV] {
   \includegraphics[width=0.43\textwidth]{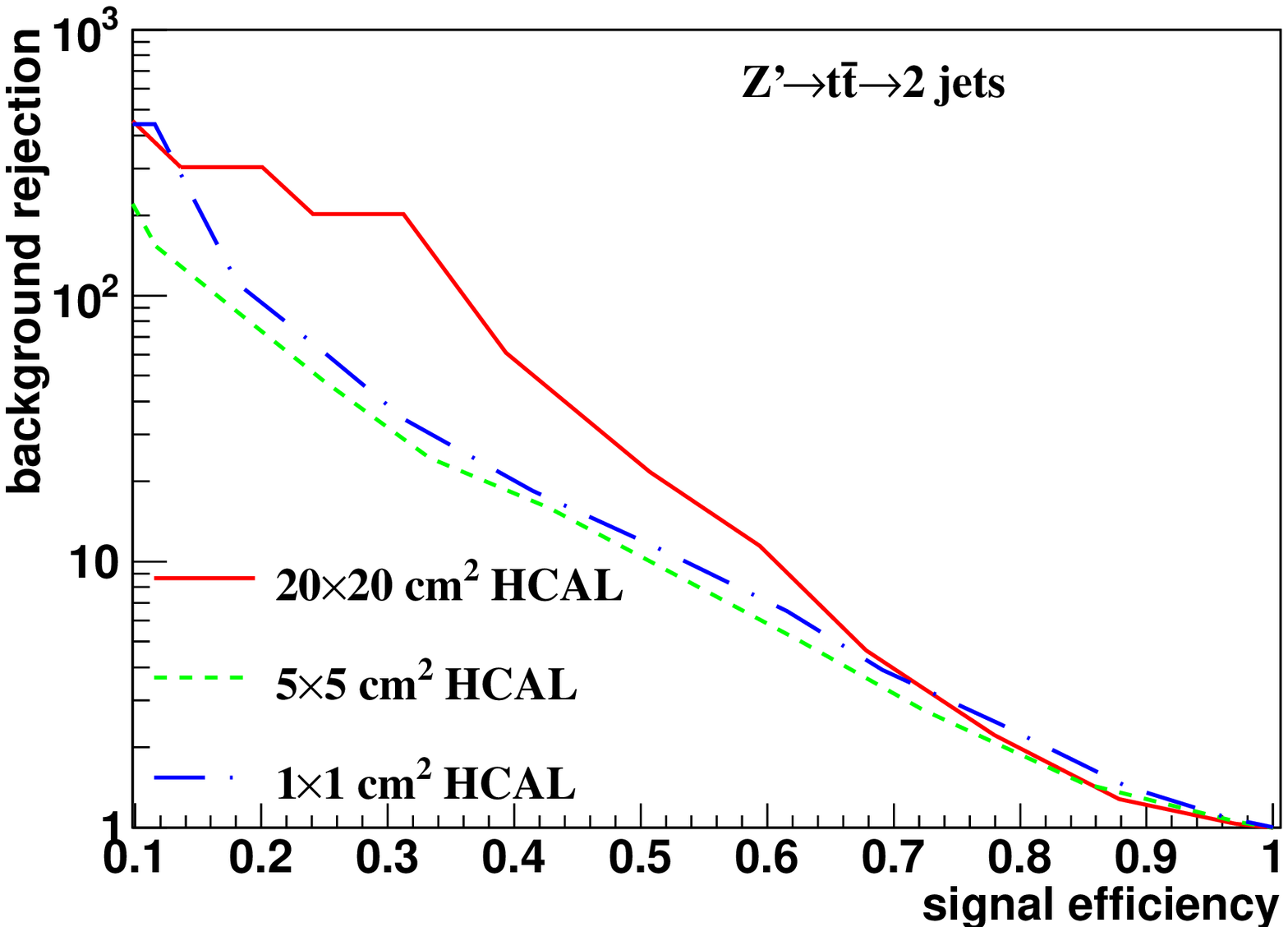}\hfill
   }
   \subfigure[$M(Z')=10$~TeV] {
   \includegraphics[width=0.43\textwidth]{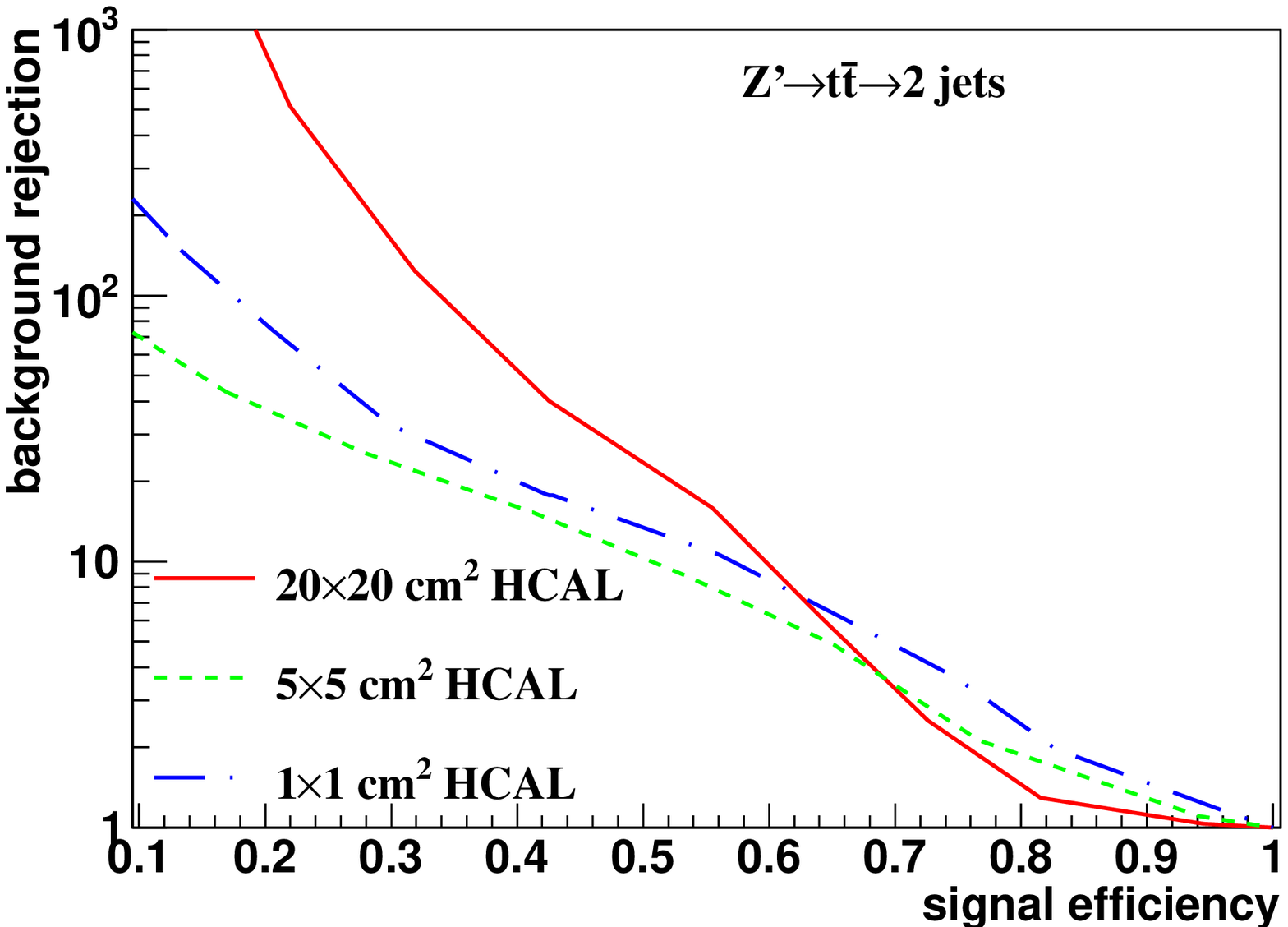}
   }
   \subfigure[$M(Z')=20$~TeV] {
   \includegraphics[width=0.43\textwidth]{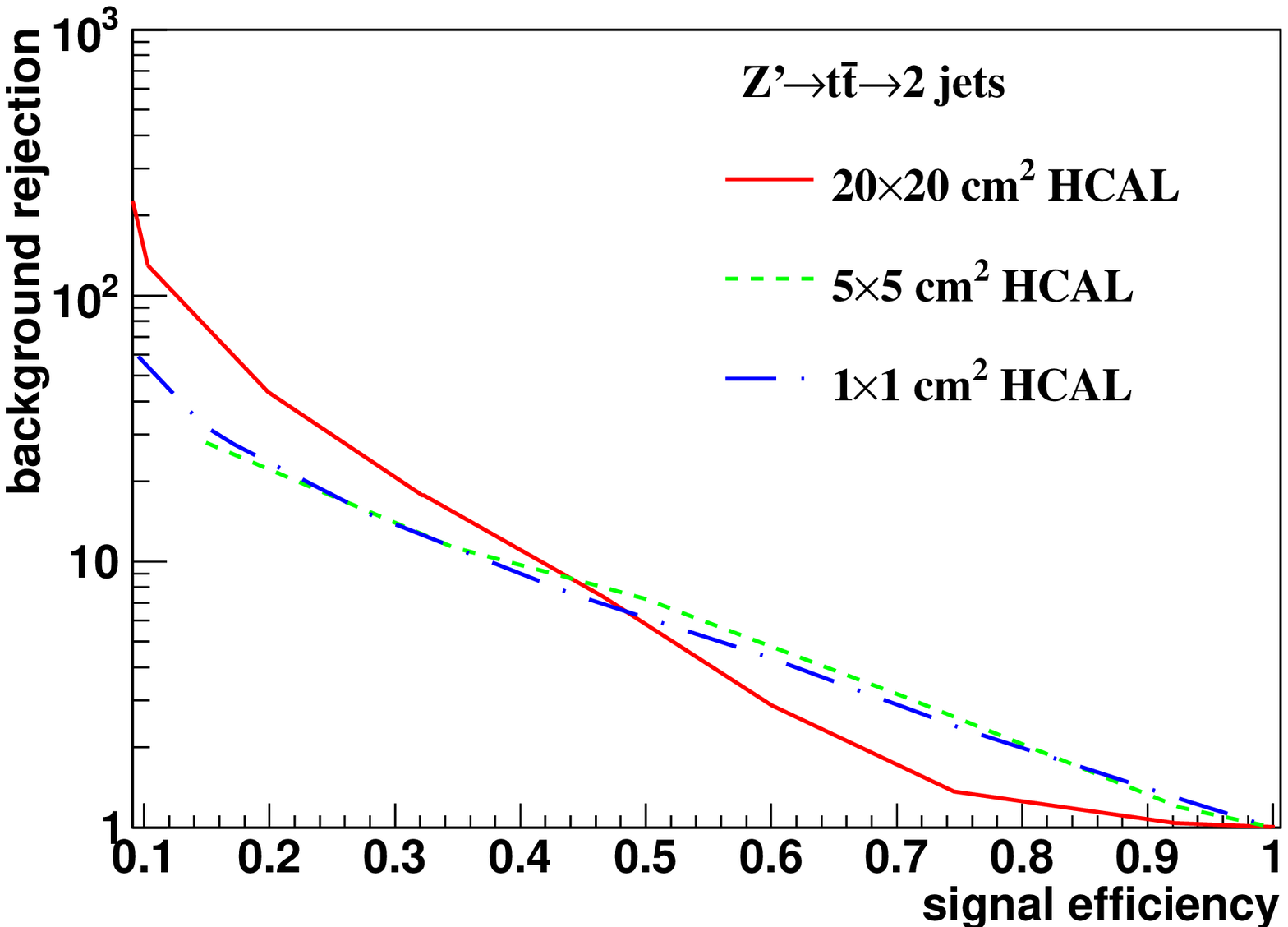}
   }
   \subfigure[$M(Z')=40$~TeV] {
   \includegraphics[width=0.43\textwidth]{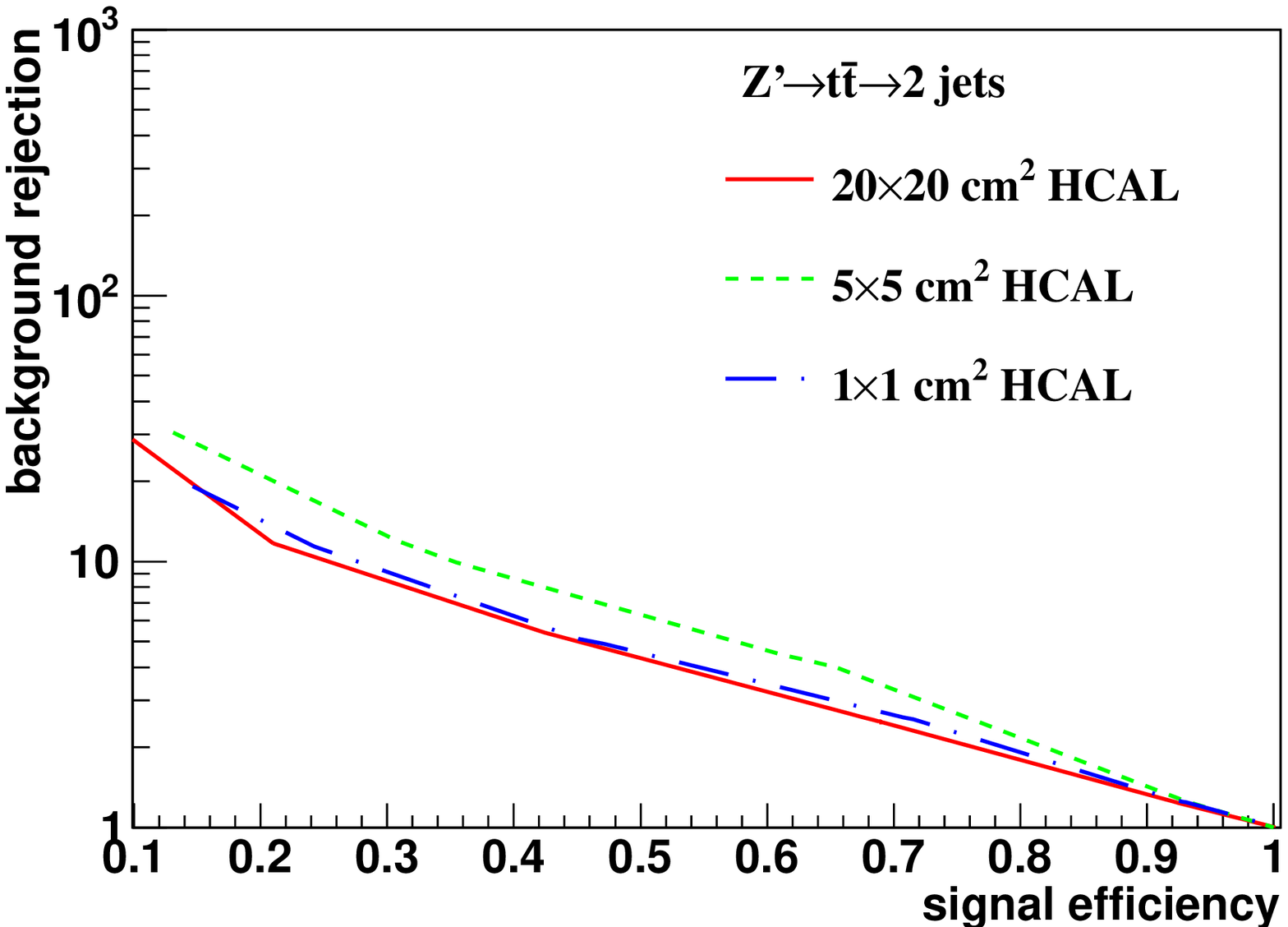}
   }
\end{center}
\caption{Signal efficiency versus background rejection rate using $\tau_{32}$. 
Resonance masses of (a) 5~TeV, (b) 10~TeV, (c) 20~TeV and (d) 40~TeV are shown 
here. In each figure, the three ROC curves correspond to different HCAL cell 
sizes.
\label{fig:Rawhit_05GeV_tau32_ROC}
}
\end{figure}

Figures~\ref{fig:Rawhit_05GeV_tau21_Dis} and~\ref{fig:Rawhit_05GeV_tau32_Dis} 
show the distributions of $\tau_{21}$ and $\tau_{32}$ for $M(Z')=20$~TeV 
after applying the requirement on the soft drop mass. The signals considered are 
the $Z'\rightarrow WW$ (for $\tau_{21}$) and 
$Z' \rightarrow t\bar{t}$ (for $\tau_{32}$) processes. 
Figures~\ref{fig:Rawhit_05GeV_tau21_ROC} and~\ref{fig:Rawhit_05GeV_tau32_ROC} 
present the ROC curves from different detector cell sizes and resonance masses, 
respectively.  The ROC curves are computed with finely-binned histograms; the latter are rebinned coarsely for display purpose only.

We find that the performance of the $1\times1~\mathrm{cm}^2$ and $5\times5~\mathrm{cm}^2$ cell sizes is similar for both the $ \tau_{21} $ and the $ \tau_{32} $ variables, for
 all resonance masses in the 5-40~TeV range. These smaller cell sizes yield a higher performance than 
 the $20\times20~\mathrm{cm}^2$ cell size when using the $ \tau_{21} $ variable, for resonance masses of 5, 10 and 20 TeV in the $WW$ final state. In the case of the $ \tau_{32} $ variable, 
  the results are ambiguous, as the $20\times20~\mathrm{cm}^2$ cell size is more (less) performant for low (high) efficiency selection criteria. 


\subsection{Energy correlation function \label{sec:ecf}}
The energy correlation function (ECF)~\cite{Larkoski:2013eya} is defined as follows: 
\begin{equation} \label{eq:ECF_Modified}
ECF(N,\beta)=\sum_{i_{1}<i_{2}<....<i_{N}\in J} \left(\prod_{a=1}^{N}p_{\mathrm{T}ia}\right)\left(\prod_{b=1}^{N-1}\prod_{c=b+1}^{N} R_{i_{b}i_{c}}\right)^{\beta},
\end{equation}
where the sum is over all constituents in jet $J$, $\pt$ is the transverse 
momentum of each constituent, and $R_{mn}$ is the distance between two constituents $m$ and $n$ in the $y$-$\phi$ plane.  
In order to use a dimensionless variable, a parameter $r_{N}$ is defined:
\begin{equation} \label{eq:ECF_ratio}
r_{N}^{(\beta)}\equiv\frac{ECF(N+1,\beta)}{ECF(N,\beta)}.
\end{equation}

\begin{figure}
\centering
\begin{center}
   \subfigure[20~$\times$~20~cm$^2$] {
   \centering
   \includegraphics[width=0.3\textwidth]{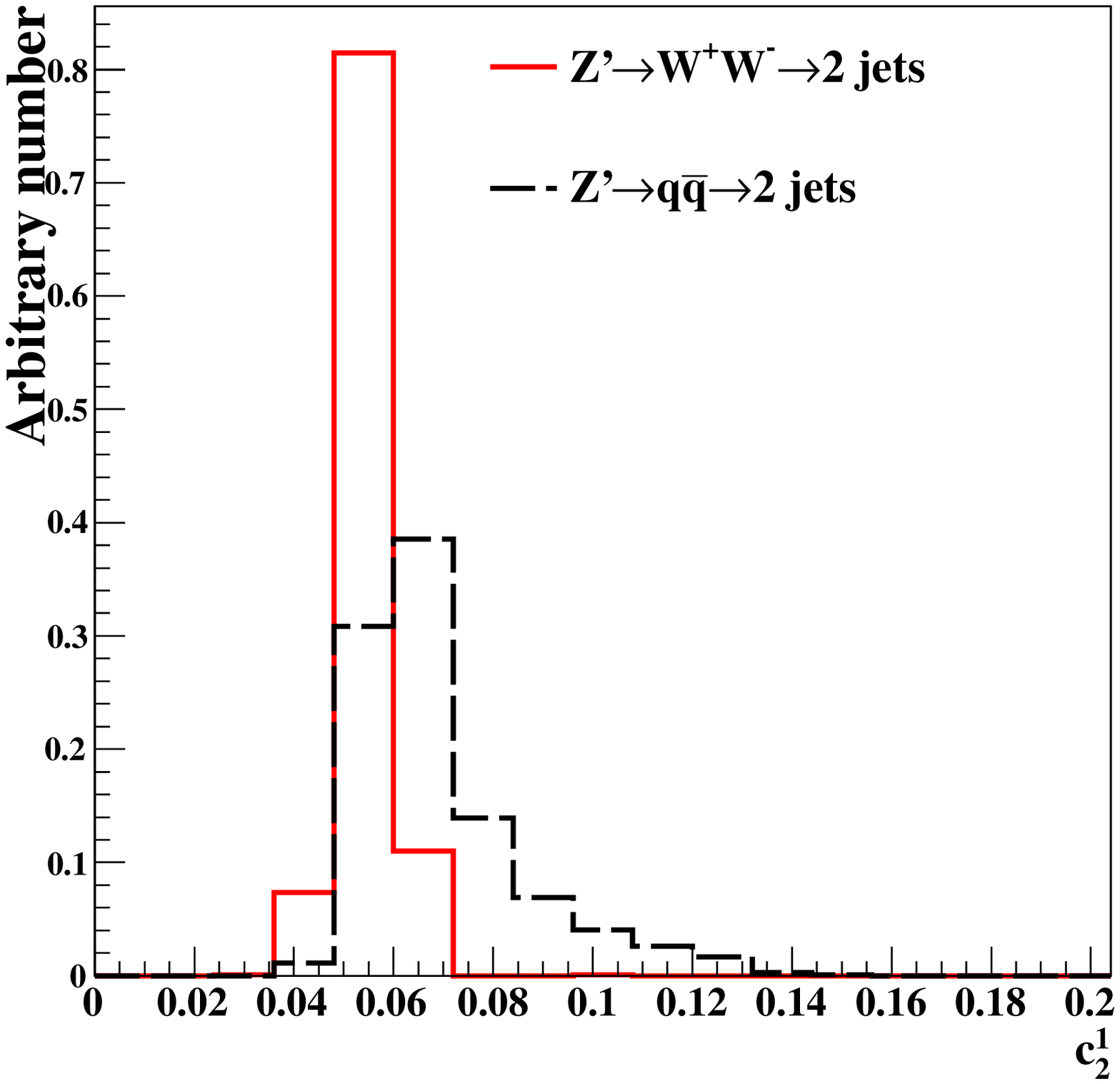}
   }
   \subfigure[5~$\times$~5~cm$^2$] {
   \centering
   \includegraphics[width=0.3\textwidth]{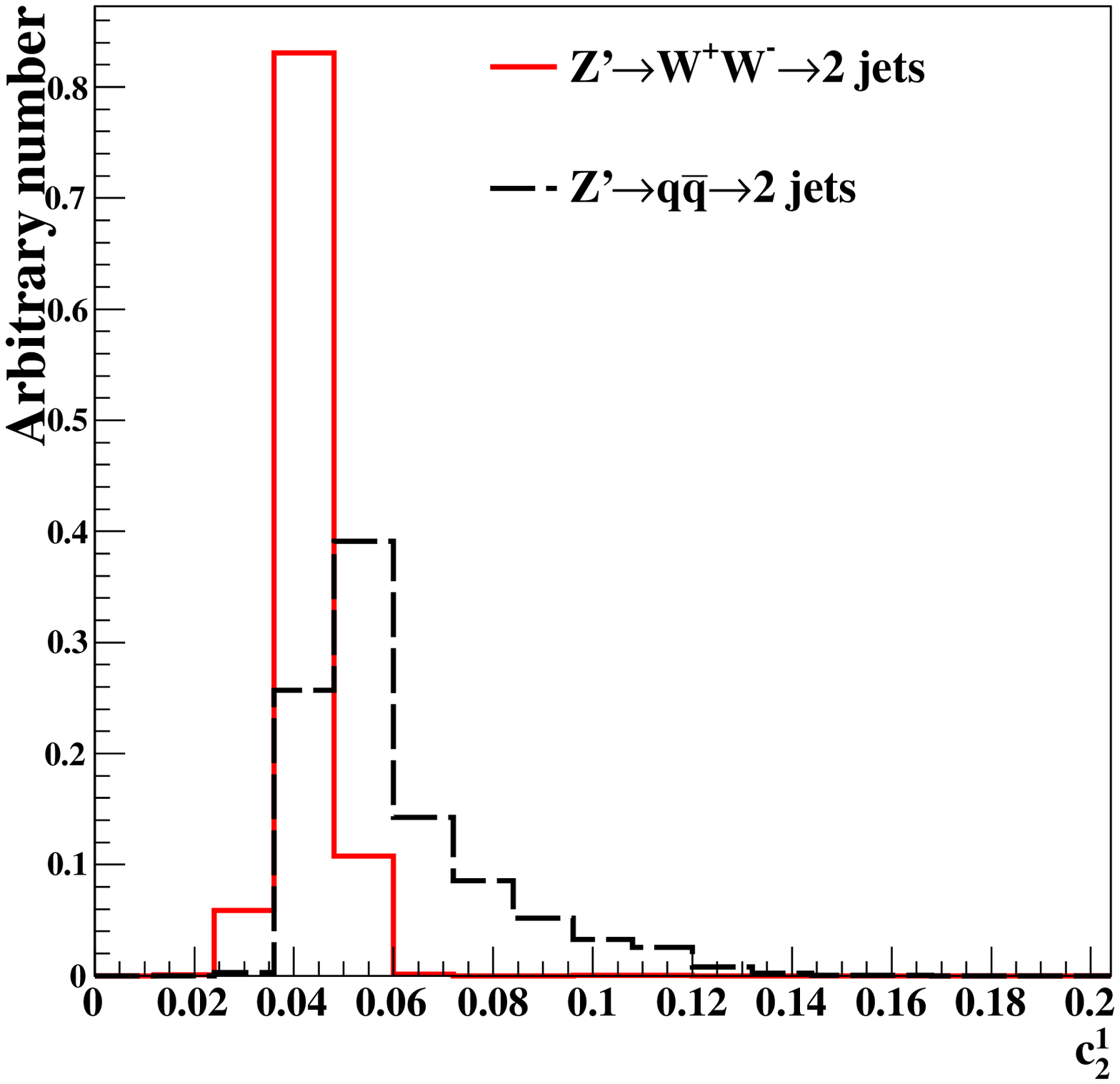}
   }
   \subfigure[1~$\times$~1~cm$^2$] {
   \centering
   \includegraphics[width=0.3\textwidth]{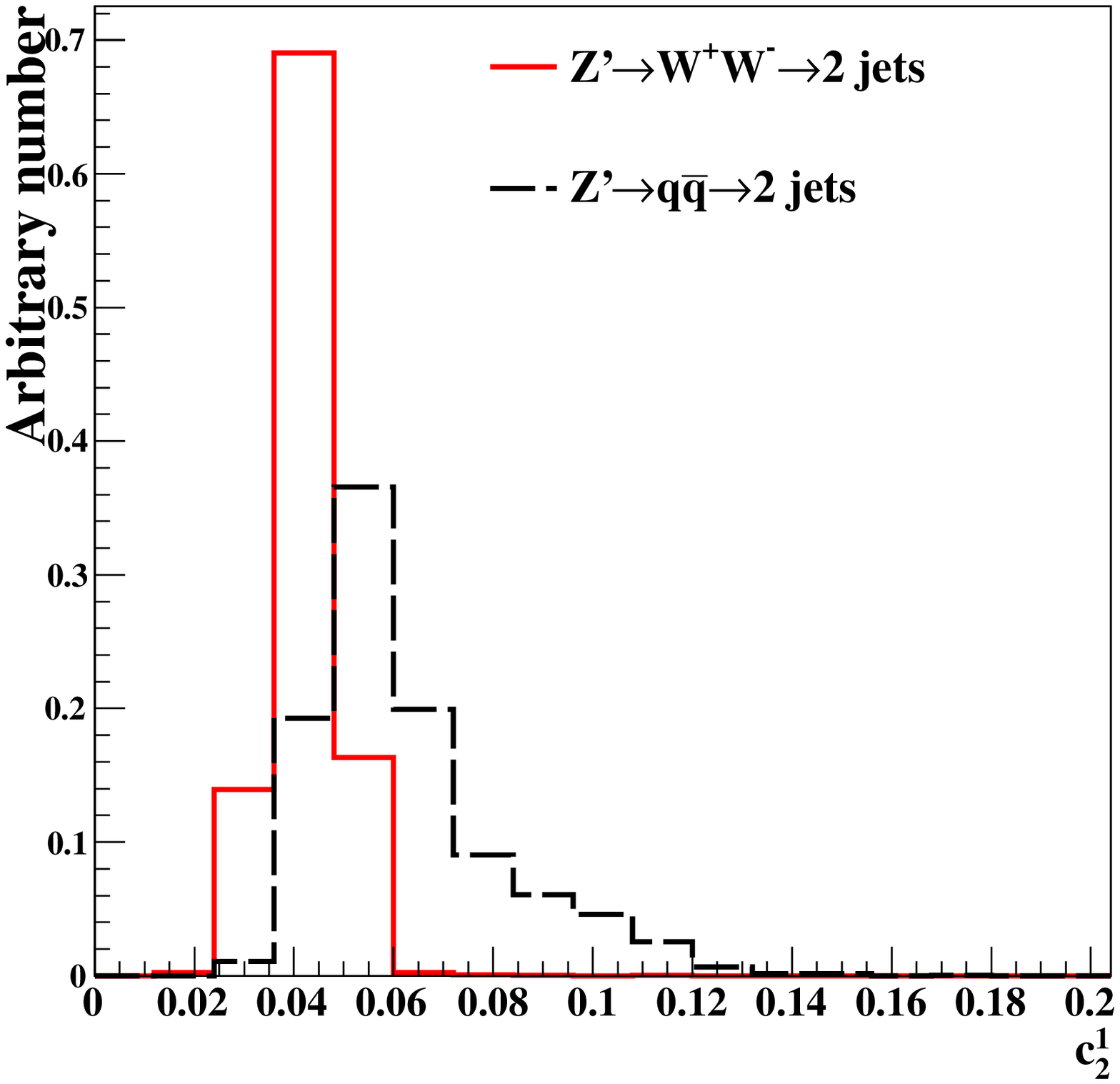}
   }
\end{center}
\caption{Distributions of $C_2^1$ with $M(Z')= 20$~TeV for different 
detector granularities. Cell sizes of 20~$\times$~20, 5~$\times$~5, and 1~$\times$~1~cm$^2$ 
are shown here.}
\label{fig:Rawhit_05GeV_c2b1_Dis}
\end{figure}

\begin{figure}
\begin{center}
   \subfigure[$M(Z')=5$~TeV] {
   \includegraphics[width=0.43\textwidth]{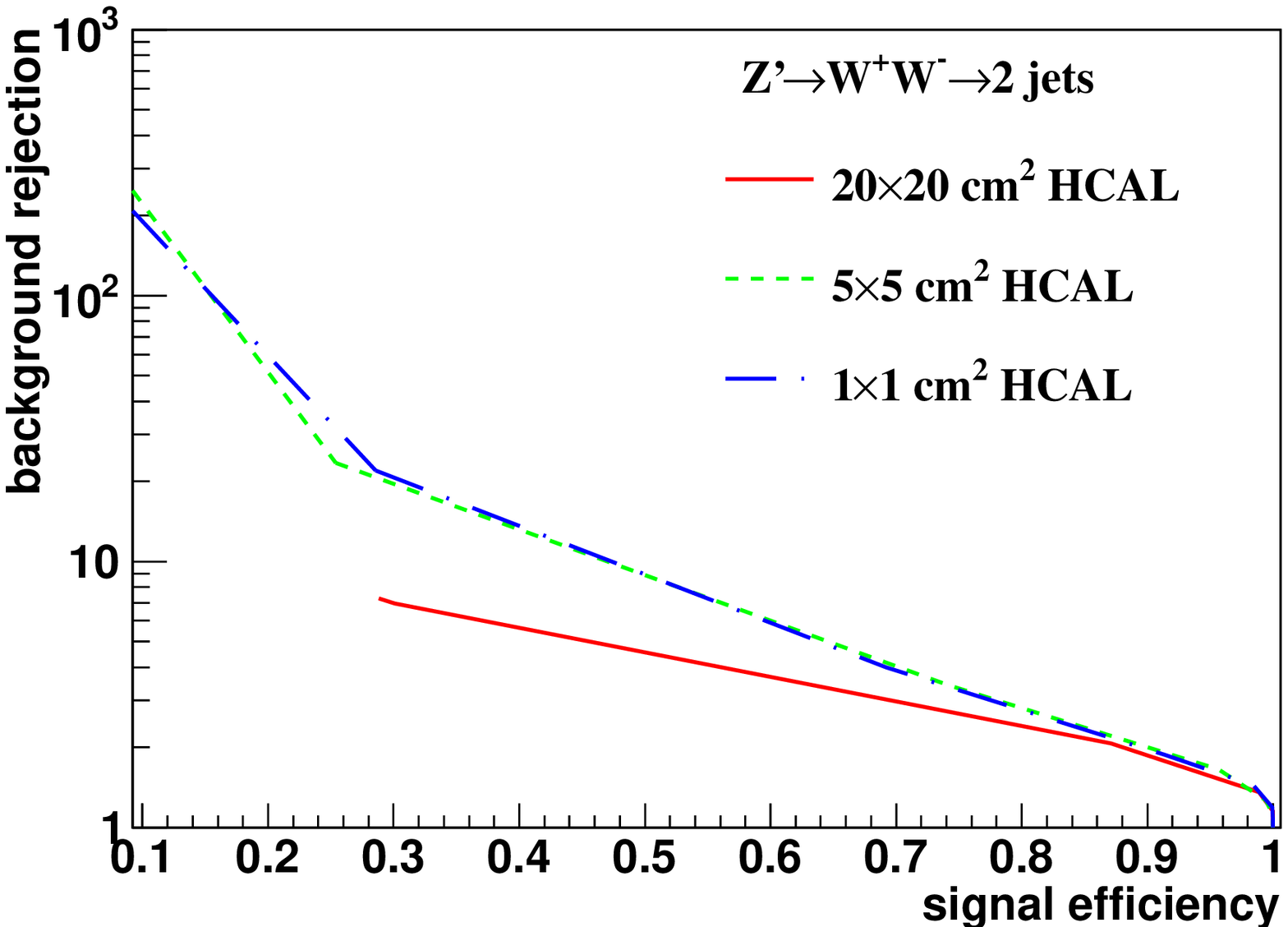}\hfill
   }
   \subfigure[$M(Z')=10$~TeV] {
   \includegraphics[width=0.43\textwidth]{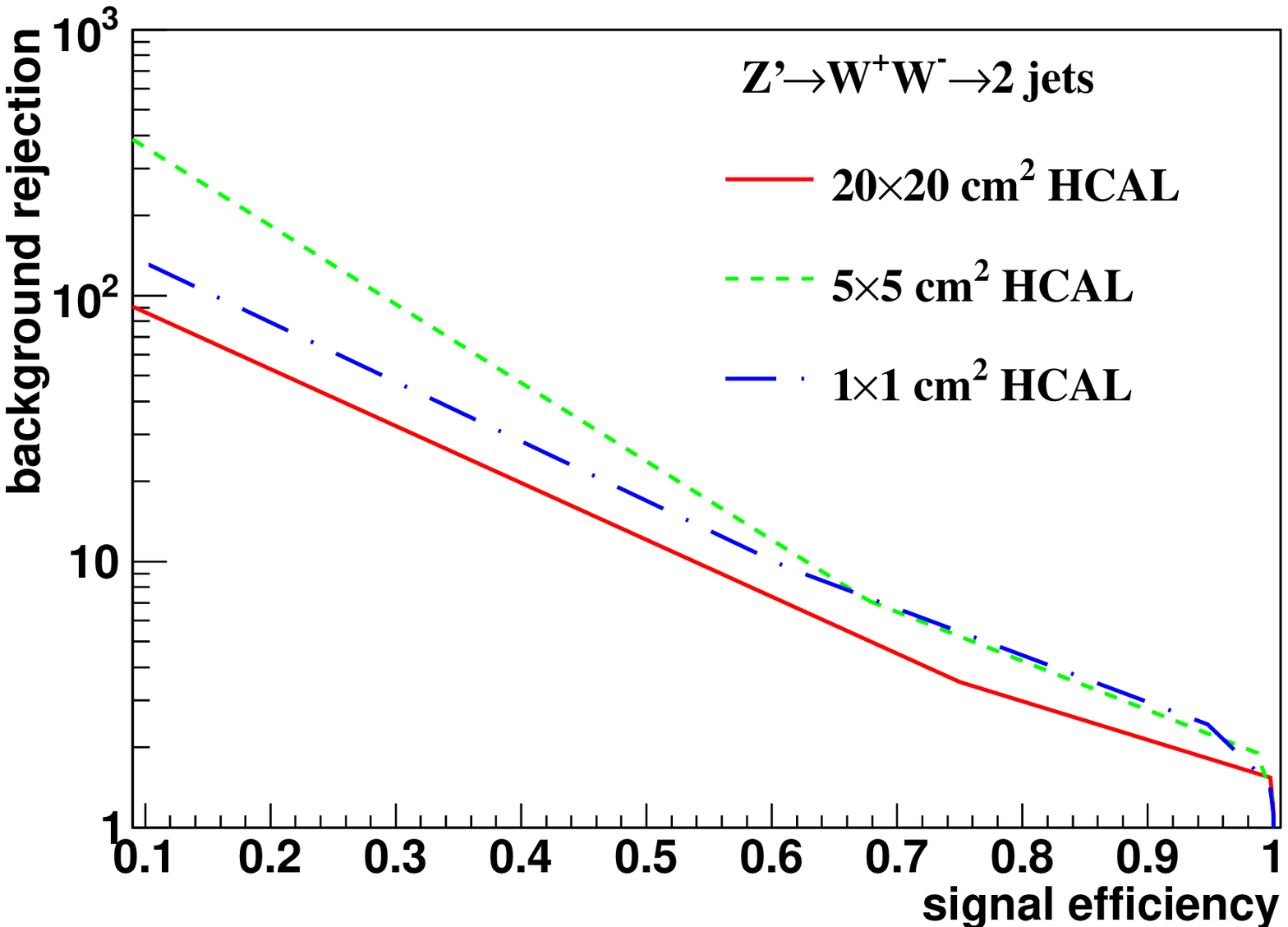}
   }
   \subfigure[$M(Z')=20$~TeV] {
   \includegraphics[width=0.43\textwidth]{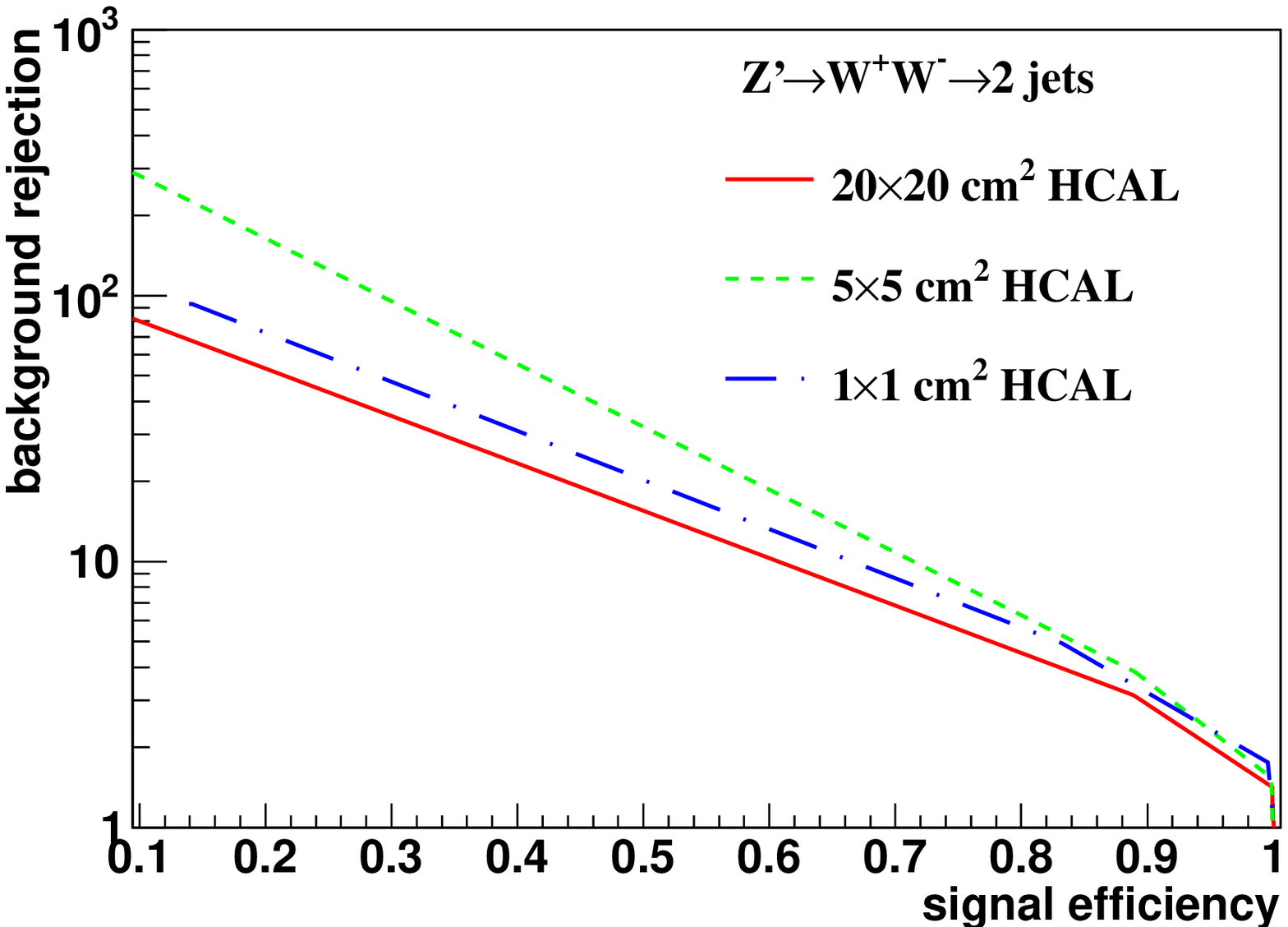}
   }
   \subfigure[$M(Z')=40$~TeV] {
   \includegraphics[width=0.43\textwidth]{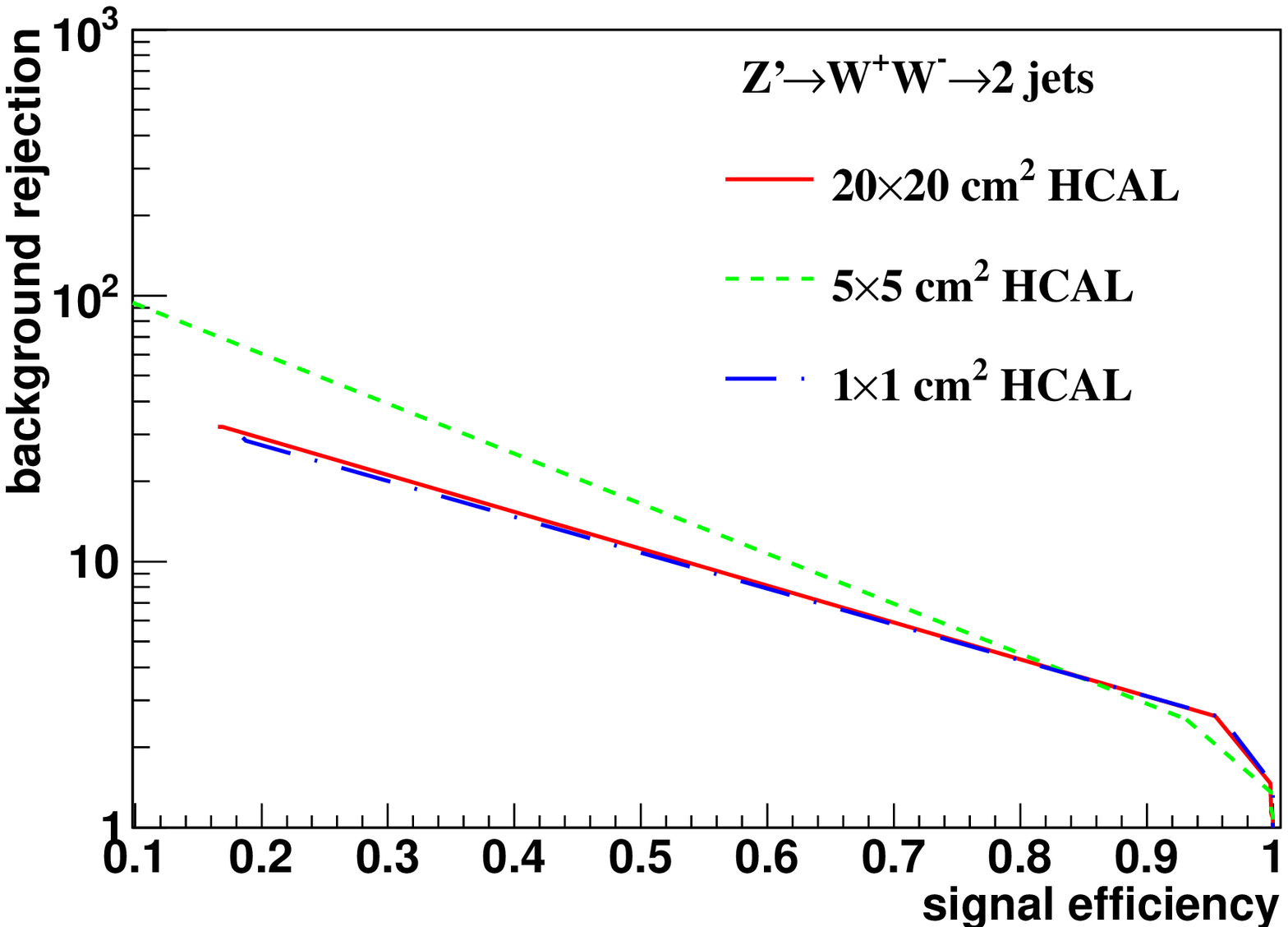}
   }
\end{center}
\caption{Signal efficiency versus background rejection rate using $C_{2}^{1}$. 
The resonance masses of (a) 5~TeV, (b) 10~TeV, (c) 20~TeV, and (d) 40~TeV are shown 
here. In each figure, the three ROC curves correspond to different detector 
sizes.}
\label{fig:Rawhit_05GeV_c2b1_ROC}
\end{figure}

The idea of $r_N$ comes from $N$-subjettiness $\tau_N$. Both $r_N$ and $\tau_N$ 
are linear in the energy of the soft radiation for a system of $N$ partons  accompanied 
by  soft radiation. In general, if the system has $N$ subjets, $ECF(N+1,\beta)$ 
should be significantly smaller than $ECF(N,\beta)$. Therefore, we can use this
 feature to distinguish jets with different numbers  of subjets. 
As in Sect.~\ref{sec:nsub}, the ratio $r_N/r_{N-1}$, denoted by $C_N$, 
(double-ratios of ECFs) is used to study the detector performance: 
\begin{equation}
C_{N}^{(\beta)}\equiv\frac{r_{N}^{(\beta)}}{r_{N-1}^{(\beta)}}=\frac{ECF(N-1,\beta)ECF(N+1,\beta)}{ECF(N,\beta)^2}.
\end{equation}
In our analysis, we set $N=2$ and $\beta=1$ ($C_2^1$).

Figure~\ref{fig:Rawhit_05GeV_c2b1_Dis} presents the histograms of $C_{2}^{1}$ 
with $M(Z')=20$~TeV after making the requirement on the soft drop mass. 
The signal considered is the $Z' \rightarrow WW$ process. 
Figure~\ref{fig:Rawhit_05GeV_c2b1_ROC} shows the ROC curves from different 
detector cell sizes for each resonance mass. One can see that 
the $5 \times 5$~$\mathrm{cm}^2$ cell size improves upon the $20 \times 20$~$\mathrm{cm}^2$ cell size, and either matches or improves upon the $1 \times 1$~$\mathrm{cm}^2$ cell size,
   for all resonance masses. 

\section{Conclusions}
The studies presented in this paper show that the reconstruction of jet substructure 
variables for future particle colliders will benefit from small cell sizes of the hadronic calorimeters. 
This conclusion was obtained using the realistic \GEANTfour\ simulation of calorimeter response combined with reconstruction of 
calorimeter clusters used as inputs for jet reconstruction. 
Hadronic calorimeters that use the cell sizes of 20~$\times $~20~cm$^2$ ($\Delta \eta \times \Delta \phi = 0.087\times 0.087$) 
are least performant for almost every 
substructure variable considered in this analysis, for jet transverse momenta between 2.5 and 10~TeV. 
Such cell sizes are similar to 
those used for the ATLAS and CMS detectors at the LHC. 
In terms of reconstruction of physics-motivated quantities  
used for jet substructure studies, the  performance 
of a  hadronic calorimeter  with 
$\Delta \eta \times \Delta \phi = 0.022\times0.022$ ($5 \times 5$~$\mathrm{cm}^2$ cell size) is, in most cases,
better than for a detector with  $0.087\times 0.087$ cells.

Thus this study confirms the  HCAL geometry of the SiFCC detector~\cite{Chekanov:2016ppq},
with the $\Delta \eta \times \Delta \phi = 0.022\times0.022$ HCAL cells.
It also confirms the HCAL design of the baseline FCC-hh~\cite{fcc1,fcc2} detector with
$\Delta \eta \times \Delta \phi = 0.025\times0.025$ HCAL cells.

It is interesting to note that,  for very boosted jets with transverse momenta close to 20~TeV, further decrease of cell size to $\Delta \eta \times \Delta \phi = 0.0043\times0.0043$ did not 
 definitively show a further improvement in performance. It should be noted that the clustering algorithm used for reconstruction
of clusters from the calorimeter hits created by the Geant4 simulation 
has been tuned to reflect the small cell sizes of the SiD calorimeter. 
However, the energy range of the hits for the tens-of-TeV jets studied 
in this paper may not be optimal for this algorithm. Therefore, this 
result needs to be understood in terms of various types of simulations 
and different options for reconstruction of the calorimeter clusters. 
The effect of changing parameters of this clustering algorithm will be 
our essential step for future studies. Even more, the complex 
circumstance of adding the pileup could be studied to understand the 
realistic data-taking  conditions for future 100 TeV colliders.

\section*{Acknowledgements}
This research was performed using resources provided by the Open Science Grid,
which is supported by the National Science Foundation and the U.S. Department of Energy Office of Science. 
We gratefully acknowledge the computing resources provided on Blues, 
a high-performance computing cluster operated by the Laboratory Computing Resource Center at Argonne National Laboratory.
Argonne National Laboratory is supported by the U.S. Department of Energy, Office of Science, Office of High Energy Physics  under contract DE-AC02-06CH11357. The Fermi National Accelerator Laboratory (Fermilab) is operated by Fermi Research Alliance, LLC under Contract No. DE-AC02-07CH11359 with the United States Department of Energy.
This research used resources of the National Energy Research Scientific  Computing Center (NERSC), a U.S. Department of Energy Office of Science  User Facility operated under Contract No. DE-AC02-05CH11231. We gratefully acknowledge Ministry of Science and Technology and Ministry of Education in Taiwan.

\newpage
\section*{References}

\bibliographystyle{elsarticle-num}
\def\bibname{\Large\bf References}
\def\refname{\Large\bf References}
\pagestyle{plain}
\bibliography{biblio}

\end{document}